\documentclass[superscriptaddress,twocolumn]{revtex4-2}
\pdfoutput=1
\usepackage{xcolor}
\usepackage{hyperref}
\hypersetup{
    colorlinks=true,       
    linkbordercolor=red,
    citebordercolor=green,
    urlbordercolor=cyan, 
}

\usepackage{amsmath,amssymb}
\usepackage{graphicx}
\usepackage{geometry}
\usepackage{array}
\usepackage{enumitem}
\setlist{noitemsep}

\usepackage{graphbox}

\renewcommand{\vec}[1]{\boldsymbol{\mathbf{#1}}}

\renewcommand{\L}{\mathcal{L}}
\renewcommand{\O}{\mathcal{O}}
\newcommand{\p}{\partial}

\begin{document}
\title{How production networks amplify economic growth}

\author{James McNerney}
\email{james\_mcnerney@hks.harvard.edu}
\affiliation{Institute for Data, Systems, and Society, Massachusetts Institute of Technology, Cambridge, MA 02139, USA}
\affiliation{Center for International Development, Kennedy School of Government, Harvard University, Cambridge, MA 02139, USA}

\author{Charles Savoie}
\affiliation{Institute for New Economic Thinking at the Oxford Martin School, University of Oxford, Oxford OX2 6ED, UK}

\author{Francesco Caravelli}
\affiliation{Center for Nonlinear Studies, Los Alamos National Laboratory, Los Alamos, NM 87545, USA}

\author{Vasco M. Carvalho}
\affiliation{University of Cambridge, Cambridge CB2 1TN, UK}
\affiliation{The Cambridge-INET Institute, Faculty of Economics, Cambridge CB3 9DD}
\affiliation{The Alan Turing Institute, London NW1 2DB, UK}
\affiliation{Center for Economic and Policy Research, London EC1V 0DX, UK}

\author{J. Doyne Farmer}
\affiliation{Institute for New Economic Thinking at the Oxford Martin School, University of Oxford, Oxford OX2 6ED, UK}
\affiliation{Santa Fe Institute, Santa Fe, NM 87501, USA}

\begin{abstract}
\noindent
Technological improvement is the most important cause of long-term economic growth.  We study the effects of technology improvement in the setting of a production network, in which each producer buys input goods and converts them to other goods, selling the product to households or other producers.  We show how this network amplifies the effects of technological improvements as they propagate along chains of production.  Longer production chains for an industry bias it towards faster price reduction, and longer production chains for a country bias it towards faster GDP growth. These predictions are in good agreement with data and improve with the passage of time, demonstrating a key influence of production chains in price change and output growth over the long term.
\end{abstract}

\maketitle

\section{Introduction}
Economic output is the result of a network of industries that buy goods from one another, convert them to new goods, and sell the output to households or other industries.  Since work by Leontief \cite{Leontief1936,Leontief1986} increasingly rich data have become available to study these networks, and research has revealed characteristics that hold across diverse economies, such as their link weight and industry size distributions \cite{Xu2011,Atalay2011,Jones2011,McNerney2013,Carvalho2008}, community structure \cite{McNerney2013}, and path-length properties \cite{Antras2012}. Economies typically have a few highly central industries that are strong suppliers to the rest of the network \cite{Carvalho2008,Bloechl2011,Jones2011}, a feature that has been incorporated into models where short-term fluctuations in output are generated by shocks to individual industries \cite{Horvath2000,Carvalho2008,Foerster2011,Gabaix2011,Acemoglu2012,Alatriste2014}. 

In this paper we study how the network structure of production affects an economy's long-term growth.   Our argument proceeds in two steps. First we show that the rate of change of an industry's price is a function of its position in the production network. This happens because productivity improvements accumulate along supply chains.  As a result, industries that rely on longer supply chains experience stronger price declines than others. Second, we show how this observation can help explain cross-country differences in economic growth. Because an industry's position in the production network and the industrial composition of a country are slow-moving variables, aggregate growth can be predicted from the structure of a country's production network. Intuitively, countries whose final demand relies relatively more on industries with longer supply chains should grow more quickly.  We find that detailed observations across industries and countries are consistent with both predictions, and help explain why some countries grow faster than others.

A large literature stresses that technological improvements are the main driver of long-term growth \cite{Solow1956,Jorgenson1987}.  Over time, improvements to productivity -- the amount of output that can be made with a given amount of inputs -- significantly alter prices and production flows in an economy. Classic work by Domar \cite{Domar1961} and Hulten \cite{Hulten1978} showed that as an industry's productivity improves, the presence of intermediate input trade -- i.e. goods and services flowing through a production network -- amplifies the aggregate benefit for an economy. Productivity growth in an industry not only reduces the price and raises the output of its goods, but some of this output can be used as inputs by other industries, enabling further increases in output, and so on.  

However, other predictions about the role of production networks have escaped notice. Using a simple model, we show that as the effects of productivity changes propagate, each industry's price declines at a rate that depends on its network position.  An industry’s price should fall in proportion to its output multiplier, a centrality metric that can be understood as the average length of an industry’s production chains where every production path is weighted by the relative size of the expenditures it represents.  An industry benefits from both its own productivity growth and from the accumulation of productivity improvements in its upstream suppliers. As a result, the longer its chains of production, the faster its expected rate of price reduction.

The connection to output multipliers is significant because these variables convey structural information about an economy. Particular industries, especially in manufacturing, are known to have larger output multipliers, while others, especially in services, tend to have smaller ones \cite{Park1989}.  This is largely because manufacturing typically devotes a greater fraction of expenses to intermediate goods and a smaller fraction to labor than services do.  Output multipliers can change with time as prices and technology evolve, and as industries substitute some inputs for others.  But output multipliers change much more slowly than other key variables in our analysis, in particular productivity growth rates and price changes (Supplementary Information).  This conforms with the idea that output multipliers capture a hard-wired aspect of production.  A producer of fabricated metal parts, for example, will largely remain the supplier to an automobile maker, and not the other way around, even if the detailed pattern of input flows changes with time. 

The relative persistence of output multipliers means that the predicted price changes noted above should correlate with enduring features of network structure.  In particular, it suggests that output multipliers should be able to predict industry price changes over long horizons.  The mechanism we study (the passing of the benefits of productivity improvement along production chains) carries other implications as well.  We derive a number of predictions that are implied by production network models, including predictions for the cross-industry variation of price changes around the expected value.

We compare these predictions with data on output multipliers and prices from 35 industry categories and 40 countries (1400 industries in total) from the World Input Output Database (WIOD) \cite{Timmer2015}. First we verify the basic mechanism of the model, observing the price reduction that industries inherit through reductions in the prices of inputs.  We document a remarkable fact -- not only do inherited price reductions contribute significantly, but for the majority of industries, inherited price reductions exceed those originating locally in the network from the productivity growth of an industry. For most industries the better part of the explanation of price reduction lies in processes happening outside the industry, in other parts of the network.

We then test predictions related to output multipliers. We do our exercises under the assumption of constant output multipliers, holding values fixed in an initial year, and studying subsequent price changes. The data agree with predictions for both the expectation value of price changes and cross-industry variation around it.  This variation shrinks with time, causing  predictions based on the expected value to become more accurate and making the output multiplier more relevant as one looks further into the future.  This means that our results also enable a simple method to  forecast changes in prices.

We then explore macro-level implications of the network's influence on prices.  We show that a consequence of the relationship between prices and output multipliers is that a country's GDP is predicted to grow at a rate in proportion to the average of its industries' output multipliers. Intuitively, falling prices translate into economic growth to the extent that economies enjoy price reductions by consuming more.  Production network models thus predict that, all else equal, a country's rate of growth will be higher the longer its production chains are.  To test the macro-level predictions we again turn to WIOD data.  We show that a country's average output multiplier is, like industry-level output multipliers, a slow-moving variable.  This is not surprising, as episodes of structural transformation and large-scale reorganization of production play out over many years. This in turn implies that initial cross-country variations in average output multipliers can be used to predict cross-country differences in future growth.

Taken together, the results suggest that the network structure of production plays a major role in the long-term evolution of economies.  We relate the results to two longstanding observations.  First, a well-known observation about technology evolution is that while most industries gain in productivity over long periods, some industries, especially manufacturing, improve more quickly than others \cite{Baumol1967}. Over time, this difference causes price increases in slower-improving industries, an effect known as Baumol's cost disease. The findings here provide a reason why some industries would sustain faster improvement than others over long periods. Second, the results suggest that production chains are an important factor in the process of structural change, in which economies undergo large-scale shifts in production activity over time, often from agriculture to manufacturing to services \cite{Kuznets1957}. If a shift from traditional agriculture into manufacturing increases the overall length of an economy's production chains, then the predictions here imply a natural mechanism for growth to accelerate as a country industrializes, and to move toward secular stagnation as it shifts into services. We discuss these implications further after presenting our results.

\section{Output multipliers and production chain length}
We first review some known facts about output multipliers, whose structural meaning underpins the intuition for results presented later.  Assume each industry makes only one good.  Let $a_{ji}$ denote an input coefficient, the fraction of good $j$ in industry $i$'s expenditures, and let $A = [a_{ji}]$ denote the matrix of these coefficients.  The output multipliers of an economy are given by the vector 
\begin{equation}
\vec{\L} = (I - A^T)^{-1} \vec{1},
\label{outputMultiplierDefinition}
\end{equation}
where $\vec{1}$ is a vector of 1s.  The matrix $(I - A)^{-1}$ is known as the Leontief inverse in input-output economics (e.g. \cite{Miller2009}) and as the fundamental matrix in the theory of Markov chains \cite{Kemeny1960}.  The output multiplier is also known as the total backward linkage \cite{Hirschman1958} or downstreamness \cite{Miller2015} of an industry, and is an instance of Katz centrality \cite{Katz1953}.  The Supplementary Information discusses the various mathematical representations of the output multiplier and their connections.

The structural meaning of output multipliers has been emphasized in recent studies of global value chains (e.g. \cite{Fally2012,Miller2015}). Two other ways of expressing the output multiplier highlight this connection.  Industries purchase goods from other industries as well as primary inputs (e.g. labor) from households.  Letting $\tilde{\ell}_j$ be the share of industry $j$'s expenditures that go to households, an output multiplier can be written as a sum over path lengths, $\L_i = \sum_{k=1}^\infty k \sum_{j=1}^n \tilde{\ell}_j (A^{k-1})_{ji}$.  Regarding the elements $a_{ji}$ and $\tilde{\ell}_j$ as transition probabilities \cite{Leontief1993}, an output multiplier $\L_i$ gives the average path length $k$ of all production chains that end at industry $i$, following each path backward through inputs until it reaches households.  Thus a longer supply chain length $\L_i$ captures a higher direct and indirect dependency on intermediate inputs.

\begin{figure}[t!]
\centering
\includegraphics[hsmash,width=.85\linewidth]{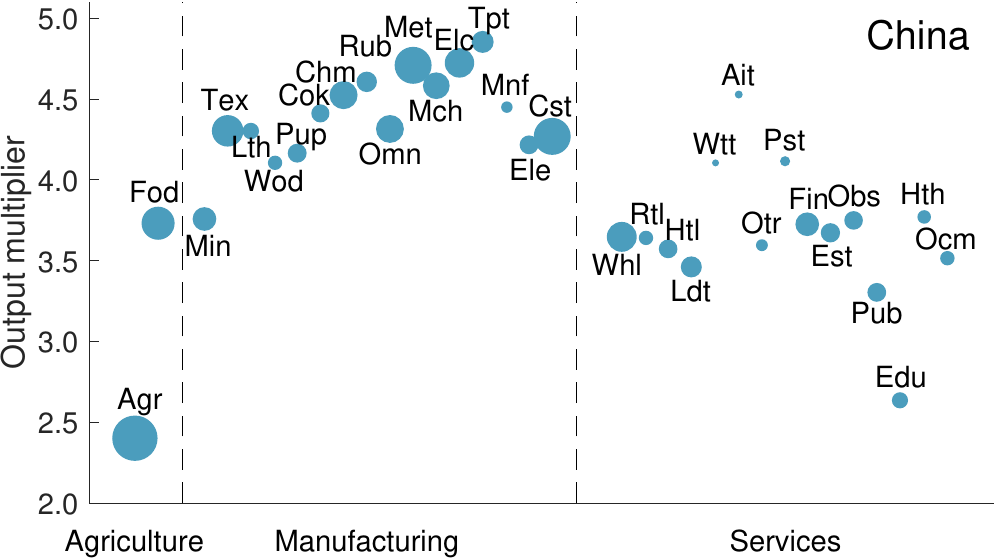}\\
\vspace{8pt}
\includegraphics[hsmash,width=.85\linewidth]{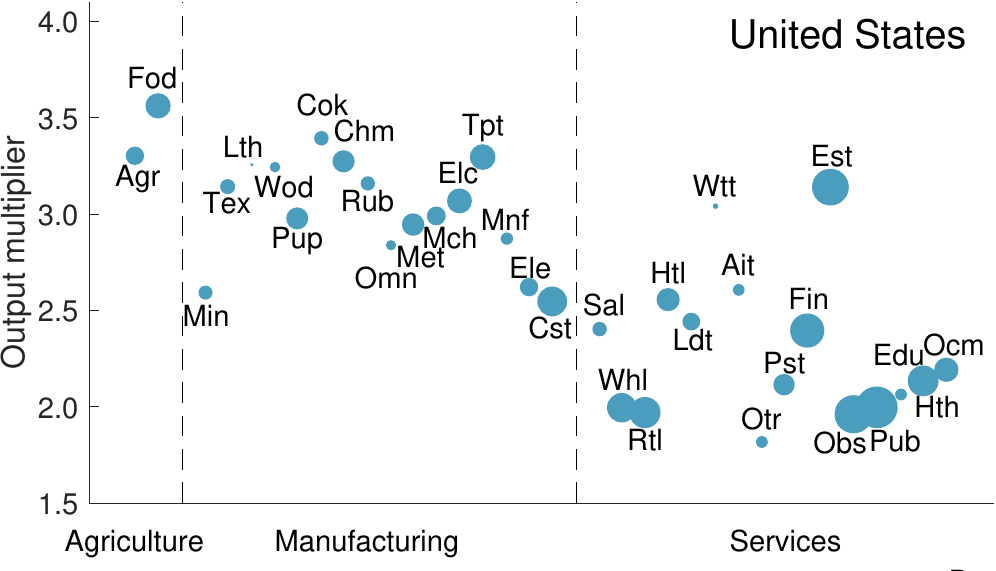}
\caption{\textbf{Output multipliers in Chinese and U.S. economies.} Node size corresponds to  gross output.  Industry codes are given in Supplementary Information Table S1.  All values are for 1995. Data source: World Input Output Database \cite{Timmer2015}.}
\label{fig_trophic_levels_depths}
\end{figure}

Output multipliers can also be expressed recursively as $\L_i = \sum_j \L_j a_{ji} + 1$.  One can think of this form in terms of an analogy with trophic structure, an organizing principle of ecology.  In an ecosystem the trophic level of a species is informally its position on a food chain \cite{Post2002}; a simple ecosystem with grass, zebras, and lions would result in grass (the species at the bottom of the food chain) having a trophic level of one, zebras a trophic level of two, and lions three.  Real ecosystems often have complex network structure, which include cycles and overlapping levels, and trophic levels are typically not integers but must be computed from a formula.  One such formula \cite{Levine1980} takes the recursive form above; letting $a'_{ji}$ be the fraction of species $j$ in the diet of $i$, the trophic level of species $i$ is $\L'_i = 1 + \sum_j \L'_j a'_{ji}$, i.e. 1 plus the average trophic level of the species it eats.  Similarly the output multiplier of an industry is 1 plus the average output multiplier of the industries from which it buys inputs. 

The recursive form makes clear that the output multiplier of an industry is influenced by two factors: the fraction of expenditures paid directly to households, and the output multipliers of the other inputs it buys.  Higher labor expenditures make it more likely that a dollar spent will go directly to the household node, realizing the shortest possible path length of 1, and lowering the output multiplier.  Similarly, dollars spent on goods from producers with high output multipliers will take more steps to reach the household node than dollars spent on goods with low output multipliers.  In the special case where an industry $i$ buys no intermediate inputs, $i$ realizes the smallest possible production path length $\L_i = 1$.

Examples of output multipliers for China and the United States are shown in Fig. \ref{fig_trophic_levels_depths}.  Each economy emphasizes different industries, but in both, manufacturing industries tend to have larger output multipliers than services (consistent with observations by Park and Chan \cite{Park1989}).  In services, humans typically provide a larger share of inputs relative to intermediate goods.  As a result, services may be expected to have shorter production chains.  Output multipliers in China tend to be higher than in the U.S. because China's household share of gross expenditures is lower.  The differences in agriculture in the two countries is illustrative.  In the U.S., agriculture is highly mechanized.  Agricultural industries depend heavily on intermediate goods relative to capital and labor inputs.  These industries have high output multipliers comparable to manufacturing.  In China agriculture is more labor-intensive, giving it a comparatively low output multiplier.

Output multipliers have long been used to project the impacts of a change in final demand, such as a government stimulus (see e.g. \cite{Miller2009}).  Additional final demand for a good requires the industry making it to buy more inputs, increasing the production of the industries that supply its inputs, and setting off a ripple effect that raises the gross output of the economy.  This amplification is greater when production chains are longer.  This represents a different process from the one we study here.  Nevertheless, the same network metric appears in both places in part because both processes involve a propagation of effects along production chains.

\section{Results}

\subsection{Network model of productivity improvement}
Our baseline model uses basic results of productivity accounting and the assumption that the price of an industry's good equals its marginal cost of production.  Industry $i$ uses $\phi_{ij}$ of good $j$ and $\ell_i$ of labor per unit of output.  Neglecting markups, the price $p_i$ of good $i$ equals its unit cost of production $p_i = \sum_j \phi_{ij} p_j + \ell_i w$ where $w$ is the wage rate.  This equation determines prices, so as the matrix of input needs $\phi_{ij}(t)$ and $\ell_i(t)$ evolves, prices change accordingly.  As shown in the Methods, the results can also be obtained in a general equilibrium framework (e.g. \cite{Long1983,Balke2000,Acemoglu2012}).  Here the key assumptions are that industries are price-takers who maximize profits at prevailing prices, subject to a production function with constant returns to scale, that consumers maximize utility subject to a budget constraint, and that prices instantaneously equilibrate supply and demand for all goods and labor.  A key point of our baseline model is that we do not need to take a stand on the functional forms of utility and production functions.  An extended presentation of this model can be found in the Supplementary Information.

Let $\widehat{\phi}_{ij} \equiv \dot{\phi}_{ij} / \phi_{ij}$ and $\widehat{\ell}_i \equiv \dot{\ell}_i/\ell_i$ denote the growth rates of $i$'s use of good $j$ and labor respectively.  An industry's improvement can be captured by its productivity growth rate $\gamma_i$ \cite{Jorgenson1987}, which can be expressed as a cost-weighted average of the rates of change of its input uses: $\gamma_i = -(\sum_j \widehat{\phi}_{ij} a_{ji} + \widehat{\ell}_{i}\tilde{\ell}_i )$.  The minus sign reflects the fact that a reduction in input use corresponds to an increase in productivity.  Let $r_i$ denote the log rate of change for the real (i.e. inflation-adjusted) price of industry $i$.  To deflate prices, the wage rate in a country was computed as the ratio of the total labor income earned to total hours worked by industries in the country, and then the rate of change of the real price of industry $i$ in country $c$ was computed as $r_i = r_i' - \rho_{c(i)}$, where $r_i'$ is the log change in nominal price and $\rho_{c(i)}$ is the log change in the wage rate of the country $c$ to which industry $i$ belongs.  Price changes can be expressed as
\begin{align}
r_i = -\gamma_i + \sum_j r_j a_{ji}.
\label{eq_priceDynamicsRecursive}
\end{align}
The first term captures industry $i$'s productivity improvement.  The second is the rate of price change for the inputs $i$ purchases, and is the growth rate of a Divisia price index.  Eq. \eqref{eq_priceDynamicsRecursive} simply says that the change in the price of good $i$ equals the change in the cost of $i$'s inputs, minus an extra term that captures $i$'s technological improvement.  Eq. \eqref{eq_priceDynamicsRecursive} represents what is known as a dual approach to productivity analysis \cite{Jorgenson1987}.  Typically, this approach is used to estimate productivity improvement, while here we are focused on modeling its effects.

\begin{figure}[t]
\center
\hspace{-3in}
{\sf\large A}\includegraphics[hsmash=r,width=.8\linewidth,align=t]{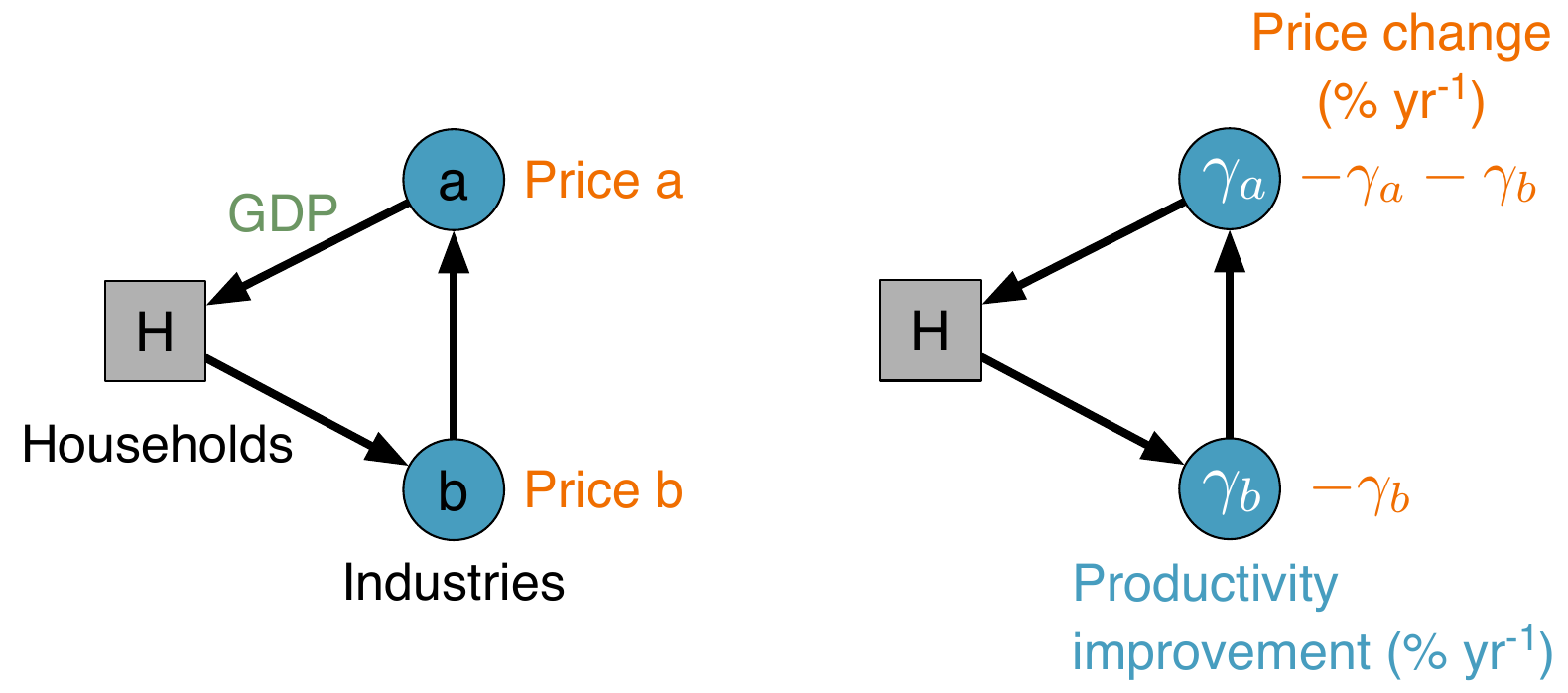}\\
\hspace{-3in}
{\sf\large B}\includegraphics[hsmash=r,width=.8\linewidth,align=t]{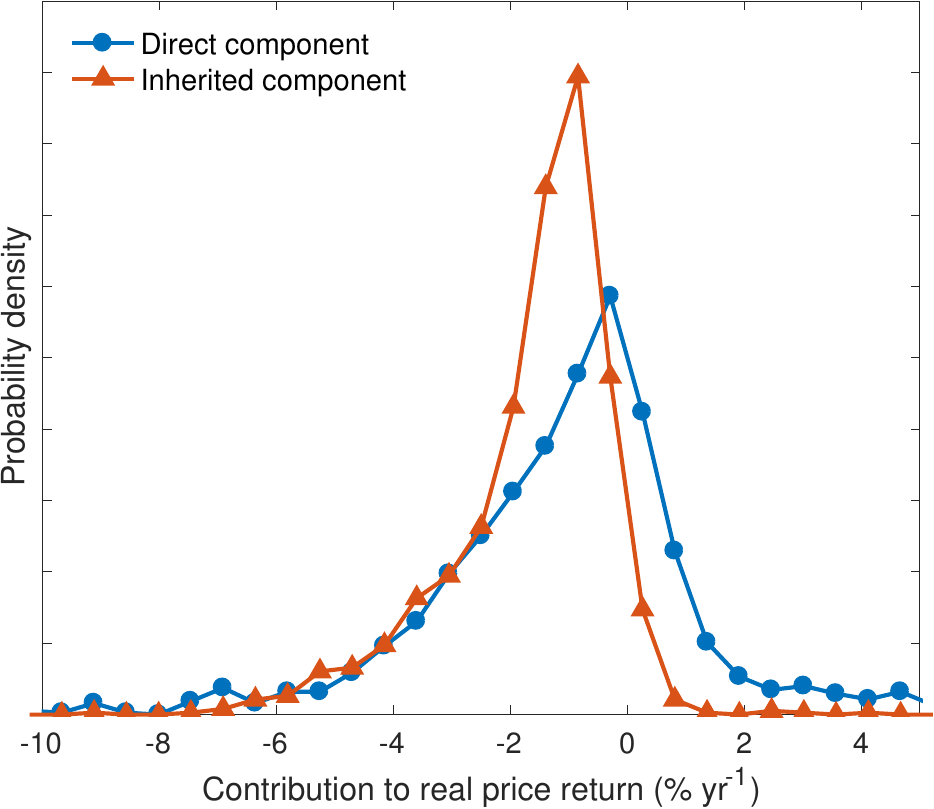}
\caption{\textbf{Productivity growth and prices.} (A) A bare-bones production network. Arrows show direction of goods.  Second diagram shows the price effects of productivity changes.  Productivity growth rates $\gamma_a$ and $\gamma_b$ lead to rates of price change $r_b = -\gamma_b$ and $r_a = -\gamma_a - \gamma_b$. (B)  The price change $r_i$ of an industry can be decomposed as $r_i = -\gamma_i + \sum_j r_j a_{ji}$, where $-\gamma_i$ is $i$'s direct contribution to its price reduction and $\sum_j r_j a_{ji}$ is the contribution from price changes passed to $i$ through input goods.  Each distribution is a histogram of 30 bins.  The direct component has larger variance, but the indirect component has lower mean, causing the larger share of price changes for a typical industry to be driven by its supply chain.}
\label{fig_local_v_nonlocal_contributions}
\end{figure}

The dynamics generated by this model are demonstrated in Fig. \ref{fig_local_v_nonlocal_contributions}A.  It depicts a circular economy that begins and ends with households.  Households sell labor to industry $b$, which makes an intermediate good that is sold to industry $a$, which makes a final good sold to households.  (For simplicity industry $a$ purchases no labor.)  When the productivity of either industry rises it needs less input per unit of good produced, causing its price to fall due to a lowered cost of production.  While industry $b$ only benefits from its own improvement, lower costs in this industry are passed on to $a$.  The result is that good $a$'s price reduction is the sum of both improvement rates.

\subsection{The importance of inherited improvement} 
Eq. \eqref{eq_priceDynamicsRecursive} can also be taken as a decomposition of an industry's price reduction.  The term $-\gamma_i$ accounts for the direct benefits of $i$'s technology improvement, while the sum $\sum_j r_j a_{ji}$ accounts for the total effect of all other productivity gains in the network.  This can be seen by writing Eq. \eqref{eq_priceDynamicsRecursive} as $r_i = -\gamma_i - \sum_{k=1}^\infty \sum_j [(A^T)^k \vec{\gamma}]_i$ where $\vec{\gamma}$ is the vector of productivity growth rates; see Supplementary Information.  The second component accumulates productivity improvements in upstream industries over production paths of all lengths $k$.  In this sense this term captures \emph{inherited} improvement -- accumulated productivity gains that are effectively passed to $i$ through reductions in input prices.

How much does inherited improvement contribute to price reduction in economies?  Fig. \ref{fig_local_v_nonlocal_contributions}B shows the distributions of the two components in the WIOD data.  Industry price changes are highly correlated with both components, with a correlation of 0.92 to the direct component and 0.71 to the inherited component.  (See the Supplementary Information for an extended discussion of these correlations.)  The direct component has a broader distribution, and as a result explains more of the variation in price changes.  

Nevertheless, the magnitudes of the two components (rather than the correlations) bear out a remarkable aspect of price evolution.  On average, inherited price reductions contribute more to price reduction (mean value -1.65\% yr$^{-1}$) than the direct components do (mean value -1.06\% yr$^{-1}$).  The average inherited cost reduction is about 1.6 times larger than the average direct component.  Considering industries individually, the inherited component contributes the better part of an industry's cost reduction in 64\% of industries in the WIOD.  As an example, from 1995 to 2009, the average price of electrical and optical equipment in China fell about 10\% per year; a rapid rate of improvement.  Out of this 10\% though, 6.2\% per year was inherited through the industry's inputs.

It is not simply the case that inherited price reductions matter to fully account for price changes in an economy.  Rather, \emph{most} price reduction comes through lower prices in purchased inputs.   This point also underscores a benefit of studying long-term price change in a production network setting, as price outcomes can be related to technology improvements in seemingly unrelated parts of the economy.

\subsection{The output multiplier bias in price evolution} 
These observations highlight the ability of production networks to accumulate the effects of productivity improvements.  How much this occurs for an individual industry depends on its position in the network.  Solving Eq. \eqref{eq_priceDynamicsRecursive} in vector form leads to $\vec{r} = -H^T \vec{\gamma}$, where $\vec{r}$ is the vector of price changes, $\vec{\gamma}$ is the vector of productivity growth rates, and $H \equiv (I - A)^{-1}$ is the Leontief inverse.  This network relationship between prices and productivity growth rates follows immediately from dual approaches to productivity growth \cite{Jorgenson1987} and has been exploited in models (e.g. \cite{Balke2000,Carvalho2019}).

Let $\bar{\gamma}$ be the average productivity growth rate across industries, and write industry $i$'s productivity growth rate as the sum of the average and a deviation, $\gamma_i = \bar{\gamma} + \Delta \gamma_i$.  The expected value of $r_i$ conditioned on the output multiplier is $E[r_i | \L_i] = - \bar{\gamma} \L_i - \sum_j E\left[\Delta \gamma_j H_{ji} \big| \L_i\right]$.  Empirically the correlations of $\Delta \gamma_j$ with the matrix elements $H_{ji}$ are low (see Supplementary Information).  Assuming $\Delta \gamma_j$ and $H_{ji}$ are uncorrelated, the second term is equal to zero, and we have
\begin{align}
E[r_i | \L_i] = - \bar{\gamma} \L_i.
\label{eq_expectedPriceChange}
\end{align}
Eq. \eqref{eq_expectedPriceChange} says the expected price change in industries with output multiplier $\L_i$ is proportional to $\L_i$.  This is because an industry benefits from both its own productivity improvement and the accumulation of upstream improvements.  As a result, industries with longer production chains will be biased to experience faster price reduction.  Eq. \eqref{eq_expectedPriceChange} indicates that the appropriate measure of production chain length for this mechanism is the output multiplier $\L_i$.  This simple relationship, which is readily obtained from standard results, places emphasis on the output multipliers as network measures.

\begin{figure*}[t!]
\center
\hspace{-7in}
\includegraphics[width=0.95\textwidth]{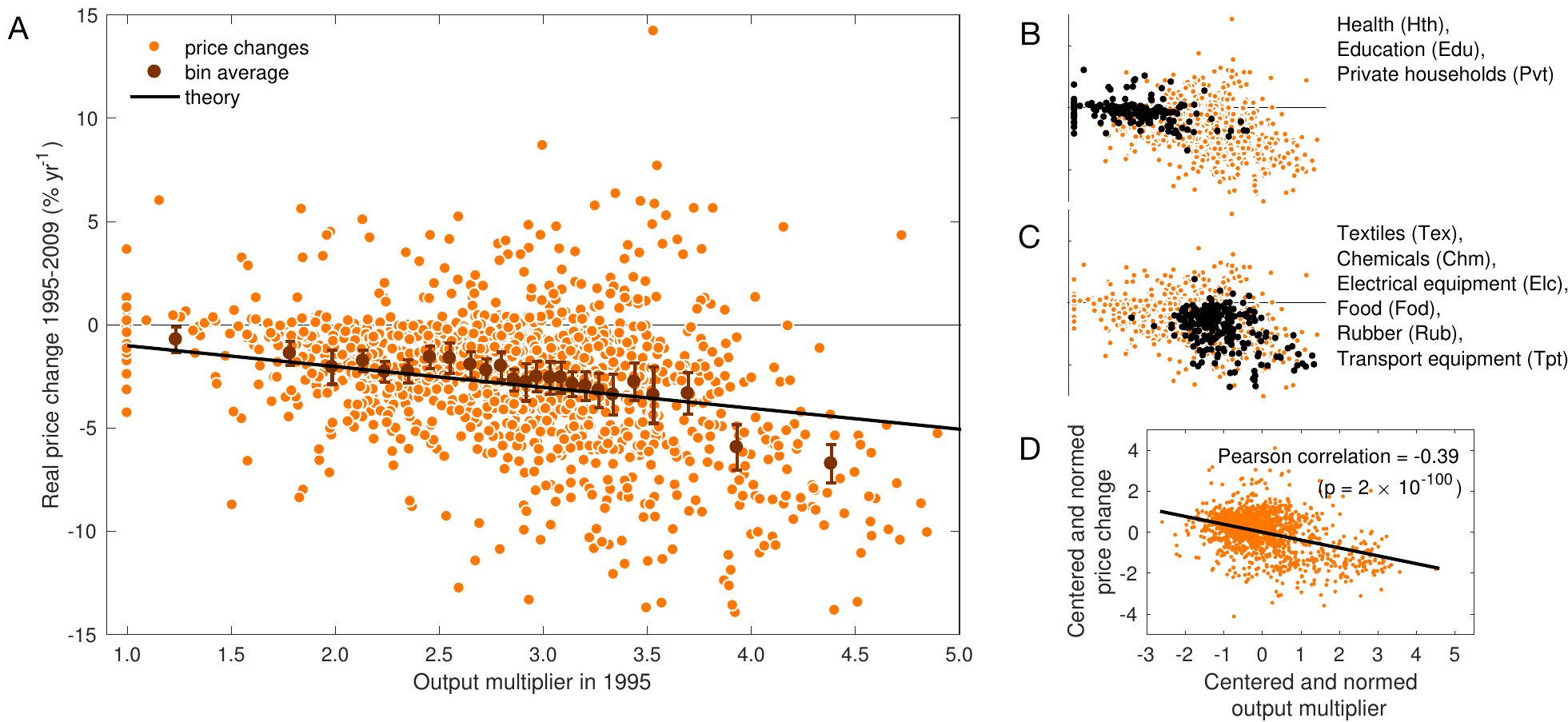}
\caption{\textbf{The output multiplier bias in price evolution.} (A) Rates of price change for 1400 world industries versus industry output multipliers (orange dots).  Industries with similar output multipliers are grouped into bins of $\sim$45 points each and the average price change is computed for each bin (brown dots).  Vertical lines give error bars of two standard deviations around the bin mean.  The black line is the theoretical prediction for the bin averages based on the relation $E[r_i | \L_i] = - \bar{\gamma} \L_i$ with $\bar{\gamma}$ equal to the mean of productivity growth rates over 1995 - 2009. (B-C)  Price changes and output multipliers with a selection of services and manufacturing industries highlighted by black dots.  (D)  Variation in output multipliers predicts variation in price changes within industry categories.  We remove dependence on industry category by standardizing observations by their cross-country averages and standard deviations, and comparing centered-and-normed price changes to centered-and-normed output multipliers.  Black line is a regression fit.}
\label{fig_priceChanges_v_OMs}
\end{figure*}

We test this prediction by looking at price changes for the 1400 industries (40 countries $\times$ 35 industry categories) in the WIOD.  We compute rates of real price change for the period 1995 - 2009 and compute output multipliers in 1995.  Comparing these (Fig.~\ref{fig_priceChanges_v_OMs}A) shows the predicted, systematic deviation of the expected price change with the output multiplier; industries with larger output multipliers are biased to realize faster price reduction.  Binning price changes by industry output multipliers and computing the average change in each bin gives an empirical estimate of the conditional expectation $E[r_i | \L_i]$.  Regressing the bin averages on the output multipliers gives a slope $-1.6\%$ per year ($p \sim 10^{-8}$, $R^2 = 0.75$).  

Alternatively, we can use Eq. \eqref{eq_expectedPriceChange} to make a prediction of $E[r_i | \L_i]$ without free parameters.  We estimate $\bar{\gamma} = 1.0\%$ per year using the productivity growth rates of industries over the period 1995 - 2009.  Using this value in Eq. \eqref{eq_expectedPriceChange} yields the theory line in Fig. \ref{fig_priceChanges_v_OMs}A.  In either approach we fix the output multipliers at their values at the start of the period.  Output multipliers help characterize network structure, and this probes the idea that this structure is sufficiently stable to approximate the accumulation of productivity improvement as a process on a static network.  We see that output multipliers fixed at their initial values can predict subsequent changes in price.  (We also find that using time-averages of the output multipliers yields very similar results, see Supplementary Information.)

The difference between the observed regression slope $-1.6\%$ per year and the predicted slope $-\bar{\gamma} = -1.0\%$ per year stems from a positive correlation between output multipliers $\L_i$ and productivity improvement rates $\gamma_i$, which have a Pearson correlation 0.11 ($p \sim 10^{-5}$).  Productivity improvement rates tend to be greater for industries with higher output multipliers, increasing the magnitude of the slope in Fig. \ref{fig_priceChanges_v_OMs}A.  This correlation is outside the model of Eq. \eqref{eq_priceDynamicsRecursive}, but is not inconsistent with it -- the model does not say what determines productivity.  To see whether this correlation drives the relationship between price changes and output multipliers, we shuffle improvement rates across industries to remove the correlation with the output multipliers (Supplementary Information Fig. S2), finding that the output multipliers retain a highly significant correlation with price changes even with this effect removed.  

Manufacturing industries are known to have higher output multipliers as well as faster price reduction (Fig. \ref{fig_priceChanges_v_OMs}B-C), suggesting this could drive their correlation. But even within an industry category, a higher output multiplier predicts faster relative price reduction.  Comparing centered-and-normed price changes with centered-and-normed output multipliers (i.e. applying fixed effects at the industry level) reveals a strong negative correlation of $-0.39$ ($p \sim 10^{-100}$) (Fig. \ref{fig_priceChanges_v_OMs}D).  This relationship also holds when we examine industries individually (Supplementary Information Table S3).  We also divide industries into the broad groupings of manufacturing, services, and agriculture, and compare the predictive ability of these labels to that of output multipliers, finding the latter to be much better predictors of price change (Supplementary Information Table S4).  While not central to our results, the correlation of price movements with network structure suggests there will also be structural correlations in the price movements of different industries with each other.  We examine this possibility in the Supplementary Information, finding good agreement with the model here as well.

\subsection{Increasing relevance of output multipliers with time}
In Fig. \ref{fig_priceChanges_v_OMs}A the price changes of industries have considerable dispersion around the expected value $E[r_i|\L_i]$.  The characteristics of this dispersion are also predicted by Eq. \eqref{eq_priceDynamicsRecursive}.  Let $r_i(t,t+T)$ be the average rate of change of price $i$ from $t$ to $t + T$.  We study the behavior of $\sigma_{r_i(t,t+T) | \L_i}
$, the standard deviation of $r_i(t,t+T)$ across industries with a given output multiplier.  In the Supplementary Information, we compute the variance of Eq. \eqref{eq_priceDynamicsRecursive}, factoring out the dependence on the output multiplier, and approximate $\sigma_{r_i(t,t+T) | \L_i}$ as
\begin{align}
\sigma_{r_i(t,t+T) | \L_i} \approx \frac{1}{\sqrt{T}} \Big[ \sigma_{\gamma,\text{direct}}
+ \rho \sigma_{\gamma,\text{inherited}} (\L_i - 1)
\nonumber\\
+ \frac{1}{2} \frac{\sigma_{\gamma,\text{inherited}}^2}{\sigma_{\gamma,\text{direct}}} (\L_i - 1)^2 \Big].
\label{eq_priceSigmaWithTime}
\end{align}
Here, $\sigma_{\gamma,\text{direct}} = (\text{Var}[\gamma_i | \L_i])^{1/2}$ is the standard deviation of the direct productivity benefit $\gamma_i$ across industries with output multiplier $\L_i$, $\sigma_{\gamma,\text{inherited}} (\L_i - 1)$ is the standard deviation of the inherited productivity benefit $(\vec{\gamma}^T A + \vec{\gamma}^T A^2 + \cdots)_i$ across industries with output multiplier $\L_i$, and $\rho$ is the Pearson correlation between direct and inherited benefits.  (The coefficient $\frac{1}{2} \frac{\sigma_{\gamma,\text{inherited}}^2}{\sigma_{\gamma,\text{direct}}}$ is small compared with $\rho \sigma_{\gamma,\text{inherited}}$, and as a result the contribution of the quadratic term will be small.)

\begin{figure}[t!]
\center
\hspace{-3.3in}
{\sf\large A }\includegraphics[hsmash=r,width=0.9\linewidth,align=t]{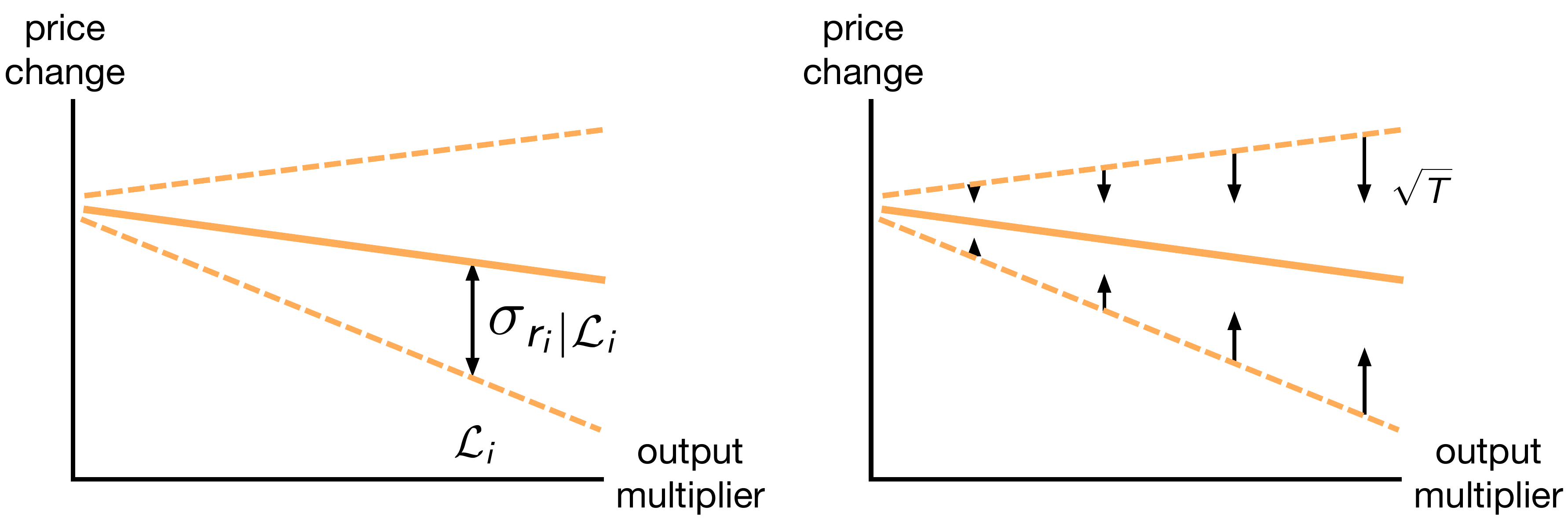}\\
\hspace{-3.3in}
{\sf\large B}\includegraphics[hsmash=r,width=0.9\linewidth,align=t]{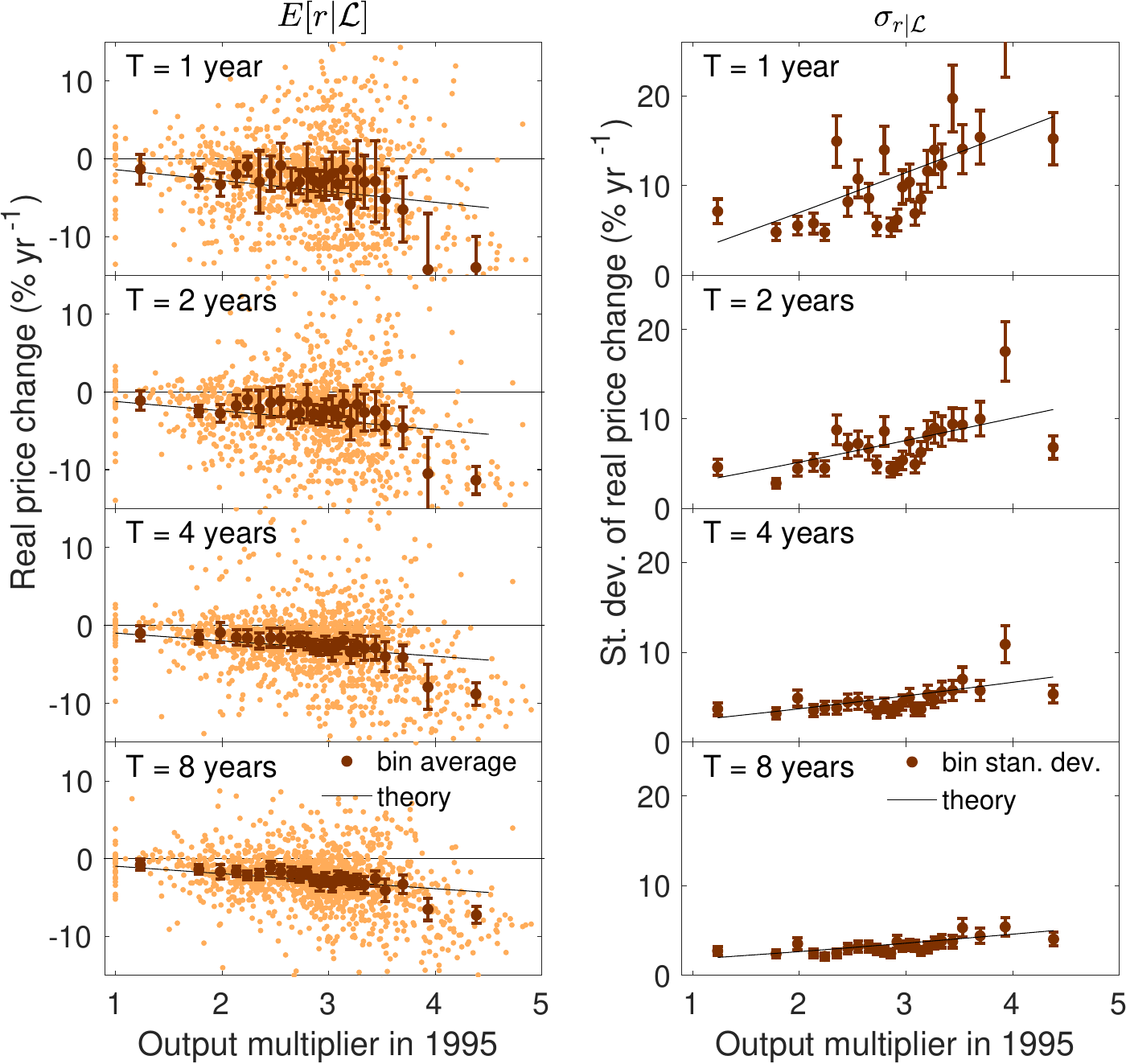}\\
\hspace{-3.3in}
{\sf\large C}\includegraphics[hsmash=r,width=0.85\linewidth,align=t]{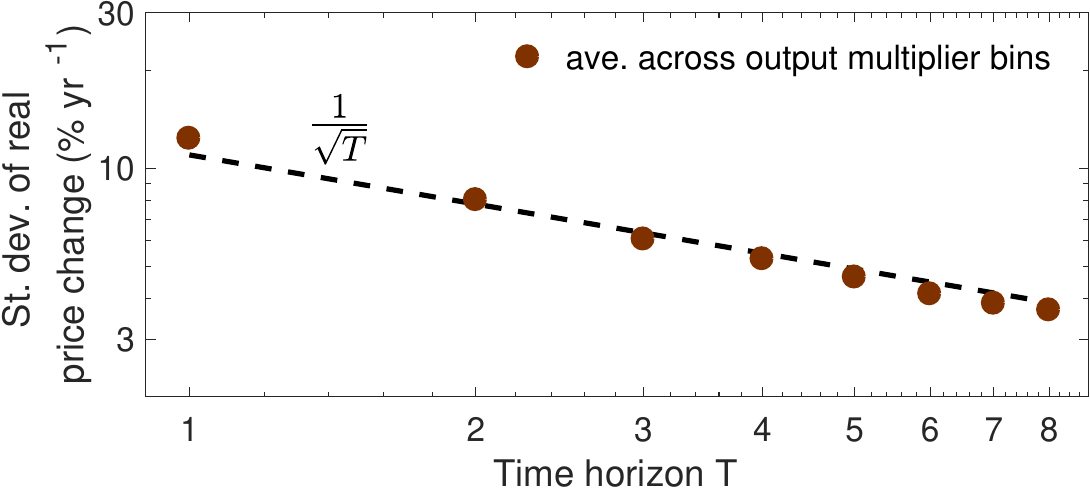}
\caption{\textbf{Cross-industry dispersion in price changes over different time horizons.} (A) Predictions of Eq. \eqref{eq_priceSigmaWithTime} for cross-industry dispersion in rates of price change. (B) We examine average rates of price change $r_i(t,t+T)$ over horizons of $T$ = 1, 2, 4, and 8 years.  Left panels show price changes for individual industries (orange dots). The average price change for industries in a given output multiplier bin (brown dots) gives an empirical estimate of the expected price reduction $E[r_i|\L_i]$.  The black line is the theory prediction for this expected value.  Right panels shows the standard deviation of price changes for industries with a given output multiplier (brown dots) and the theory prediction (black line).  (C) Standard deviation of $r_i(t,t+T)$ for all industries.  The dashed line is a guide to the eye and is proportional to $1/\sqrt{T}$.}
\label{fig_timeHorizon}
\end{figure}

Eq. \eqref{eq_priceSigmaWithTime} makes two predictions (Fig. \ref{fig_timeHorizon}A).  First, in any given time period, variation in price change is greater for industries with large output multipliers.  Second, for any given output multiplier, this variation shrinks with time like $1/\sqrt{T}$.  The second prediction means that dispersion in price changes around the expected value shrinks as the prediction horizon gets longer.  As a result, the output multiplier of an industry becomes increasingly relevant for its price reduction over time.

We test these predictions with observations from the WIOD over the time horizons $T =$ 1, 2, 4, and 8 years (Fig. \ref{fig_timeHorizon}B-C).  We again hold the output multipliers fixed at their values in the year 1995, exploiting the relatively slow rate of change of output multipliers over time.  In addition to the dependence of $\sigma_{r_i(t,t+T) | \L_i}$ on $\L_i$ predicted by Eq. \eqref{eq_priceSigmaWithTime}, we find an additional effect in which industries with larger output multipliers have larger variation in productivity improvement, leading to a dependence of $\sigma_{\gamma,\text{direct}}$ on $\L_i$.  Similar to the correlation between productivity improvement and output multipliers, this correlation lies outside the theory here, though is not inconsistent with our findings.  To take this correlation into account, for each time horizon we build a linear model of $\sigma_{\gamma,\text{direct}}$'s dependence on $\L_i$.  This model by itself (without the additional terms in Eq. \eqref{eq_priceSigmaWithTime}) explains roughly 60\% of the slope of $\sigma_{r_i(t,t+T) | \L_i}$, and thus by itself yields a poor fit to the data.   We use this linear model within Eq. \eqref{eq_priceSigmaWithTime} to obtain the prediction for $\sigma_{r_i(t,t+T) | \L_i}$ for each time horizon and obtain a much better prediction, shown in Fig. \ref{fig_timeHorizon}B.

Over any given time horizon, price changes vary more among industries with large output multipliers (Fig. \ref{fig_timeHorizon}B).  As time passes, this dispersion across industries shrinks at a rate in good agreement with the $1/\sqrt{T}$ prediction (Fig. \ref{fig_timeHorizon}C).  (See also Supplementary Information and Fig. S3 for further discussion.) Note that the higher dispersion in price changes for industries with large output multipliers accounts for the triangular shape of price changes in Figs. \ref{fig_priceChanges_v_OMs}-\ref{fig_timeHorizon}.  Over time this triangle narrows, with the expected value Eq. \eqref{eq_expectedPriceChange} becoming an increasingly good predictor, i.e. an industry's price evolution becomes better predicted by a long-term behavior based on its output multiplier.

\subsection{The average output multiplier and economic growth}
We now consider the implications of the results above for economic growth.  To the extent that an economy enjoys real price reductions by consuming more, falling prices will translate into greater output.  The relationship of output multipliers with price reductions thus suggests a relationship with growth as well.  Aggregating across price reductions in all industries, it can be shown (see e.g. \cite{Acemoglu2012} and Supplementary Information) that the real growth rate $g$ of a closed economy depends on productivity growth rates as $g = \vec{\theta}^T H^T \vec{\gamma}$.

\begin{figure*}[t]
\centering
\hspace{-3in}
{\sf\large A }\includegraphics[hsmash=r,width=0.4\linewidth,align=t]{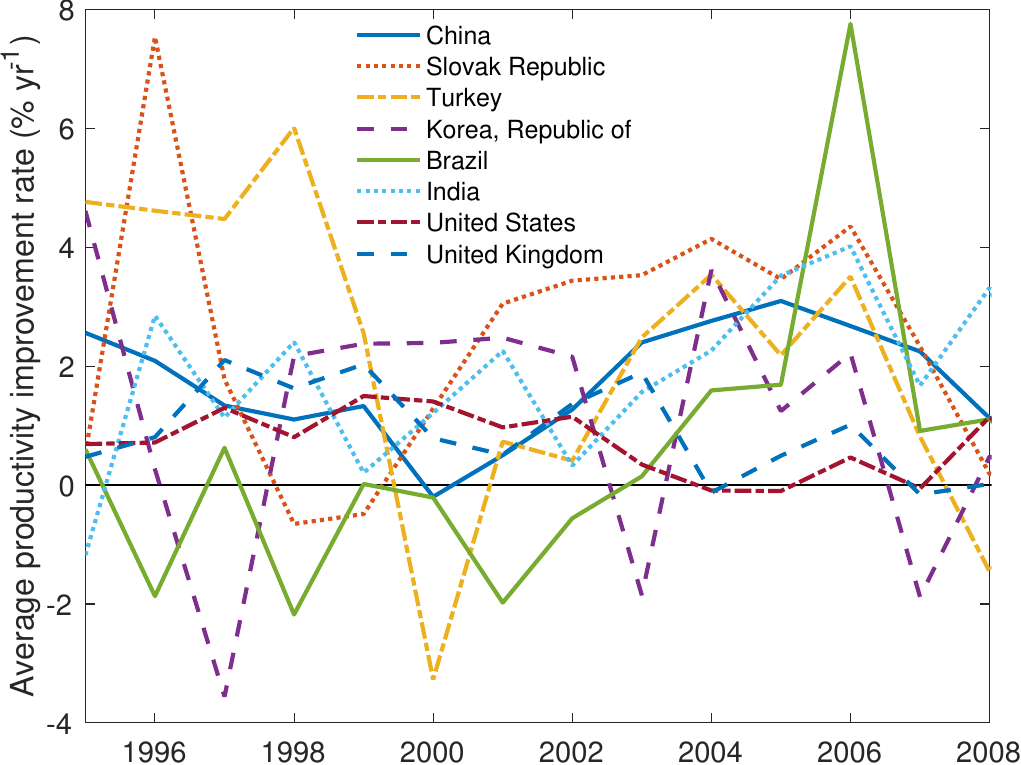}
\hspace{3in}
{\sf\large B }\includegraphics[hsmash=r,width=0.4\linewidth,align=t]{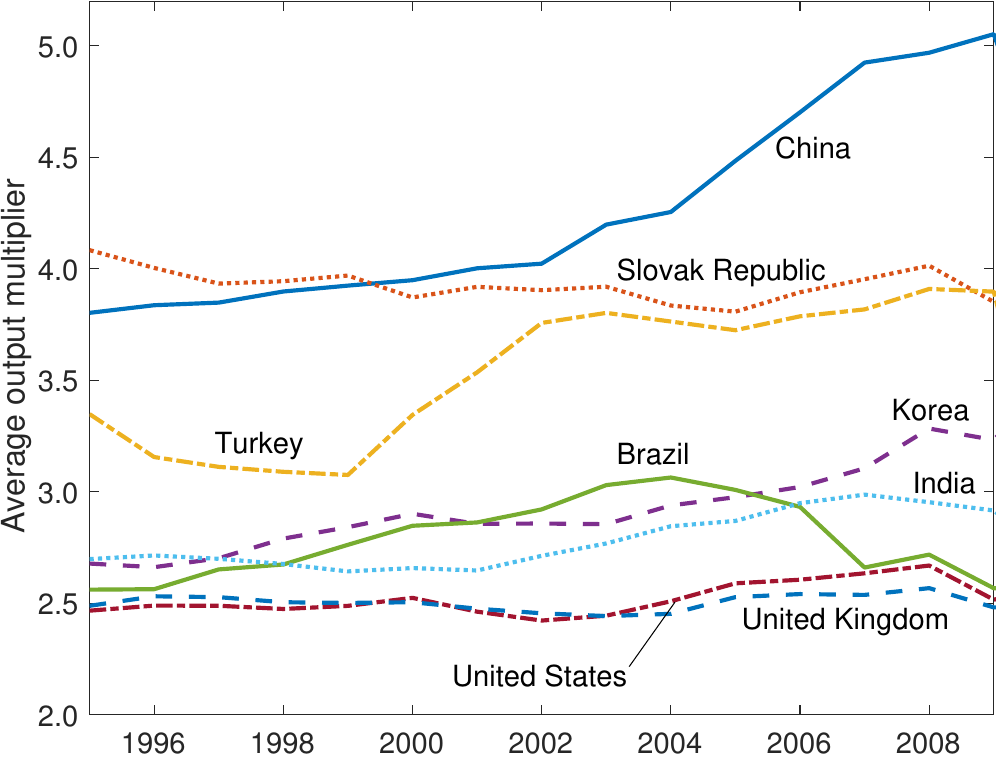}
\caption{\textbf{Productivity improvement and average output multipliers over time.}  (A) Average productivity improvement rate ($\tilde{\gamma}_c$) and (B) average output multiplier ($\bar{\L}_c$) for a selection of countries.}
\label{fig_trophic_depth_over_time}
\end{figure*}

The rate of growth can be readily recast (see Supplementary Information) in terms of output multipliers as well, yielding
\begin{align}
g = \tilde{\gamma} \bar{\L}.
\label{eq_growth_result}
\end{align}
Here $\tilde{\gamma} \equiv \sum_i \eta_i \gamma_i$ is the average rate of productivity improvement across a country's industries with weights $\eta_i$ giving the share of industry $i$ in gross output (total revenue of all industries).  The factor $\bar{\L} \equiv \sum_i \theta_i \L_i$ is the weighted average of industries' output multipliers with $\theta_i$ giving the share of industry $i$ in final output (GDP).  Eq. \eqref{eq_growth_result} predicts that the GDP growth of a country is proportional to the product of its average productivity improvement and its average output multiplier.  It factors GDP growth into two parts, one that depends on productivity, and another that is purely a structural property of the economy's production network.  Thus, similar to Eq. \eqref{eq_expectedPriceChange}, standard results can be manipulated to give a simple but novel expression that relates growth with production chain length and communicates readily with data.

The dynamics of $\tilde{\gamma}$ and $\bar{\L}$ differ in character (Fig. \ref{fig_trophic_depth_over_time}).  The average improvement rate fluctuates considerably from year to year; the average output multiplier varies much more slowly.  One way to quantify this difference is to compare the time variation in output growth, productivity improvement, and output multipliers.  For each variable $X(t)$ we compute the coefficient of variation (CV) $\sigma_X / \mu_X$ where $\sigma_X$ is the time standard deviation (volatility) and $\mu_X$ is the time average from 1995 to 2009.  Typical CVs (geometric mean across countries) are 1.4 for output growth, 2.3 for average productivity growth, and 0.041 for the average output multiplier.  By this measure, average output multipliers have about $1.4/0.041 \sim 34$ times less variation over time than growth rates do, and $2.3/0.041 \approx 56$ times less variation than productivity improvement rates.  (We similarly find that industry-level output multipliers have $2.47/0.057 \approx 43$ times less variation than price changes and $3.8/0.057 \approx 66$ times less variation than productivity improvement.)  As with individual industries, this relative persistence at the aggregate level makes intuitive sense given that underlying production relationships take time to change.  A difference at the aggregate level is that $\bar{\L}$ may change because of shifts in an economy's final output mix, even if its industry-level output multipliers $\L_i$ were to remain the same, a point we revisit in the Discussion.

\begin{figure*}[t]
\centering
\hspace{-3in}
{\sf\large A }\includegraphics[hsmash=r,width=0.37\linewidth,align=t]{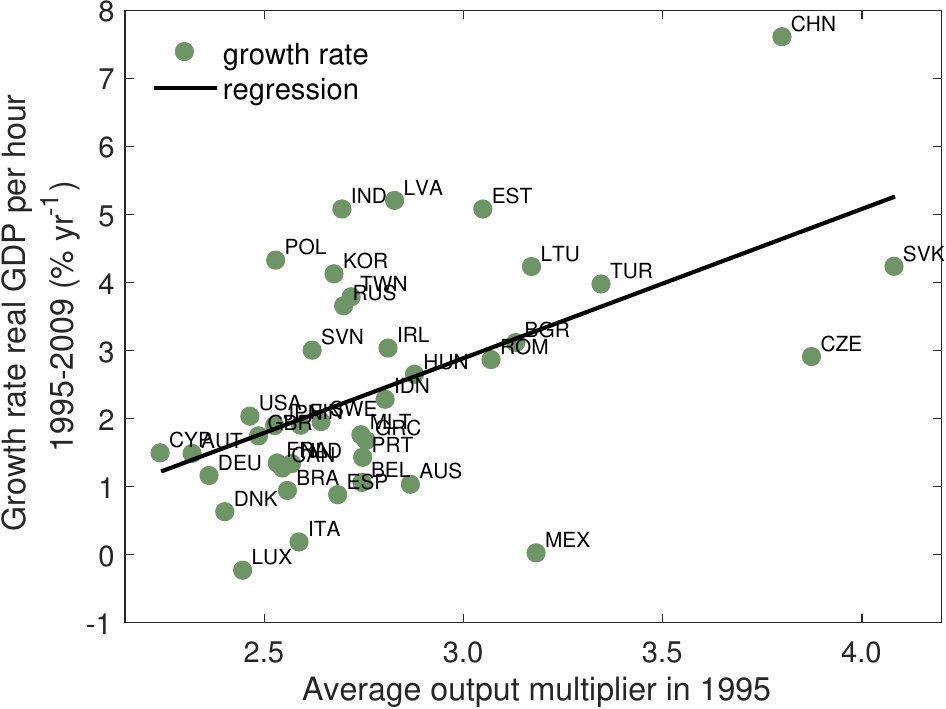}\hspace{10pt}
\hspace{3in}
{\sf\large B }\includegraphics[hsmash=r,width=0.40\linewidth,align=t]{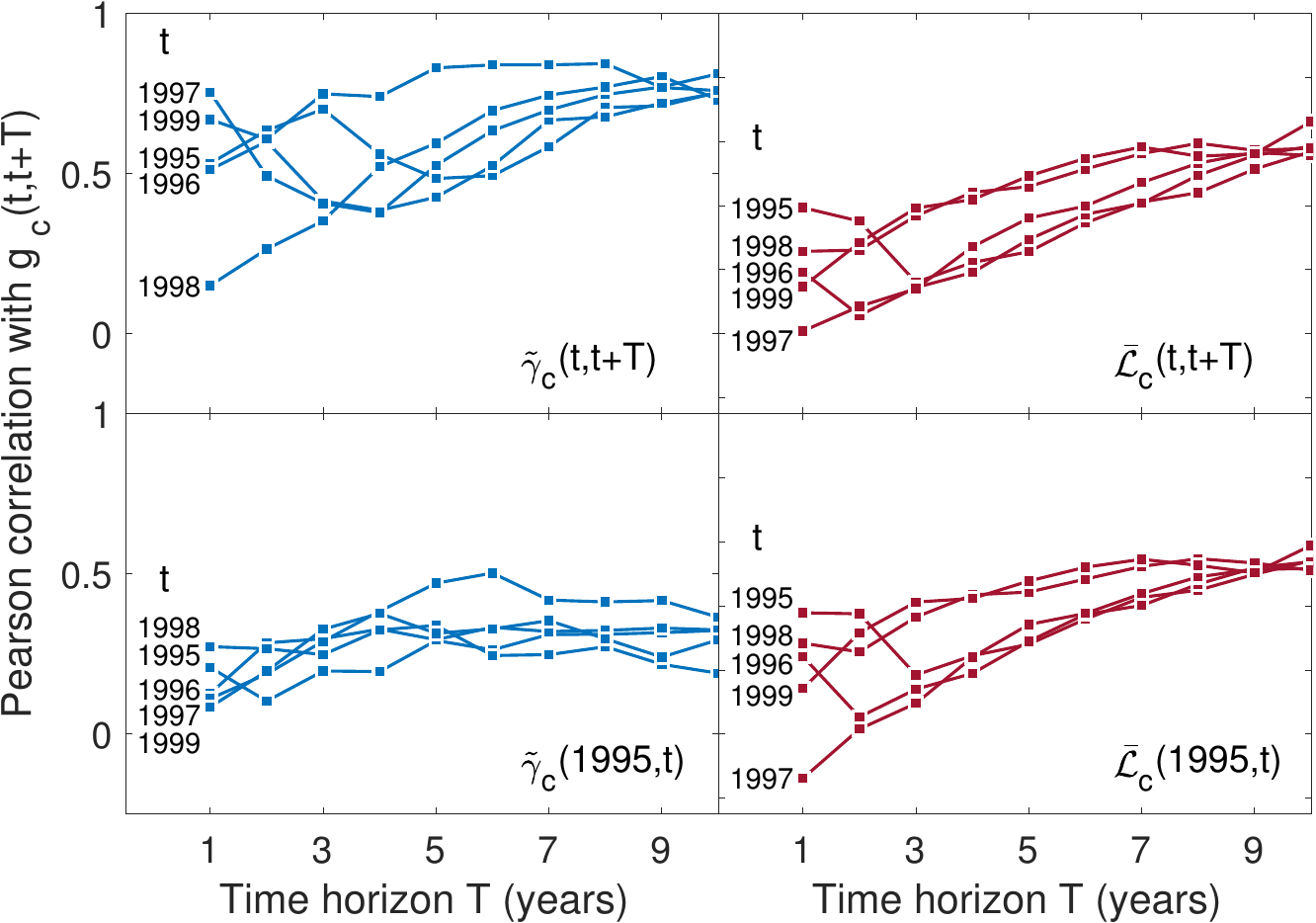}\\
\hspace{-6in}
{\sf\large C }\includegraphics[hsmash=r,width=0.76\linewidth,align=t]{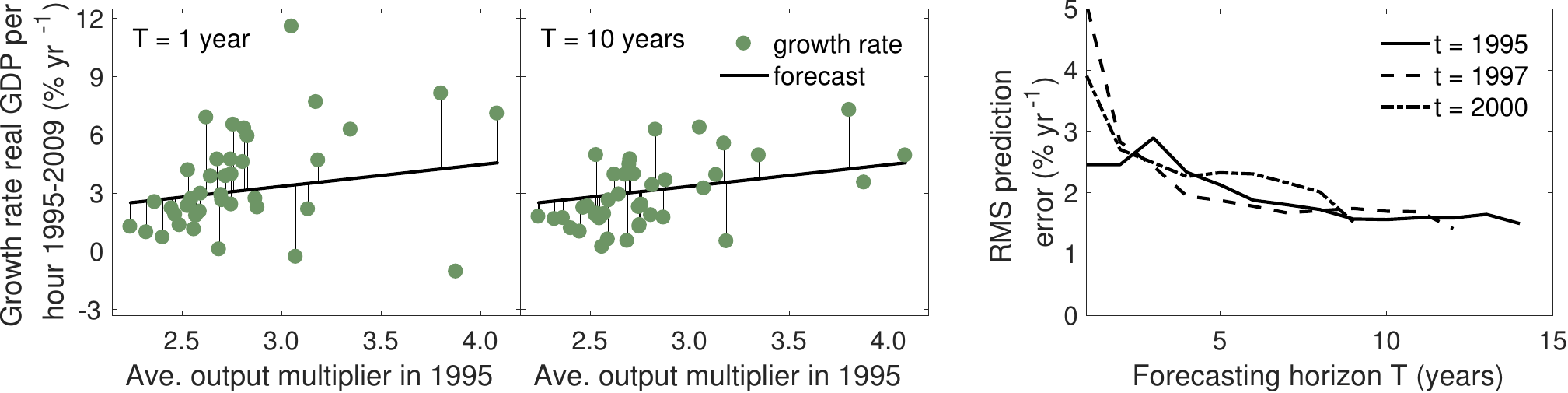}
\caption{\textbf{Country growth and the average output multiplier.} (A)
Growth rates of real GDP per hour over 1995 - 2009 for 40 WIOD countries versus $\bar{\L}$ in 1995.  Solid line is an OLS regression fit.  Country codes are given in Supplementary Information Table S2.  (B) Correlation of growth with average productivity growth rate and average output multiplier as time passes.  For each country $c$ we compute the time-average growth rate $g_c$(t,t+T) from year t to year t+T.  Top panels show the correlation of $g_c$(t,t+T) (across countries) with time-average productivity improvement ($\tilde{\gamma}_c$(t,t+T)) and time-average output multiplier ($\bar{\L}_c$(t,t+T)) over the same period (t,t+T).  Bottom panels show the correlation of growth in (t,t+T) with productivity improvement and average output multipliers from the prior period (1995,t).  (C) Simple forecast of country growth rates from multiplying countries' average output multipliers in a starting year t by the mean improvement rate across all countries and years of our data prior to t.  Left panels show forecasted and actual growth for starting year t = 1995.  Right panel shows root mean square prediction error over a range of horizons with starting years 1995, 1997, and 2000.}
\label{fig_growths_v_trophicDepth}
\end{figure*}

The persistence of $\bar{\L}$ in Eq. \eqref{eq_growth_result} suggests that a country's average output multiplier should not only correlate with its current growth rate but also with its growth rate for some time into the future.  In the WIOD data (Fig. \ref{fig_growths_v_trophicDepth}A) the growth rates of real GDP per hour over 1995 - 2009 have a Pearson correlation $\rho = 0.53$ ($p = 4 \times 10^{-4}$) with countries' average output multipliers in the initial year 1995.  For longer production chains to result in faster growth, the average rate of productivity improvement of an economy must not decrease as the average output multiplier gets larger.  In fact we observe the opposite tendency; the positive correlation noted earlier between productivity growth and output multipliers at the industry level now appears as a positive correlation at the aggregate level ($\rho = 0.45$) between $\tilde{\gamma}$ and $\bar{\L}$.  We do not attempt to explain the correlation here, though it is plausible that factors such as investment would increase both the length of production chains and the rate of technological improvement at the same time.  This correlation means that the regression relation between growth and output multipliers in Fig. \ref{fig_growths_v_trophicDepth}A reflects two effects: the theoretical prediction that, all else equal, countries with larger average output multipliers should grow faster, and the empirical observation that countries with larger average output multipliers tend to have higher average productivity growth rates.

While $\tilde{\gamma}$ and $\bar{\L}$ enter Eq. \eqref{eq_growth_result} symmetrically, these variables have fundamentally different characters.  The average productivity growth rate is a rate-of-change, measured between two end times, while the average output multiplier is a state variable measured at a point in time.  The former is noisy and the latter persistent.  As theory predicts, both factors have high correlations with growth that rise with the length of the period one examines (Fig. \ref{fig_growths_v_trophicDepth}B, top panels), but the reasons for this rise differ for each factor.  For $\tilde{\gamma}$, examining a longer period means integrating over more of a country's history of productivity gains; whether an economy realized years of rapid or sluggish improvement significantly influences whether these were also years of rapid growth.  For $\bar{\L}$ however, examining a longer time period does not improve its correlation with growth by averaging over its own history, but by giving time for the fluctuations in the productivity factor to average out.  Unlike the productivity growth rate the correlation of $\bar{\L}$ with growth is not sensitive to using values from the period in which the growth is observed (Fig. \ref{fig_growths_v_trophicDepth}B, bottom panels).

These observations and theory suggest that the output multiplier should be able to forecast growth.  We examine this possibility with an extremely simple forecasting model.  From the perspective of an observer in year $t$, we conduct a forecast for the next $T$ years, in which we multiply the average output multiplier $\bar{\L}_c$ of every country in year $t$ by a representative guess of $\tilde{\gamma}$ for the future.  Here we use the average of $\tilde{\gamma}$ across all countries in the years of our data leading up to $t$.  By doing this we are also removing all cross-country variation in productivity growth to focus solely on the predictive power of the output multiplier.  Remarkably, and despite the clear simplicity of this forecasting model, prediction error falls as the forecasting period grows (Fig. \ref{fig_growths_v_trophicDepth}C).  As noted already this occurs because output multipliers are persistent quantities and longer horizons provide more averaging over productivity growth fluctuations.

Strictly, this analysis does not identify the output multiplier as a causal factor; rather it demonstrates that the structure of the production network helps forecast growth.  Yet there are at least four reasons to interpret it causally: (i) These forecasts are consistent with a clear causal mechanism generically implied by standard production network models; (ii) they emerge directly from the aggregation of the price effects documented earlier; (iii) they come with no free parameters; and (iv) are consistent with not just the sign of the effect, but its predicted functional form.

Finally, we note that the structural significance of $\bar{\L}$ is enhanced by a remarkable theorem  by Fally (2012) \cite{Fally2012} and Finn (1976) \cite{Finn1976}.  Data on production networks vary in level of aggregation from a few industries to hundreds, raising the concern that the average output multiplier varies with the granularity of the data.  But it can be shown that the average output multiplier of a closed economy is independent of aggregation and equal to the ratio $\bar{\L} = \mathcal{O}/Y$ where $\mathcal{O}$ is gross output and ${Y}$ is net output (GDP) \cite{Fally2012}.  In the Supplementary Information we discuss this point further, and also examine it in the practical context of an open economy.

This result is important also because it implies that economies with very different production networks, but the same ratio $\O/Y$, would experience the same amplification to growth. Three comments are in order here. First, we note that the independence of the rate of growth from network structure is an approximation, reflecting the first order nature of our growth result. When we consider second order effects on growth (36), the details of the network can become relevant through the effects that relative price changes have on consumption shares and input shares. (Though we find evidence that such effects are modest for the time scales we study here, see Supplementary Information). Second, the fact that there exist macro-level sufficient statistics to characterize growth amplification does not imply there is no value in understanding its emergence as a network outcome. In particular, it takes a well-specified model to understand whether and how these sufficient statistics provide good approximations or whether they fail under certain circumstances. Third, in addition, our production network environment delivers testable micro-level predictions on sectoral relative price evolution that are important in themselves. The fact that these micro-predictions are in good agreement with data suggests that production networks can provide a causal mechanism (as opposed to reduced form, sufficient statistics) for differences in growth across countries.

\section{Discussion}
Economics has emphasized the outsized role of productivity in explaining cross-country differences in growth, and our theory features this.  But production network models predict that variation in the output multiplier matters as well.  These models generate a variety of predictions for how the output multiplier should shape price evolution and growth, with which observations from data are in good agreement.  Recent studies \cite{Fally2012,Miller2015} have emphasized that the output multiplier can be understood as the average length of an industry's production chains.  This leads to an intuitive mechanism driving our results: An industry benefits from its own productivity gains, and that of upstream suppliers, and so the effects of productivity improvement accumulate along production chains.  As a result, two countries realizing the same average micro-level productivity improvement can, if their production networks differ in depth, experience different aggregate growth rates.




At a micro level, the results suggest one reason why some technologies improve more rapidly than others, especially a version of this question that arises in Baumol's classic observation. As confirmed many times since, in the 1960s Baumol observed that some industries, particularly manufacturers, realize productivity gains more rapidly than others \cite{Baumol1967,Nordhaus2008}, and that over long periods such sustained differences will significantly impact prices.  The relative prices of quickly-improving industries will naturally fall, but those of slower-improving industries (including many services) will increase, even if these industries are realizing some improvement \cite{Baumol1967}.  This effect has often been cited as a cause of long-term price increases in health care \cite{Bates2013} and education \cite{Baumol1996}.  The results here point to a reason why manufacturing would be able to sustain faster improvement for long periods.  Observations of output multipliers \cite{Park1989} reinforce the intuitive idea that manufacturing industries tend to have longer production chains than services.  Manufacturing industries are advantaged by their network positions to benefit more greatly from productivity gains across the network.  The results here suggest a nuanced way to help distinguish fast and slow segments of an economy, in that part of what helps define the fast-improving segment is the set of industries with large output multipliers rather than manufacturers per se.  In particular, the fact seen earlier that output multipliers retain their predictive power within broad industries suggests that our analysis provides a more operational way to distinguish fast- and slow-improving industries.

At the aggregate level, the results suggest a new perspective on the long process of structural change \cite{Kuznets1957}, emphasizing changes to the length of an economy's production chains.  One expects an undeveloped economy to have short chains of production.  As manufacturing becomes more prominent, the average output multiplier increases.  Finally, as service industries become more prominent, the average output multiplier decreases.  The predictions here suggest that, all else equal, an economy will accelerate its growth during the manufacturing stage and relax to slower growth when it becomes more developed.  In Fig.~\ref{fig_growths_v_trophicDepth}a, developed economies have low average output multipliers and low growth rates while economies that are developing a strong manufacturing sector, such as China or Slovakia, tend to have high average output multipliers and high growth.  The WIOD \cite{Timmer2015} does not contain data for undeveloped countries and we cannot confirm that their average output multipliers are low, though it would be surprising if it were otherwise.

To get a feel for the amplification to growth associated with production chains we consider some rough figures.  In Fig. \ref{fig_trophic_depth_over_time}, China's average output multiplier increases from 3.8 to 5 from 1995 to 2009.  Had it not done this, Eq. \eqref{eq_growth_result} implies that its growth rate at the end of the period would have been reduced by a factor $3.8/5 \approx 0.76$.  More drastically, suppose China had the output multiplier of the United States of about 2.5, and as a rough figure take China's output multiplier over the period to be its midpoint around 4.4.  If China otherwise had the same average productivity improvement, its growth would have been smaller by a factor $2.5/4.4 \approx 0.57$, i.e. close to half of its growth during the period would have been lost.

The results suggest that differences in the average output multiplier have been an important factor in the large income differences that exist across countries.  These income differences originated in changes that economies underwent during industrialization \cite{Bairoch1982,Galor2011}, when we would  expect production chains to have started becoming more developed.  The potential for the average output multiplier to help explain disparate income levels has not gone unnoticed.  Jones \cite{Jones2011,Jones2013} notes that accounting for intermediate goods in models of aggregate production can generate large multiplier effects, with values that help explain observed differences in income.

Our results suggest that policy-makers may want to design network-aware industrial policies, targeting particular nodes in the production network.  Our results per se do not rationalize policy intervention, but could be combined with a theory of distortions in the network such as sector-specific credit market distortions, product or input market imperfections, or, more generally, wedges.  We conjecture that in certain settings policy-makers would target nodes relying on longer chains of production, though network-targeting prescriptions may depend in detail on the nature of the distortions (e.g. size, functional form, the loss function to be minimized, and so on.)  Arguments for such network targeting policies are developed further in other works (e.g. \cite{Liu2019,bigio2020distortions,baqaee2020productivity}).

Our results show that the structure of a production network, taken as given, can serve as a proximate cause of growth differences across countries. A natural follow-up question is how the production network evolves and how two-way causal relations between growth and production networks function. For example, growth over long periods usually comes with shifts in the consumption basket (43), which will slowly change the output multiplier of an economy as noted earlier. Furthermore, one may also expect rising real wages to drive innovations that reduce labor needs relative to intermediate goods \cite{Kennedy1964}. Such changes will tend to lengthen production chains over time. Finally, international trade, by inducing changes in both production and final demand patterns, offers another potential source of dynamics in the production network. For example, in our simple framework, when two countries open to trade their average output multipliers will become more similar; see Supplementary Information. This suggests that trade openness induces cross-country convergence in growth rates through changes in the production network, a result echoing previous arguments in the economic growth literature \cite{BenDavid1993}. In all, the results here call for further investigations that include endogenizing the slow evolution of production networks over the growth process, and further exploring the role of production networks in long-term growth.

\section{Materials and methods}

\subsection{Data} 
We use data from the World Input-Output Database (WIOD)  \cite{Timmer2015}, which contains worldwide input-output tables across time for 35 industries in 40 countries, together accounting for around 88\% of world GDP.  Our analysis covers the period 1995 - 2009.  We excluded 2010 and 2011 from analysis because a large number of countries lacked data on labor compensation required to compute the output multipliers (see below).  The data also include  production price indices from which we computed rates of price change.  Since the period 2007-2009 may be regarded as exceptional because of the Great Recession we also checked the effect of excluding these years, finding very similar results (Supplementary Materials).

\subsection{Output multipliers}
We treated the world as one large economy and examined the $1400 \times 1400$ matrix $A$ of input coefficients corresponding to all industries in all countries.  We took the Leontief inverse and computed the $1400$-dimensional vector $\vec{\L} = (I - A^T)^{-1} \vec{1}$ whose elements give the output multiplier of each industry in each country.  Industries and their output multipliers are listed in Supplementary Information Table~S1.

We considered two ways of computing industry output multipliers based on two interpretations of the labor coefficient $\tilde{\vec{\ell}}$.  In the first case we interpret $\tilde{\vec{\ell}}$ to account for all factor payments to households (value added, row code r64).  In the second case we interpret $\tilde{\vec{\ell}}$ to account for households' labor income only (using the labour compensation field and WIOD exchange rates to convert to U.S. dollars).  In each case the total expenditures $M_j$ of each industry $j$ were computed, either including or excluding the non-labor factor payments of $j$, and then its input coefficients computed as $a_{ij} = M_{ij} / M_j$ where $M_{ij}$ is industry $j$'s payments to industry $i$.  The results were qualitatively similar either way, and the results reported here use the latter calculation.  The main difference between the approaches is that the output multipliers are smaller when including all payments to households.  This increases the share of payments made to the household sector, thus shortening average path lengths.

WIOD did not contain data for labor and capital income separately for the Rest Of the World (ROW) region.  We compared the results of excluding ROW altogether with including it using an assumed fraction of value added to represent labor income.  We found very similar results either way.  Results shown are based on including ROW with an assumed labor fraction 0.5, similar to the global average (0.57 in 2009) computed across the WIOD countries.

The average output multiplier of each country is the weighted sum of the output multipliers of its industries.  The weight of industry $i$ in country $c$ was given by the share of $i$ in $c$'s contribution to world final demand, i.e. $Y_i/\sum_{j\in c} Y_j$ where $Y_i$ is the world final demand for good $i$.  The final demand $Y_i$ accounts for consumption and investment payments by all countries (i.e. column codes c37-c42, summed over countries) and excludes net exports, since in WIOD the latter are accounted for within the input-output table.  Countries and their average output multipliers are listed in Supplementary Information Table S2.

\subsection{Price changes and productivity growth rates}
The change $r_i'$ in the nominal price of industry $i$ was computed as the annual log changes in the industry's gross output price index.  The wage rate in a country was computed as the ratio of the total labor income earned to total hours worked by industries in the country, and the log change in the wage rate was computed to give $\rho_c$.  The rate of change of the real price of industry $i$ in country $c$ was then computed as $r_i = r_i' - \rho_{c(i)}$ where $c(i)$ is the country $c$ to which industry $i$ belongs.

Productivity improvements rates were estimated as $\vec{\hat{\gamma}} = (A^T - I)\vec{r}$.  This method represents a dual approach to estimating productivity changes \cite{Jorgenson1987,OECDProductivityHandbook2001}, in which the average growth rate of an industry's input prices is computed and the growth of its output price is subtracted off, with the difference ascribed to improvements by the industry.  An industry $j$'s productivity improvement $\gamma_j$ is its direct component of cost reduction, while the indirect component is the cost reduction $\sum_i r_i a_{ij}$ it inherits through inputs.  At the country level, the average productivity improvement rate $\tilde{\gamma}_c$ for country $c$ was estimated as $\hat{\tilde{\gamma}}_c = \sum_{i \in c} \eta_i \,\hat{\gamma}_i$, where $\eta_i$ is the share of industry $i$ in the gross output of country $c$ to which it belongs.


\subsection{Growth rates}  We computed country growth rates with data from the World Bank's World Development Indicators \cite{WorldBank}.  GDP in current local currency was deflated using each country's GDP deflator, then divided by hours worked by persons engaged across all industries using WIOD socioeconomic data.  Annual growth rates were then computed as the log change between consecutive years.  Data for Taiwan was unavailable from the World Development Indicators, and we instead used data from the Penn World Tables \cite{PennWorldTables}.

\subsection{General equilibrium framework for Eqs. (3)-(5)}  
In our baseline model (presented in the Supplementary Information) accounting identities and input-output relationships are used as the starting point to derive Eq. \eqref{eq_priceDynamicsRecursive}, from which the predictions Eqs. \eqref{eq_expectedPriceChange} - \eqref{eq_growth_result} follow.  This approach has the advantage of enabling powerful results within the simple framework of accounting relationships.  However, Eq. \eqref{eq_priceDynamicsRecursive} can also be readily obtained in a general equilibrium framework following e.g. Long and Plosser \cite{Long1983}, Acemoglu et al. \cite{Acemoglu2012}, Baqaee and Farhi \cite{Baqaee2019}.  This reinforces the fact that, for the central predictions of this paper, we do not need to take a stand on the functional form of utility and production functions.  In this section our objective is to show how Eq. \eqref{eq_priceDynamicsRecursive} can be obtained in a general equilibrium framework.  From either framework, the predictions Eqs. \eqref{eq_expectedPriceChange} - \eqref{eq_growth_result} follow as shown in the Supplementary Information.

Let $\vec{X}_i \equiv (X_{i1}, \,\ldots,\, X_{iN}, L_i)$ denote the vector of input rates to industry $i$, and assume non-joint production.  Producer $i$ has a production function $f_i(\vec{X}_i,t)$ that at any given time $t$ characterizes the best available (i.e. Pareto efficient) production methods.  

Industries are assumed to be price-takers who maximize profits at prevailing prices.  The demand for inputs by industry $i$ is
\begin{align}
\vec{X}_i(\vec{p},t) \equiv \arg\max_{\vec{X}_i} \;f_i(\vec{X}_i,t) p_i - \vec{X}_i \cdot \vec{p},
\label{eq_profit_maximization}
\end{align}
where $\vec{p} = (p_1, \ldots, p_N, w)$ is the vector of prices.  Households are assumed to maximize a utility function $U(\vec{C})$ subject to the budget constraint $\vec{C}\cdot \vec{p} = L w$, yielding the vector $\vec{C}(\vec{p},L)$ of households' demand functions for each good.  

At equilibrium, prices $\vec{p}$ are such that all markets (goods and labor) clear:
\begin{align}
\sum_j X_{ji}(\vec{p},t) + C_i(\vec{p},L) &= f_i\big(\vec{X}_i(\vec{p},t), t\big)		\quad\text{for all $i$} \label{eq_goods_clearing}\\
\sum_i L_i(\vec{p},t) &= L \label{eq_labor_clearing}.
\end{align}
Assume that industries always have production possibilities $f_i(\vec{X}_i,t)$ characterized by constant returns to scale.  Under these conditions, industries earn no economic profit at equilibrium, and activities earning deficits are not operated.  Without loss of generality, let $i$ index only industries with positive output levels.  Since these producers earn zero profit at equilibrium, revenues and expenditures satisfy the balance relation
\begin{align}
\sum_j X_{ji} p_i + C_i p_i = \sum_j X_{ij} p_j + L_i w		\qquad\text{for all $i$}.
\label{eq_producer_income_balance}
\end{align}
This is the accounting identity for industry $i$ in an input-output table, with the balancing item of valued added corresponding to the sum over primary input payments.  The model above thus gives these observed payments an interpretation as an outcome of an economy in general equilibrium.

Technology improvement, as captured by productivity growth, is represented by the advance of the Pareto frontier of the best available production methods.  This is given by the partial derivative of the production function $f_i(\vec{X}_i, t)$ with respect to time while holding inputs to $i$ fixed.  For convenience we use the hat notation to denote the growth rate of a variable, e.g. $\widehat{X}_i = \dot{X}_i / X_i$.  Taking the time-derivative of $\ln f_i$ and solving for the partial derivative with respect to time leads to $i$'s productivity growth rate,
\begin{align}
\gamma_i \equiv \frac{\p \ln f_i}{\p t} = \widehat{X}_i  - \sum_j \epsilon_{ij} \widehat{X}_{ij} - \epsilon_{iL} \widehat{L}_i,
\label{eq_productivityAlways}
\end{align}
where $\epsilon_{ij} \equiv \p \ln f_i/\p \ln X_{ij}$ is $i$'s output elasticity with respect to input $j$.

When producers are profit-maximizers under perfect competition, and production functions have constant returns to scale, the share of expenditures $a_{ji}$ that industry $i$ spends on input $j$ equals the output elasticity $\epsilon_{ij}$, a condition known as allocative efficiency.  This follows from the first-order conditions of Eq. \eqref{eq_profit_maximization}, which lead to $p_j/p_i = \p f_i/\p X_{ij}$ for all inputs $j$.  Multiplying by $X_{ij}/X_i$ gives
\begin{align}
a_{ji} = \frac{X_{ij} p_j}{X_i p_i} = \frac{\p \ln f_i}{\p \ln X_{ij}} = \epsilon_{ij}.
\label{eq_aji_and_eij}
\end{align}
Using this, Eq. \eqref{eq_productivityAlways} becomes
\begin{align}
\gamma_i = \widehat{X}_i  - \sum_j \widehat{X}_{ij} a_{ji} - \widehat{L}_i \ell_i.
\label{eq_productivityResidual}
\end{align}
Eq. \eqref{eq_productivityResidual} is the residual expression of productivity improvement, in which the  growth in output not explained by the average growth in input usage is attributed to productivity growth.

To see how productivity growth affects prices, we take the time-derivative of the logarithm of Eq. \eqref{eq_producer_income_balance}.  Noting that $\sum_j X_{ji} + C_i$ is $i$'s total production $X_i$, this leads to
\begin{align}
\widehat{X}_i + \widehat{p}_i
= \sum_j \left(\widehat{X}_{ij} + \widehat{p}_j \right) a_{ji} + (\widehat{L}_i + \widehat{w}) \ell_i.
\end{align}
Using the fact that $\ell_i = 1 - \sum_j a_{ji}$, after rearrangement we have 
\begin{align}
\widehat{p}_i - \widehat{w}
= -\left(\widehat{X}_i - \sum_j \widehat{X}_{ij} a_{ji} - \widehat{L}_i  \ell_i \right) + \sum_j (\widehat{p}_j - \widehat{w}) a_{ji}.
\end{align}
The term in parentheses is the productivity growth rate of $i$, Eq. \eqref{eq_productivityResidual}.  Defining the real price changes $r_i \equiv \widehat{p}_i - \widehat{w}$, we then have
\begin{align}
r_i = -\gamma_i + \sum_j r_j a_{ji}
\end{align}
which is Eq. \eqref{eq_priceDynamicsRecursive}.  From here, the predictions Eqs. \eqref{eq_expectedPriceChange} - \eqref{eq_growth_result} follow as described in the Supplementary Information.

\section*{Acknowledgments}
We thank Jerry Silverberg, Dario Diodato, Frank Neffke, Sultan Orazbayev, Sergey Paltsev, David Pugh, Ricardo Hausmann, Fulvio Castellacci, Ariel Wirkierman, Robert Axtell, Francois Lafond, and two anonymous referees for valuable feedback.  This work was enabled by the National Science Foundation under grant SBE0738187 and in part by the European Commission project FP7-ICT-39 2013-611272 (GROWTHCOM) and a Leading Technology Policy Fellowship from the MIT Institute for Data, Systems, and Society.  VMC gratefully acknowledges support from the Philip Leverhulme Prize, the ERC Consolidator Grant MICRO2 MACRO (\#GAP-101001221) and The Productivity Institute.

\bibliography{MasterBibliography.bib}
\bibliographystyle{naturemag}

\end{document}


\title{\bf How production networks amplify economic growth (Supplementary Materials)}
\author{James McNerney, Charles Savoie,\\Francesco Caravelli, Vasco Carvalho, J. Doyne Farmer}
\date{}
\maketitle

\clearpage
\tableofcontents

\section{Extended model description \& predictions for prices and growth}\label{sec_modelDescription}
In this section we derive the predictions of the main text.

\subsection{Background on production networks} \label{ref_background}
The predictions here can be obtained in either of two closely related modeling approaches.  One approach, which we call the baseline model, emphasizes relations familiar in productivity accounting and input-output economics, while the other uses a general equilibrium model.  The baseline model can be combined with relevant observations about input coefficients (Section \ref{sec_persistence}) and household consumption (Section \ref{sec_consumptionGrowth_v_priceReturns}) and enables powerful results in a simple framework of accounting relationships.  The general equilibrium model lets us view the production network as an optimization outcome from a set of underlying production and utility functions.  Each  approach has advantages, and either is sufficient to obtain the predictions here.

This section develops the baseline model, which is obtained by differentiating basic accounting identities. The general equilibrium model is described in the Methods section of the main text, and we also remark on it below and make contact between the two approaches as we go.  We first review notation relevant for both models, then briefly discuss the general equilibrium approach.  We then develop the description of technology dynamics and its effects in the network.  The final result of this section is the recursion relation for price changes Eq. \eqref{eq_priceChangeRecursive} (Eq. (2) in the main text), from which the predictions for price evolution and growth are then developed in Sections \ref{sec_evolutionPriceChanges} - \ref{sec_covariancePrediction}.

\paragraph{\it\normalsize Notation for production network transactions.} In this section (written for both an economics and a general scientific audience) we review background on production networks needed for our model.  We consider a simple closed economy (i.e. having no imports or exports) with no government or financial sector.  The economy consists of a set of nodes representing industries and households, and a set of directed, weighted links representing their transactions.  One can think of these links either in physical terms, with edges corresponding to the flows of goods, or in monetary terms, with edges corresponding to the flows of money.  Goods and money flow in opposite directions.  Each industry consumes a set of input goods and transforms them into a single output good.%
\footnote{Since there is one good for each industry, a node can represent either an industry or the good it produces.} %
One node represents the household sector, which uses final consumption goods made by industry nodes and produces labor.

Industries are indexed by $i$ and $j$.  Goods flows (per year, say) are denoted by $X$ and money flows by $M$,
\begin{align}
X_{ij} = \text{flow of goods from $j$ to $i$}
\qquad\text{and}\qquad
M_{ij} = \text{flow of money from $j$ to $i$},
\end{align}
which are related by the identity $M_{ij} = X_{ji} p_i$, where $p_i$ is the nominal price of good $i$.

We neglect capital accumulation, investment, savings and taxes, and assume that all money flowing into a node is immediately spent.  The money flowing into node $i$ then equals the money flowing out.  For industries, this can be expressed as
\begin{align}
\sum_j X_{ji} p_i + C_i p_i = \sum_j X_{ij} p_j + L_i w		\qquad\text{for all $i$},
\label{eq_producerBalanceRelation}
\end{align}
where $C_i$ is the flow of good $i$ to households, $L_i$ is the labor supplied to industry $i$, and $w$ is the price of labor, i.e. the wage rate.  For the household node, the balance of payments can be expressed as $Lw = \sum_j C_j p_j$, where $L = \sum_i L_i$ is the total labor supplied to all industries.

\paragraph{\it\normalsize A general equilibrium framework.}  Eq. \eqref{eq_producerBalanceRelation} can be studied through the lens of a general equilibrium model.  Let $\vec{X}_i \equiv (X_{i1}, \,\ldots,\, X_{iN}, L_i)$ denote the inputs to industry $i$.  Assume non-joint production, and let $f_i(\vec{X}_i,t)$ denote the production function of industry $i$, which characterizes the best available (i.e. Pareto efficient) production methods at a given time $t$.  Taking industries to be price-takers who maximize profits at prevailing prices, the demand for inputs by industry $i$ is
\begin{align}
\vec{X}_i(\vec{p},t) \equiv \arg\max_{\vec{X}_i} \;f_i(\vec{X}_i,t) p_i - \vec{X}_i \cdot \vec{p},
\label{eq_profit_maximization_SM}
\end{align}
where $\vec{p} = (p_1, \ldots, p_N, w)$ is the vector of prices.  Households consume a basket of goods $\vec{C}$, and are assumed to maximize a utility function $U(\vec{C})$ subject to the budget constraint $\vec{C}\cdot \vec{p} = L w$, yielding the vector $\vec{C}(\vec{p},L)$ of households' demand functions for each good.  

At equilibrium, prices $\vec{p}$ are such that all markets (goods and labor) clear:
\begin{align}
\sum_j X_{ji}(\vec{p},t) + C_i(\vec{p},L) &= f_i\big(\vec{X}_i(\vec{p},t), t\big)		\quad\text{for all $i$} \label{eq_goods_clearing_SM}\\
\sum_i L_i(\vec{p},t) &= L \label{eq_labor_clearing_SM}.
\end{align}
Assume that industries perpetually have production possibilities $f_i(\vec{X}_i,t)$ characterized by constant returns to scale.  Under these conditions, industries earn no economic profit at equilibrium, and activities earning deficits are not operated.  Without loss of generality, let $i$ index only industries with positive output levels.  Since these producers earn zero profit at equilibrium, revenues and expenditures satisfy the balance relation $f_i(\vec{X}_i,t) p_i = \vec{X}_i \cdot \vec{p}$, or
\begin{align}
\sum_j X_{ji} p_i + C_i p_i = \sum_j X_{ij} p_j + L_i w		\qquad\text{for all $i$},\nonumber
\end{align}
which is the accounting identity for industry $i$, Eq. \eqref{eq_producerBalanceRelation}.  Thus, this model gives observed payments an interpretation as the outcome of an economy in a general price equilibrium across its markets.

\paragraph{\it\normalsize Physical and monetary input coefficients.}  In the baseline model we treat the accounting identity Eq. \eqref{eq_producerBalanceRelation} as a given.   We assume that all variables change smoothly and differentiate this expression with respect to time, leading to a relationship between price changes and changes in production dependencies (Eq. \eqref{eq_priceChangeRecursive}).  To describe these dependencies it is useful to define the \textit{physical input coefficients} of an industry, which give the amount of each input good needed per unit of output:
\begin{align}\label{eq_usage_ratio_definition}
\phi_{ij} \equiv \frac{X_{ij}}{X_i}
\qquad\text{and}\qquad
\ell_{i} \equiv \frac{L_i}{X_i}.
\end{align}
As we discuss below, technology changes can be expressed in terms of changes to these coefficients.  It is also useful to define the corresponding set of monetary \emph{input coefficients}, which give the shares of a producer's total payments spent on each input,
\begin{align}
a_{ji} \equiv \frac{X_{ij} p_j}{X_i p_i}
\qquad\text{and}\qquad
\tilde{\ell}_{i} \equiv \frac{L_i w}{X_i p_i}.
\end{align}
These shares sum to 1 for each $i$, $\sum_j a_{ji} + \tilde{\ell}_i = 1$, which can be shown by dividing Eq. \eqref{eq_producerBalanceRelation} by $i$'s revenues $(\sum_j X_{ji} + C_i) p_i = X_i p_i$.  From their definitions, it can be seen that the two sets of coefficients are related by $a_{ji} = \frac{1}{p_i} \phi_{ij} p_j$ and $\tilde{\ell}_i = \frac{1}{p_i} \ell_i w$.\footnote{Nearly all data on inter-industry transactions is gathered initially in terms of monetary payments.  The monetary input coefficients are thus more easily observed than the physical ones.  However, if a set of price indices can be constructed for goods (as with the WIOD data we use here), the physical input coefficients can also be estimated.}

The two sets of coefficients can be compactly gathered in the matrices
\begin{align}
\Phibar \equiv \begin{pmatrix} \Phi & \vec{\ell} \\ {\bf c} & 0 \end{pmatrix}
\qquad\text{and}\qquad
\bar{A} \equiv \begin{pmatrix} A & {\bf \tilde{c}} \\  \vec{\tilde{\ell}} & 0 \end{pmatrix}.
\end{align}
Here $\Phi$ and $A$ are $N \times N$ matrices collecting the $\phi_{ij}$ and $a_{ij}$.  The $N \times 1$ vector $\vec{\ell}$ collects the $\ell_i$ and the $1 \times N$ vector $\vec{\tilde{\ell}}$ collects the $\tilde{\ell}_i$.  We define $c_i = C_i / L$, household consumption of good $i$ per unit of labor provided, and $\tilde{c}_i = p_i C_i/Lw$, the share of household expenditures devoted to good $i$, and collect these into the vectors $\vec{c}$ and $\vec{\tilde{c}}$.  Using $\Phibar$, Eq. \eqref{eq_producerBalanceRelation} can be compactly written as $\vec{p} = \Phibar \vec{p}$.  Since each industry's expenditure shares sum to 1, $\bar{A}$ is a column-normalized stochastic matrix.  The elements of $\bar{A}$ can be thought of as transition probabilities for money flows in a Markov chain.  The submatrix $A$ has columns that sum to numbers less than 1, and can be understood as the substochastic matrix of an absorbing Markov chain.  This fact is useful in the various interpretations of the output multiplier (see page \pageref{sec_connectionEcology}).

\paragraph{\it\normalsize Technology improvement and changing physical input coefficients.}  Physical input coefficients change with time as technology develops and new production processes become available.  When these developments represent improvements it means that these processes are able to yield a higher quantity or quality of output from a given amount of inputs.  This means that the physical input coefficients $\phi_{ij}$ have an overall tendency to become smaller with time.  This process is noisy, as some coefficients of an industry may increase by large factors while others shrink, as innovations cause some inputs to be substituted for others.  Letting $\widehat{X} \equiv \dot{X}/X$  denote the growth rate of a variable, a simple measure of improvement in $i$ is the average growth rate of input usage across all inputs, with each input weighted by its share of expenses:
\begin{align}
\gamma_i \equiv -\left(\sum_j \widehat{\phi}_{ij} a_{ji} + \widehat{\ell}_{i}\tilde{\ell}_i\right).
\label{eq_gammaAverage}
\end{align}
The minus sign expresses the fact that improvement is positive when $i$ reduces its use of inputs.  The improvement rate $\gamma_i$ is local to node $i$, describing improvement in $i$'s production processes independent of improvements made in other industries.

\paragraph{\it\normalsize Technology improvement and productivity growth.}  Like Eq. \eqref{eq_producerBalanceRelation}, 
Eq. \eqref{eq_gammaAverage} of the baseline model can be viewed through the lens of a general equilibrium framework, in which $\gamma_i$ is producer $i$'s productivity improvement.  Technology improvement, as captured by productivity growth, is represented by shifts in the Pareto frontier of the best available production methods, which is given by the partial derivative of the production function $f_i(\vec{X}_i, t)$ with respect to time while holding inputs to $i$ fixed.  Taking the time-derivative of the logarithm of $f_i$ and solving for the partial derivative with respect to time leads to $i$'s productivity growth rate
\begin{align}
\gamma_i \equiv \frac{\p \ln f_i}{\p t} = \widehat{X}_i  - \sum_j \epsilon_{ij} \widehat{X}_{ij} - \epsilon_{iL} \widehat{L}_i,
\label{eq_productivityAlways_SM}
\end{align}
where $\epsilon_{ij} \equiv \p \ln f_i/\p \ln X_{ij}$ is $i$'s output elasticity with respect to input $j$.  When producers are profit-maximizers under perfect competition and production functions have constant returns to scale, the share of expenditures $a_{ji}$ devoted to input $j$ equals the output elasticity $\epsilon_{ij}$, a condition known as allocative efficiency.  This can be shown from the first order conditions of Eq. \eqref{eq_profit_maximization_SM}, which lead to $p_j/p_i = \p f_i/\p X_{ij}$ for all inputs $j$.  Multiplying by $X_{ij}/X_i$ then gives
\begin{align}
a_{ji} = \frac{X_{ij} p_j}{X_i p_i} = \frac{\p \ln f_i}{\p \ln X_{ij}} = \epsilon_{ij},
\label{eq_aji_and_eij_SM}
\end{align}
and using this Eq. \eqref{eq_productivityAlways_SM} becomes
\begin{align}
\gamma_i = \widehat{X}_i  - \sum_j \widehat{X}_{ij} a_{ji} - \widehat{L}_i \tilde{\ell}_i.
\label{eq_productivityResidual_SM}
\end{align}
Eq. \eqref{eq_productivityResidual_SM} is the residual expression of productivity improvement, in which the  growth in output not explained by the average growth in input usage is attributed to productivity growth.  Finally, to see that Eq. \eqref{eq_productivityResidual_SM} equals the average rate of coefficient change Eq. \eqref{eq_gammaAverage}, note that since $\phi_{ij} \equiv X_{ij}/X_i$ and $\ell_i \equiv L_i/X_i$ we have $\widehat{\phi}_{ij} = \widehat{X}_{ij} - \widehat{X}_i$ and $\widehat{\ell}_i = \widehat{L}_i - \widehat{X}_i$.  Plugging in then converts one expression to the other.

\paragraph{\it\normalsize Productivity improvement and prices.} We now show how productivity growth affects prices by taking the time-derivative of the logarithm of Eq. \eqref{eq_producerBalanceRelation} (viewed as either an accounting identity in the baseline model or as a result of an economy in general equilibrium, following the discussion on page \pageref{eq_profit_maximization_SM}.)  Noting that $X_i = \sum_j X_{ji} + C_i$ is $i$'s total production, this leads to
\begin{align}
\widehat{X}_i + \widehat{p}_i
= \sum_j \left(\widehat{X}_{ij} + \widehat{p}_j \right) a_{ji} + (\widehat{L}_i + \widehat{w}) \ell_i.
\end{align}
Since expenditure shares sum to 1, we can write $\ell_i$ as $1 - \sum_j a_{ji}$.  Plugging this in, after rearrangement one gets
\begin{align}
\widehat{p}_i - \widehat{w}
= -\left(\widehat{X}_i - \sum_j \widehat{X}_{ij} a_{ji} - \widehat{L}_i  \ell_i \right) + \sum_j (\widehat{p}_j - \widehat{w}) a_{ji}.
\label{eq_priceChangeRecursive_almost}
\end{align}
The term in parentheses is the productivity growth rate of $i$, Eq. \eqref{eq_productivityResidual_SM}.  Taking the wage rate as the numeraire, the difference $r_i \equiv \widehat{p}_i - \widehat{w}$ denotes the rate of change of the real (i.e. inflation-adjusted) price of good $i$.  We can thus write Eq. \eqref{eq_priceChangeRecursive_almost} simply as
\begin{align}
r_i = -\gamma_i + \sum_j r_j a_{ji},
\label{eq_priceChangeRecursive}
\end{align}
which is the recursion equation for price changes shown as Eq. (2) in the main text.

\subsection{Industry price changes}\label{sec_evolutionPriceChanges}
In vector form Eq. \eqref{eq_priceChangeRecursive} is $\vec{r} = -\vec{\gamma} + A^T \vec{r}$, and after solving for $\vec{r}$ we have
\begin{align}
\vec{r} = - H^T \vec{\gamma}.
\label{eq_realPriceChanges_gammaVec}
\end{align}
where $H = (I - A)^{-1}$ is the \textit{Leontief inverse}, which appears ubiquitously in input-output economics \cite{Miller2009}.

Without  loss of generality, any realized value of $\gamma_i$ can be decomposed into a sum of its average value across industries $\bar{\gamma}$ and a deviation $\Delta \gamma_i$.  Substituting this into Eq. \eqref{eq_realPriceChanges_gammaVec}, in index form we have
\begin{align}
r_i = - \bar{\gamma} \sum_j H_{ji} - \sum_j \Delta \gamma_j H_{ji} = - \bar{\gamma} \L_i - \sum_j \Delta \gamma_j H_{ji},
\label{eq_r}
\end{align}
where the quantity $\L_i = \sum_j H_{ji}$ is the \textit{output multiplier} of industry $i$, also called $i$'s total backward linkage \cite{Rasmussen1957} or downstreamness \cite{Miller2015}.  An output multiplier $\L_i$ gives the average path length of all production chains that end at industry $i$, following each path backward through inputs until it reaches households.  Under an analogy to food webs, $\L_i$ corresponds to the trophic level of industry $i$ \cite{Levine1980}.  See page \pageref{sec_connectionEcology} for a discussion of different interpretations of the output multiplier.  Computing the expectation value of $r_i$ across industries conditioned on the output multiplier, we have
\begin{align}
E[r_i | \L_i] = - \bar{\gamma} \L_i - \sum_j E\left[\Delta \gamma_j H_{ji} \big| \L_i\right].
\label{eq_r2}
\end{align}
If $\gamma_j$ is uncorrelated with $H_{ji}$, then $E[\Delta \gamma_j H_{ji} | \L_i] = E[H_{ji} | \L_i]\, E[\Delta \gamma_j | \L_i] = 0$, since by construction $E[\Delta \gamma_j] = 0$ and the lack of correlation with $H_{ji}$ also implies that $\Delta \gamma_j$ is uncorrelated with $\L_i$.%
\footnote{Empirically these correlations are low.  Looking across industries $i$, the correlation of $\vec{\gamma}$ with the $i$th column of $H$ is less than 0.07 for 95\% of industries.} %
In this case Eq. \eqref{eq_r2} reduces to
\begin{align}
E[r_i | \L_i] = - \bar{\gamma} \L_i,
\label{eq_r3}
\end{align}
i.e. the expected real price change of industry $i$ is proportional to $\L_i$, with proportionality constant $-\bar{\gamma}$.

Thus, if the correlations between the industry improvement rates and the elements of the Leontief inverse are sufficiently low, over timescales where the output multiplier remains roughly constant, we expect the long-term decline of the real prices of an industry to be proportional to its output multiplier.  This is a striking result because it connects the long-run rate at which the cost of a product falls with a purely structural property of the economy.  The output multiplier is a structural property in the sense that it depends only on the network of production relationships.

The proportionality between the expected price changes of industries and their output multipliers given in Eq. \eqref{eq_r3} is exact when improvement rates across industries are uncorrelated with the elements of the Leontief inverse.  As a counterexample, suppose all improvements rates were zero except for that of industry $k$, $\gamma_i = \gamma_0 \delta_{ik}$.  Then instead of Eq. \eqref{eq_r3} one would obtain $r_i = - \gamma_0 G_{ik}$, i.e. price changes would depend on the $k^{th}$ row of the Leontief inverse, and we would not expect a relationship between price changes and output multipliers.  As noted earlier though, the correlations between improvement rates and the Leontief inverse are weak, and Fig. 3 of the main text shows a strong relationship between expected price changes and output multipliers.

\subsection{Country growth rates}\label{sec_evolutionGrowthRates}
We now derive a relationship to GDP growth.  Balance of payments at the household node means that revenues $L w$ are equal to expenditures $\vec{C} \cdot \vec{p}$, and dividing by $L$ to solve for the wage rate gives $\vec{w} = \vec{c} \cdot \vec{p}$.  Taking the time-derivative gives $\dot{w} = \dot{\vec{c}} \cdot \vec{p} + \vec{c} \cdot \dot{\vec{p}}$, or (using the definition $c_i  \equiv C_i/L$)
\begin{align}
\dot{w} = \sum_i \frac{C_i}{L}\!\left(\frac{\dot{C}_i}{C_i} - \frac{\dot{L}}{L}\right) \!p_i +  \sum_i \frac{C_i}{L} \dot{p}_i.
\label{omegaDot}
\end{align}
Here, $\dot{C}_i/C_i \equiv g_i'$ is the growth rate of household consumption of good $i$ and $\dot{L}/L \equiv h$ is the growth rate of labor provided.  In the simple economy here, the GDP is the total expenditure by households on consumption goods,  $Y \equiv \sum_i p_i C_i$.  Let $\theta_i \equiv p_i\,C_i/Y$ denote the share of GDP devoted to good $i$.  Using these definitions, Eq. \eqref{omegaDot} can be rearranged as
\begin{align}
\dot{w} = \Big(\vec{\theta} \!\cdot\! \vec{r'} + \vec{\theta} \!\cdot\! \vec{g'} - h \Big) w,
\label{eq_wdot2}
\end{align}
where $\vec{\theta} = (\theta_1, \ldots, \theta_N)$ and $\vec{g'} = (g'_1,\ldots, g'_N)$.  The term $\vec{\theta} \cdot \vec{r'} \equiv r'$ is the average rate of increase in the prices of final goods and measures the inflation rate of the economy.\footnote{$r'$ is the growth rate of a Divisia price index for final goods.  It corresponds to either the consumer price index or the GDP deflator.  These coincide in the simple economy here with no government, financial sector, imports or exports.}  
The term $\vec{\theta} \cdot \vec{g'} \equiv g'$ is the average rate of growth in the consumption of final goods.  It represents the growth rate of real (i.e. inflation-adjusted) GDP.  We let $g \equiv g'-h$ denote the  growth rate of real GDP per unit of labor, and let $r \equiv r'-\rho$ denote the average growth rate of prices after deflating by the growth in wages.  Then after rearrangement, Eq. \eqref{eq_wdot2} can be written as simply
\begin{align}
g = -r.
\label{eq_gr_relationship}
\end{align}
Eq. \eqref{eq_gr_relationship} says that the growth rate of real GDP per unit of labor is equal to the rate at which real prices decrease.  Finally, multiplying Eq. \eqref{eq_realPriceChanges_gammaVec} by $\vec{\theta}$ to obtain $r$ and plugging it into Eq. \eqref{eq_gr_relationship} gives the growth rate in terms of the local improvement rates:
\begin{align}
g = \vec{\theta}^T H^T \vec{\gamma}.
\label{eq_growthRate_gammaVec}
\end{align}

To recast this relationship in terms of output multipliers, it is useful to first derive a relationship between total output and the Leontief inverse.  The total revenue of industry $i$ equals the sum of revenues from final and intermediate consumption,
\begin{align}
M_i\equiv Y_i+\sum_j M_{ij}.
\label{eq_gross_is_final_plus_intermediate}
\end{align} 
Writing intermediate consumption payments in terms of input coefficients as $M_{ij} = a_{ij}M_j$, Eq. \eqref{eq_gross_is_final_plus_intermediate} can be written in matrix form as $\vec{M}=\vec{Y} + A \vec{M}$ and solved to give $\vec{M}^T= \vec{Y}^T (I-A^T)^{-1} = \vec{Y}^T H^T$.  Multiplying both sides by $\vec{\gamma}/Y$, and recalling that $\vec{\theta} = \vec{Y}/Y$, this becomes
\begin{align}
\vec{\theta}^T H^T \vec{\gamma} = \frac{\vec{M}^T \vec{\gamma}}Y.
\label{almostInvariant}
\end{align}
The gross output of the economy is $\O \equiv \sum_{i=1}^N M_i$ and we define the output-weighted average improvement rate $\tilde{\gamma}$ as
 \begin{align}
\tilde{\gamma} & \equiv \sum_{i=1}^N \frac{M_i \gamma_i}{\O} = \sum_{i=1}^N \eta_i \gamma_i.
\label{def:gamma_tilde}
\end{align}
Here the weights $\eta_i = M_i/\O$ give the fractions of each industry in gross output.  Then Eq. \eqref{almostInvariant} becomes
\begin{align}
\vec{\theta}^T H^T \vec{\gamma} =  \tilde{\gamma} \frac{\O}{Y}.
\label{eq_grossOuputOverNetOutput}
\end{align}
Eq. \eqref{eq_grossOuputOverNetOutput} holds for any vector of improvement rates $\vec{\gamma}$.  In particular, choosing $\vec{\gamma} = \tilde{\gamma} (1, \ldots, 1)^T = \tilde{\gamma} \vec{1}$ shows that for a closed economy the ratio $\O/Y$ is equal to the \textit{average output multiplier}, the GDP-weighted average of the industry output multipliers:\begin{align}
\bar{\mathcal{L}} \equiv \vec{\theta}^T H^T \vec{1} = \frac{\O}{Y}.
\label{eq_aggregateOutputMultiplier}
\end{align}
This also re-derives the important result shown in Fally (2012) \cite{Fally2012} that the average output multiplier is well-defined and invariant under aggregation.%
\footnote{This also relates to an empirical relationship noted by Jones (2011) \cite{Jones2011b}, who measured average output multipliers for several countries (albeit with a slightly different definition than that given here), noting that they are closely matched by a quantity that reduces to $\O/Y$.}
We can now write (\ref{eq_growthRate_gammaVec}) as
\begin{align}
g = \tilde{\gamma} \bar{\mathcal{L}} = \tilde{\gamma} \frac{\O}{Y}
\label{eq_growth}
\end{align}
Eq. \eqref{eq_growth} cleanly separates GDP growth into two terms, one which depends on improvement rates, and another that depends purely on structural properties of the production network.

\paragraph{\it\normalsize Response of households and second order effects.}\label{sec_secondOrderEffects}  Note that arriving at Eq. \eqref{eq_growth} does not require assuming a particular form for households' utility function $U(\vec{C})$.  Its independence from the choice of utility function is also reflected in Eq. \eqref{eq_growthRate_gammaVec} by its dependence only on the consumption shares $\vec{\theta}$, and not on any derivatives of $\vec{\theta}$, since any utility function can produce a given set of consumption shares with a suitable choice of share parameters.  The independence of this growth expression from the utility function may be surprising, since it is clear that consumers' responses to changing prices and increased levels of wealth could have a growth effect over time.  Changes in consumption patterns could shift the average production length of final goods delivered to households, altering both the average output multiplier and the growth rate of Eq. \eqref{eq_growthRate_gammaVec}.

To see how such effects enter, note that Eq. \eqref{eq_growthRate_gammaVec} and Eq. \eqref{eq_growth} are first order expressions in that they characterize the first time-derivative of output.  The effect of the household response on growth can be seen from taking the calculation to second order by computing the second time-derivative of output, or equivalently the first time-derivative of the growth rate.  Rewriting Eq. \eqref{eq_growthRate_gammaVec} in terms of real price changes as $g = - \vec{\theta} \cdot \vec{r}$, the time-derivative is $\dot{g} = -\dot{\vec{\theta}} \cdot \vec{r} - \vec{\theta} \cdot \dot{\vec{r}}$, where the behavior of $\dot{\vec{\theta}}$ will depend on household preferences.  To get a feel for this dependence, consider the case where households have a CES utility function $U(\vec{C}) = \left(\sum_i b_i C_i^{\frac{\sigma - 1}{\sigma}} \right)^{\frac{\sigma}{\sigma - 1}}$, where $\sigma$ is the elasticity of substitution and the $b_i$ are share parameters.  Maximizing utility subject to the budget constraint $\vec{p} \cdot \vec{C} = wL$ leads to consumption shares given by $\theta_i = b_i^\sigma p_i^{1-\sigma} / \sum_j b_j^\sigma p_j^{1-\sigma}$, whose time-derivative is $\dot{\theta}_i = (1-\sigma) \theta_i (r_i - \vec{\theta} \cdot \vec{r})$.  Plugging this into $\dot{g} = -\dot{\vec{\theta}} \cdot \vec{r} - \vec{\theta} \cdot \dot{\vec{r}}$ then leads to
\begin{align}
\dot{g} = -(1-\sigma) \text{Var}_{\vec{\theta}}(\vec{r}) - \vec{\theta} \cdot \dot{\vec{r}},
\label{eq_outputSecondOrder}
\end{align}
where $\text{Var}_{\vec{\theta}}(\vec{r}) \equiv \sum_i \theta_i r_i^2 - \left(\sum_i \theta_i r_i\right)^2$ is the consumption shares-weighted variance of the price changes across industries.  \jm{(For an expansive development of first order and second order characterizations of the effects of productivity shocks, see Ref. \cite{Baqaee2019}.)}

The first term in Eq. \eqref{eq_outputSecondOrder} shows how the rate of growth may be sped up or slowed down depending on how households adapt consumption patterns to changing relative prices.  In the Cobb-Douglas case where $\sigma = 1$, households' consumption shares remain constant as prices change.  As we would expect, the first term above vanishes in this case.  In contrast, if consumers have Leontief preferences, then $\sigma = 0$, and the growth rate is reduced over time.  This happens because households reap no benefits from substituting out goods whose real prices fall at relatively slower rates.  Since preferences are Leontief, the shares of the fastest-improving goods become steadily smaller in the consumption basket, lowering the impact of further cost reductions in these goods over time.  Intuitively, the size of this effect depends on the size of the dispersion in rates of price reduction among goods, $\text{Var}_{\vec{\theta}}(\vec{r})$.  Similarly, the growth rate could change over time due to shifts in the consumption basket as households reach higher levels of wealth.  This effect could be examined by considering non-homothetic utility functions.

In Section \ref{sec_consumptionGrowth_v_priceReturns} we also explore these issues empirically.  This includes estimating the substitution elasticity, the price change dispersion term, and the potential drag on growth from changes to consumption patterns.  The results suggest that the effect of these changes on growth over 10 years is modest.  Correspondingly, this suggests that the first order equation Eq. \eqref{eq_growth} may provide a good approximation over such periods or longer.

\subsection{Dispersion in price changes across industries}\label{sec_priceChangeVariation}
Fig. 4 of the main text shows that the price changes of individual industries vary considerably around the mean behavior $E[r_i|\L_i]$.  In this section, we study this dispersion by examining the standard deviation of price changes among industries with a given output multiplier, $\sigma_{r_i|\L_i}$.  We derive the theory's prediction for how $\sigma_{r_i|\L_i}$ should vary with the output multiplier.  We also show how the dispersion in price changes is expected to change with the passage of time. It will be seen that $\sigma_{r_i|\L_i}$ grows with $\L_i$, and shrinks at a rate $1/\sqrt{T}$ over a time horizon $T$.

First, note that the Leontief inverse $H = I + A + A^2 + \cdots$ can be written as $H = I + A(I + A + A^2 + \cdots) = I + AH$.  Plugging this into Eq. \eqref{eq_realPriceChanges_gammaVec}, in index form we have
\begin{align}
r_i = -\gamma_i  - \sum_j \gamma_j (AH)_{ji}.
\label{eq_priceChangeSeparated}
\end{align}
Writing $r_i$ in this way allows us to separate the direct effect of $i$'s productivity improvement $\gamma_i$ from all indirect effects.%
\footnote{In addition to the direct effect of $\gamma_i$ on $r_i$, $\gamma_i$ also has indirect effects though $i$'s use of its own production as an input, as well as through other cycles of production that use the production ouput of node $i$.  These indirect effects of $\gamma_i$ on $r_i$ are accounted for in the second term of Eq. \eqref{eq_priceChangeSeparated}.} %
Taking the variance of $r_i$ while conditioning on the output multiplier gives
\begin{align}
\Var [r_i | \L_i] = 
\Var [ \gamma_i | \L_i ]
+ 2\Cov \left[ \gamma_i, \sum_j \gamma_j (AH)_{ji} \Big| \L_i \right]
+ \Var \left[ \sum_j \gamma_j (AH)_{ji} \Big| \L_i \right].
\end{align}
The first term accounts for variation in the direct improvement benefits $\gamma_i$ across industries with the same output multiplier, while the last term accounts for the variation in their inherited benefits $(\vec{\gamma}^T A + \vec{\gamma}^T A^2 + \cdots)_i$.

The second and third terms above depend on $\L_i$ because the column sums of $AH$ are equal to the output multipliers minus 1, i.e. $\sum_j (AH)_{ji} = \sum_j (H - I)_{ji} = \L_i - 1$.  The larger that $\L_i$ is, the greater the elements of the $i$th column of $AH$, and the larger $\sum_j \gamma_j Z_{ji}$ will tend to be.  However, conditioning on $\L_i$ means that the $i$th column of $AH$ is constrained to equal $\L_i - 1$, allowing us to factor out the $\L_i$ dependence from the variance and covariance terms above.  To remove this dependence, we first factor $\L_i - 1$ from the elements of $AH$, i.e. $(AH)_{ji} = (\L_i  - 1) Z_{ji}$, where the elements $Z_{ji}$ vary from 0 to 1 and obey the normalization $\sum_j Z_{ji} = 1$.  This leads to
\begin{align}
\Var [r_i | \L_i] = 
\Var [ \gamma_i | \L_i ]
+ 2\Cov \left[ \gamma_i, (\L_i - 1) \sum_j \gamma_j Z_{ji} \Big| \L_i \right]
+ \Var \left[ (\L_i - 1) \sum_j \gamma_j Z_{ji} \Big| \L_i \right].
\end{align}
From this equation we immediately see one prediction about the dispersion of price changes.  Because the variance is scaled by the output multipliers, this dispersion will be greater among industries with larger output multipliers.

To write this dependence in a more interpretable form we take advantage of the fact that $\L_i$ is a constant in the variances and covariances over industries, and may be pulled outside:
\begin{align}
\Var [r_i | \L_i] = 
\Var [ \gamma_i | \L_i ]
+ 2(\L_i - 1) \, \Cov \left[ \gamma_i, \sum_j \gamma_j Z_{ji} \Big| \L_i \right]
+ (\L_i - 1)^2 \, \Var \left[ \sum_j \gamma_j Z_{ji} \Big| \L_i \right].
\label{eq_varianceOf_r_stop1}
\end{align}
We denote the variance over $\gamma_i$ as $\sigma_{\gamma,\text{direct}}^2$, and denote the  variance over $\sum_j \gamma_j Z_{ji}$ as $\sigma_{\gamma, \text{inherited}}^2$.  The covariance term can be written as $\rho \sigma_{\gamma,\text{direct}} \sigma_{\gamma, \text{inherited}}$ where $\rho$ is the correlation between $\gamma_i$ and $\sum_j \gamma_j Z_{ji}$.  Eq. \eqref{eq_varianceOf_r_stop1} then becomes
\begin{align}
\Var [r_i | \L_i]
&= \sigma_{\gamma,\text{direct}}^2
+ 2 \rho \sigma_{\gamma,\text{direct}} \sigma_{\gamma,\text{inherited}} (\L_i - 1)
+ \sigma_{\gamma,\text{inherited}}^2 (\L_i - 1)^2,
\end{align}
and the standard deviation is
\begin{align}
\sigma_{r_i|\L_i} = \left[\sigma_{\gamma,\text{direct}}^2
+ 2 \rho \sigma_{\gamma,\text{direct}} \sigma_{\gamma,\text{inherited}} (\L_i - 1)
+ \sigma_{\gamma,\text{inherited}}^2 (\L_i - 1)^2 \right]^{1/2}.
\label{eq_stdev_priceChanges}
\end{align}

In the WIOD, $\sigma_{\gamma,\text{direct}}$ is larger than $\sigma_{\gamma,\text{inherited}}$, with $\sigma_{\gamma,\text{inherited}} / \sigma_{\gamma,\text{direct}} \sim 1/4$.  Expanding Eq. \eqref{eq_stdev_priceChanges} in powers of $(\sigma_{\gamma,\text{inherited}} / \sigma_{\gamma,\text{direct}})$, $\sigma_{r_i|\L_i}$ can be approximated as 
\begin{align}
\sigma_{r_i|\L_i} \approx \sigma_{\gamma,\text{direct}}
+ \rho \sigma_{\gamma,\text{inherited}} (\L_i - 1)
+ \frac{1}{2} \frac{\sigma_{\gamma,\text{inherited}}^2}{\sigma_{\gamma,\text{direct}}} (\L_i - 1)^2.
\label{eq_stdev_priceChanges_approximate}
\end{align}
Eq. \eqref{eq_stdev_priceChanges_approximate} separates three sources of variation in price changes among industries with the same output multiplier.  The first term accounts for the variation in productivity improvement among these industries, while the second and third terms account for variation in inherited benefits.  The second term is driven by industries' self-payments, which cause a correlation $\rho$ between $\gamma_i$ and $\sum_j \gamma_j Z_{ji}$ due to the diagonal term $\gamma_i Z_{ii}$ in the latter.  An industry's consumption of its own output allows it to realize an inherited benefit from its own productivity growth, and thus variation in industries' productivity growth rates adds to the variation in inherited benefits across industries.  This self-consumption need not be large to induce a substantial correlation $\rho$, which in the WIOD has a value around 0.5.  Self-payments tend be relatively strong in input-output tables, which increases the strength of the diagonal elements $Z_{ii}$.  At the same time, while the contribution of the 1399 off-diagonal terms $\sum_{j\neq i} \gamma_j Z_{ji}$ may be larger than that of $\gamma_i Z_{ii}$, summing the off-diagonal elements has an averaging effect on their contribution, reducing its variation from one industry $i$ to another.  The two effects combine to give $\gamma_i$ and $\sum_j \gamma_j Z_{ji}$ a high correlation $\rho$.

Next, we extend Eq. \eqref{eq_stdev_priceChanges_approximate} to predict how the cross-industry dispersion in price changes shrinks with time.  Assuming independent and identically-distributed price changes in each year, the variance in the total price change will grow linearly with $T$.  Our plots in the main text instead show the average rate of price change over a period, i.e.
\begin{align}
r_i(t,t+T) \equiv \frac{1}{T} \sum_{t=1}^{T} r_i(t).
\end{align}
The variance of $r_i(t,t+T)$ is thus
\begin{align}
\sigma_{r_i(t,t+T) | \L_i}^2 
\equiv \Var\left[ \frac{1}{T} \sum_{t=1}^{T} r_i(t) \Big| \L_i \right] 
= \frac{1}{T^2} \sum_{t=1}^{T} \sigma_{r_i | \L_i}^2
= \frac{1}{T^2} T \;\sigma_{r_i | \L_i}^2,
\end{align}
and therefore
\begin{align}
\sigma_{r_i(t,t+T) | \L_i} \approx \frac{1}{\sqrt{T}} \Big[ \sigma_{\gamma,\text{direct}}
+ \rho \sigma_{\gamma,\text{inherited}} (\L_i - 1)
+ \frac{1}{2} \frac{\sigma_{\gamma,\text{inherited}}^2}{\sigma_{\gamma,\text{direct}}} (\L_i - 1)^2 \Big].
\label{eq_priceSigmaWithTime_SM}
\end{align}

Eq. \eqref{eq_priceSigmaWithTime_SM} makes two important predictions about the variation in price changes across industries that have the same output multiplier.  First, over any given period of time, this variation will be broader among industries with larger output multipliers.  Second, the variation in the time-average price change will shrink with time, at a rate $1/\sqrt{T}$.  In terms of Fig. 4 of the main text, this latter prediction means that, within any given output multiplier bin, the dispersion in price changes will decrease as the time horizon grows.  Importantly, this implies that the expectation value Eq. \eqref{eq_r3} should become an increasingly good description of price changes among industries as time passes.  In Fig. \ref{fig_priceChanges_v_OMs_full}, we examine these predictions of Eq. \eqref{eq_priceSigmaWithTime_SM} using observations from the WIOD over a range of time horizons.

\subsection{Covariance of price changes}\label{sec_covariancePrediction}
The correlation of price movements with network structure suggests that there will also be structural correlations in the price movements of different industries with one another.  While not central to our results, we develop this prediction and demonstrate that it also is supported by  WIOD data.  The equation for price changes, Eq. \eqref{eq_realPriceChanges_gammaVec}, leads to a prediction for the co-movement of prices in the network as characterized by their covariance $R_{ij} = E_t\left[ r_i(t) r_j(t) \right] - E_t[r_i(t)] E_t[r_j(t)]$, where $E_t[\cdot]$ denotes an expectation value over time.  Eq. \eqref{eq_realPriceChanges_gammaVec} gives the price change of good $i$ as $r_i = -\sum_m H_{mi} \gamma_m$.  Multiplying by $r_j$ and taking the expectation value results in $E_t\left[ r_i r_j \right] = \sum_{m,n} H_{mi} E_t\left[\gamma_m \gamma_n\right] H_{nj}$.  Subtracting $E_t[r_i] E_t[r_j] = \sum_i E_t[\gamma_m] H_{mi} \sum_i E_t[\gamma_n] H_{ni} $ then results in
\begin{align}
E_t\left[ r_i r_j \right] - E_t[r_i] E_t[r_j]
= \sum_{m,n} H_{mi} \Big( E_t\left[\gamma_m \gamma_n\right] - E_t[\gamma_m] E_t[\gamma_n] \Big) H_{nj}.
\label{eq_covariancePrediction}
\end{align}
This can be written in matrix form as $R = H^T G H$ where $R_{ij}$ are the covariances of the price changes and $G_{mn}$ are the covariances of the productivity growth rates.

The diagonal elements of $G$ are the time variances of productivity growth rates, while the off-diagonal elements are their covariances.  To study the predictions of Eq. \eqref{eq_covariancePrediction}, we decompose $G$ into its diagonal elements $D$ and off-diagonal elements $O$, $G = D + O$.  Plugging this into $R = H^T G H$ leads to a corresponding decomposition of the price change covariances,
\begin{align}
R = H^T  D H + H^T O H.
\label{eq_Rfull}
\end{align}
The first term above shows that price changes would be predicted to co-vary even if there were no covariance between productivity growth rates.  The price changes of goods depend on productivity improvement throughout the network in a way that depends on the network's structure, as captured by $H$.  Taking the expectation value of Eq. \eqref{eq_Rfull} across industries, and assuming that the expected value of the second term is zero, we have
\begin{align}
E[R] 
= H^T D H.
\end{align}
We could also consider the further simplification where all industries have the same time variance of productivity improvement rates $v_{\gamma}$.  In this case, the expression for $R$ would become especially simple, reducing to $R = v_{\gamma} H^T H$.  This increases the similarity to the earlier prediction for the expected value of price changes, in that the expected covariances become proportional to a factor that depends only on network structure, $H^T H$.  Empirically, comparing actual price covariances to predicted values for all pairs of industries leads to about 1 million observations, for which we find a slope of 1.14 with a negligible $p$-value (below our machine's precision).

\newpage
\section{Supplementary discussion}

\subsection{Mathematical representations of output multipliers}\label{sec_connectionEcology}
Output multipliers can be written in several forms, each drawing attention to different ways of viewing these quantities.  Here we discuss three of these forms and how they are connected.  One way of writing an output multiplier $\L_i$ is in the recursive form
\begin{align}
\L_i = 1 + \sum_j \L_j a_{ji}.
\label{eq_trophic_level_ecology}
\end{align}
This way of writing the output multiplier emphasizes a parallel with ecology, where Eq. \eqref{eq_trophic_level_ecology} is often used to define a species' trophic level.  In this context, $a_{ji}$ is the energy fraction of species $j$ in the diet of species $i$.  The trophic level of a species may be defined in relation to that of others, with species $i$ having a trophic level $\L_i$ that is 1 greater than the average trophic level of the species it eats \cite{Levine1980,Adams1983,Polis1991,Guimera2010}.  Similarly, in the context of production networks, $a_{ji}$ is the fraction of good $j$ in the expenditures of industry $i$, and the output multiplier of an industry is 1 greater than the average output multiplier of the industries from which it buys inputs.

Another way to write the output multiplier comes from putting (\ref{eq_trophic_level_ecology}) in vector form as $\vec{\L} = \vec{1} + A^T\vec{\L}$ and solving for $\vec{\L}$:
\begin{equation}
\vec{\L} = (I - A^T)^{-1} \vec{1} = H^T \vec{1}.
\label{eq_outputMultiplierInterpretation}
\end{equation}
This is the usual expression for output multipliers, as column sums of the Leontief inverse $H = (I - A)^{-1}$.  This form arises naturally in input-output economics, where the output multiplier $\L_i$ is used to predict the changes in the gross output of an economy that come from an increase in the final demand for good $i$ \cite{Rasmussen1957,Miller2009}.  The dynamics that lead to this prediction are different from those we study here, since the productivities of industries are taken to be constant, and changes to final demand are instead the stimulant for increases in (gross) output.  The same network metric appears in both models, reflecting the fact that both depend on the propagation of effects along production chains.  From Eq. \eqref{eq_trophic_level_ecology} and Eq. \eqref{eq_outputMultiplierInterpretation} it is also possible to see that the output multiplier is a special case of Katz centrality \cite{Katz1953}. 

A third way to write the output multipliers emphasizes their meaning as a network path length.  The elements of the input matrix $A$ and the coefficients $\vec{\tilde{\ell}}$ can be interpreted as transition probabilities in a network of money flows \cite{Leontief1993}, and the relationship of Eq. \eqref{eq_trophic_level_ecology}-Eq. \eqref{eq_outputMultiplierInterpretation} to path lengths is a classic result of the theory of absorbing Markov chains \cite{Kemeny1960}.  To show this, we first derive the identity $(I - A^T)^{-1}\vec{\tilde{\ell}} = \vec{1}$:
\begin{align}
\left[\big(I - A^T\big)^{-1}\vec{\tilde{\ell}}\right]_i &= \sum_j \tilde{\ell}_j \left[\big(I - A\big)^{-1}\right]_{ji} = \sum_{k=0}^\infty \sum_j \tilde{\ell}_j \big(A^k\big)_{ji} = 1.
\label{eq_specialIdentity_derived}
\end{align}
The last equality comes from noting that $\sum_j \tilde{\ell}_j \big(A^k\big)_{ji}$ is the probability that a unit of currency starting at industry node $i$ arrives at the household node in exactly $k+1$ steps ($k$ steps between industries, plus a final step to households). Summing this over all possible path lengths $k$ therefore gives the probability of ever reaching households, which is 1.  Next, we substitute $(I - A^T)^{-1}\vec{\tilde{\ell}} = \vec{1}$ into $\vec{\L} = (I - A^T)^{-1}\vec{1}$ to obtain $\vec{\L} = (I-A^T)^{-2}\vec{\tilde{\ell}}$, and note that the matrix $(I-A^T)^{-2}$ has an infinite series form $\sum_{k=1}^\infty k \left(A^T\right)^{k-1}$.  Using this series expansion leads to
\begin{align}
\L_i = \sum_{k=1}^\infty k \sum_{j=1}^n \tilde{\ell}_j (A^{k-1})_{ji}.
\label{eq_pathlength}
\end{align}
This expression shows that $\L_i$ is an average path length in a Markov chain.  The inner sum is the probability that a random walk starting at node $i$ will arrive at the household node after exactly $k$ steps.  Summing over $k$, Eq. \eqref{eq_pathlength} thus gives the mean number of steps a random walk starting at node $i$ takes to reach households.

\subsection{Hulten's theorem and Domar weights}\label{sec_Hulten}
The model we present here expands the scope of a classic result known as Hulten's theorem \cite{Hulten1978}, which relates the  aggregate rate of productivity change to the improvement rates of individual producers.  Hulten's theorem states that the rate of increase of total factor productivity $T$ is a weighted sum of the productivity improvement rates of industries $\gamma_i$, with weights $M_i/Y$ originally proposed by Domar \cite{Domar1961}:
\begin{align}
\frac{\dot{T}}{T} = \sum_{i = 1}^N \frac{M_i}{Y} \gamma_i.
\label{eq_TFP_growth_rate}
\end{align}
$T$ captures the amount of aggregate output produced that is not explained by amounts of primary production inputs used.  In a simple model where productivity improvements drive growth, the growth rate $g$ is equal to $\dot{T}/T$.

The Domar weights sum to a number greater than 1, reflecting a multiplier process involving intermediate goods \cite{Hulten1978}.  The amplification of aggregate output is the same one that we have derived here.  The relationship to (\ref{eq_growth}) can be seen by writing out the definition of $\tilde{\gamma}$ (Eq. \eqref{def:gamma_tilde}):
\begin{align}
g = \tilde{\gamma} \bar{\L} = \left(\sum_i \frac{M_i}{\O} \gamma_i\right) \frac{\O}{Y}
= \sum_i \frac{M_i}{Y} \gamma_i.
\label{eq_consistencyWithHulten}
\end{align}
The theory here thus reproduces Hulten's theorem as a side effect, after starting from a simple mechanism for technological improvement.  The theory also does quite a bit more, because our formulation cleanly separates structural properties from improvement rates and makes the relationship between growth and the average output multiplier clear.  Comparing Eq. \eqref{eq_TFP_growth_rate} and Eq. \eqref{eq_consistencyWithHulten} shows that $\dot{T}/T$ corresponds to $\tilde{\gamma} \bar{\L}$, i.e. the model decomposes the \emph{aggregate} productivity growth rate $\dot{T}/T$ into the \emph{average} productivity growth rate $\tilde{\gamma}$ and a factor characterizing network structure.  Most importantly, our model leads to the empirically testable price and growth predictions presented in the main text.

It is worth noting how the Domar weights, $D_i \equiv M_i/Y$, relate to the output multipliers, as these two quantities both convey information about an economy's structure.  From Eq. \eqref{eq_TFP_growth_rate}, it can be seen that the Domar weights provide the minimal statistics to aggregate the productivity growth rates of producers.  As noted earlier, the vector of industries' gross outputs equals $\vec{M} = (I - A)^{-1} \vec{Y}$, which means that the Domar weights can be written as
\begin{align}
\vec{D} = (I - A)^{-1} \vec{\theta},
\end{align}
where $\vec{\theta}$ denotes the GDP shares of industries.  Since the output multipliers are given by $\vec{\L}^T = \vec{1}^T (I - A)^{-1}$, we thus have the following set of relationships:
\begin{align}
\vec{1} \cdot \vec{D} = \vec{\theta} \cdot \vec{\L} 
= \underbrace{\vec{1}^T(I - A)^{-1}}_{\vec{\L}} \vec{\theta}
\!\!\!\!\!\!\!\!\!\!\!\!\!\!\!\!\!\!\!\!\!\!\!\!\!\!\!\!\!\!\overbrace{ \phantom{(I - A)^{-1} \vec{\theta}}}^{\vec{D}}
= \frac{\O}{Y}.
\label{eq_Domar_and_OMs}
\end{align}
Eq. \eqref{eq_Domar_and_OMs} shows two things.  First, the sum of the Domar weights equals the GDP-weighted average of the output multipliers.  Second, it shows that the Domar weights combine two kinds of information: the GDP shares, and the input relationships among producers.  In contrast, the output multipliers remove the effect of an economy's output mix, and depend only on producer input relationships.  In this sense they characterize the production network independently of what final goods an economy chooses to make.  If an economy shifts toward manufacturing, for example, the Domar weight of manufacturing industries will increase, while the output multipliers of these industries need not change.

\subsection{Fally-Finn theorem and the sufficiency of the average output multiplier}\label{sec_aggregation}

\jmTwo{In Section \ref{sec_evolutionGrowthRates} we rederive a result shown by Fally  (2012) \cite{Fally2012} and Finn (1976) \cite{Finn1976} (the latter under a different interpretation of the input-output network) that the average output multiplier is equal to the ratio of gross output to net output.  This result is remarkable in that implies the average output multiplier is independent of a wide range of network details, and correspondingly, that economies with very different production networks but the same ratio $\mathcal{O}/Y$ would experience the same amplification of growth.  This is closely related to a well-known implication of Hulten's theorem \cite{Hulten1978} that the aggregate benefits of the productivity improvement of an individual industry depend only on its Domar weight.}

\jmTwo{We see that the ratio of aggregate gross output to value added is a sufficient statistic for the amplification of growth in our framework.  Notably, this holds exactly in a Cobb-Douglas general equilibrium model.  As a caveat, in more general settings the independence of the rate of growth from network structure is an approximation, reflecting the first order nature of our growth result, and that when we consider second order effects on growth, the details of the network become relevant through the effects that relative price changes have on consumption shares and input shares (Eq. \eqref{eq_outputSecondOrder}).  Section \ref{sec_evolutionGrowthRates} explicitly works out the dependence for consumption shares, and the result for input shares would be very similar.  This is reminiscent of Baqaee and Fahri’s \cite{Baqaee2019} discussion of Hulten’s theorem as adapted to our framework.  Nevertheless, in Section \ref{sec_consumptionGrowth_v_priceReturns} we evaluate this approximation empirically for changes in consumption shares, including a first-pass exercise at estimating the substitution elasticity, the price change dispersion term, and the potential drag on growth from changes to consumption patterns under non-unitary elasticities of substitution.  The results suggest that the effect of these second order terms on growth over 10 years is modest.  Correspondingly, this suggests that the first order approach provides a good approximation over the horizons we deal with in this paper.  (This is also reminiscent of Baqaee and Fahri’s conclusion of second order effects not appreciably changing the typical business cycle volatility implied by production network models \cite{Baqaee2019}.)}

\jmTwo{The fact that there exist macro-level sufficient statistics to characterize the growth amplification, without appealing to network data, does not mean there is no value in understanding the emergence of this amplification as an equilibrium outcome.  First, it takes a well-specified model to be able to understand whether and how these sufficient statistics provide good approximations and whether they fail under certain circumstances.  Second, our production network environment delivers micro-level predictions that are important in themselves, and provide researchers with another means of testing model-specific predictions.  Indeed, it is precisely because these micro-level predictions are specific to our production network environment that we devote a large fraction of our paper to them. The fact that these micro-predictions are in good agreement with data lends further credence to the possibility that our framework can provide a causal mechanism (rather than simply sufficient statistics) for differences in cross-country growth.}

\subsection{Average output multipliers for open economies}

Here we show that trade between countries tends to bring their average output multipliers closer together.  We derive the result for the two-country case and show the generalization to an arbitrary number of countries.  Label the two countries as country 1 and country 2, each with $N$ industries. If the economies are closed to trade, the world input matrix $A$ will take the form
\begin{align}
A =
\begin{pmatrix}
A_1	&0\\
0	&A_2
\end{pmatrix},
\end{align}
where $A_1$ and $A_2$ are the two input matrices and 0 is an $N \times N$ matrix of zeros. The Leontief inverse is given by
\begin{align}
H = (I-A)^{-1} =
\begin{pmatrix}
(I_N - A_1)^{-1} & 0\\
0 & (I_N - A_2)^{-1}
\end{pmatrix},
\end{align}
where $I_N$ is the $N\times N$ identity matrix.  The output multipliers in country $c$ are $\vec{\L}_c = (I-A_c^T)^{-1} \vec{1}$, and the average output multiplier is $\bar{\mathcal{L}}_c = \vec{\theta}_c \vec{\L}_c$, where $\vec{\theta}_c$ is the vector of GDP shares for country $c$.

Trade between economies leads to non-zero elements in the off-diagonal blocks of $A$.  To see what effect this has on the average output multipliers we consider a first order perturbation in which each country starts using imported goods from the other country, with correspondingly less use of domestic goods.  We let the input coefficients of country 1's imports increase by an amount $\epsilon_{21}A_1$, and simultaneously reduce country 1's domestic input coefficients by the same amount.  Thus, total input requirements remain the same (in value terms) for each good, with a fraction $1 - \epsilon_{21}$ now spent on the domestic version of the good and a fraction $\epsilon_{21}$ spent on the foreign version.  Making a similar change to country 2's input coefficients with parameter $\epsilon_{12}$, the new matrix of input coefficients for the `open' world economy $A^O$ is
\begin{align}
A^O &= 
\begin{pmatrix}
(1-\epsilon_{21})A_1 & \epsilon_{12} A_2\\
\epsilon_{21} A_1 & (1-\epsilon_{12})A_2
\end{pmatrix}
=
\begin{pmatrix}
A_1 &0\\
0 &A_2
\end{pmatrix}
+
\begin{pmatrix}
-\epsilon_{21}A_1 	&\epsilon_{12}A_2\\
\epsilon_{21}A_1 	&-\epsilon_{12}A_2
\end{pmatrix} 
= A + S(\vec{\epsilon}),
\end{align}
where $S(\vec{\epsilon})$ is the second matrix above.  After the countries become open to trade, the Leontief inverse for the world is 
\begin{align}
H^O = \left(I-A^O \right)^{-1}
&= \left[I - A - S(\vec{\epsilon})\right]^{-1} \nonumber\\
&= \left\{(I-A) \left[I - H S(\vec{\epsilon})\right] \right\}^{-1} \nonumber\\
&= \left[ I - H S(\vec{\epsilon}) \right]^{-1} H\nonumber\\
&= \left[ I + H S(\vec{\epsilon}) + (H S(\vec{\epsilon}))^2 + \cdots \right] H.
\end{align}
Let $H_c \equiv (I_N - A_c^T)^{-1}$ denote the Leontief inverse of country $c$ when the countries were closed to trade.  Writing $H^O$ to first order in the $\epsilon$ parameters, we have
\begin{align}
H^O = 
\begin{pmatrix}
H_1 &0\\
0 &H_2
\end{pmatrix}
+
\begin{pmatrix}
-\epsilon_{21} H_1 A_1 H_1 	&\epsilon_{12} H_1 A_2 H_2\\
\epsilon_{21} H_2 A_1 H_1	&-\epsilon_{12} H_2 A_2 H_2
\end{pmatrix}
+
O(\epsilon^2).
\end{align}
The industry output multipliers in the two countries then become
\begin{align}
\begin{pmatrix}
\vec{\L}_1^O\\ \vec{\L}_2^O
\end{pmatrix}
=
\begin{pmatrix}
\vec{\L}_1\\ \vec{\L}_2
\end{pmatrix}
+
\begin{pmatrix}
\epsilon_{21} H_1^T A_1^T (\vec{\L}_2 - \vec{\L}_1)\\
\epsilon_{12} H_2^T A_2^T (\vec{\L}_1 - \vec{\L}_2)
\end{pmatrix}
+
O(\epsilon^2).
\end{align}
The average output multipliers of each country can be obtained by weighting industries by their final output shares in each country.  Doing this for country 1,  the average output multiplier after becoming open to trade, $\bar{\mathcal{L}}_1^O$, is related to its average output multiplier while it was closed to trade, $\bar{\mathcal{L}}_1$, by
\begin{align}
\bar{\mathcal{L}}_1^O = \bar{\mathcal{L}}_1+\epsilon_{21}\vec{\theta}_{1} H_1 A^T_1 \left(\vec{\mathcal{L}}_2 - \vec{\mathcal{L}}_1\right)  + O (\epsilon^2).
\end{align}

To understand this expression, suppose the industry output multipliers $\vec{\mathcal{L}}_1$ of country 1 start out generally smaller than those in country 2.  The elements of $\vec{\theta}_{1}$ and $H_1 A^T_1$ are positive, so the second term above will lead to a higher average multiplier than if the economy were closed.  The opposite effect occurs in country 2, whose average multiplier falls.  Thus, trade pulls the average output multipliers of the two countries closer together.  For $N$ countries, the calculation generalizes straightforwardly, leading to the result
\begin{align}
\bar{\mathcal{L}}_c^O &= \bar{\mathcal{L}}_c + \sum_{\substack{b\neq c}} \epsilon_{bc} \vec{\theta}_c H_c A_c^T\left(\vec{\mathcal{L}}_b - \vec{\mathcal{L}}_c\right) + O(\epsilon^2).
\end{align}
\jm{Thus, in our simple framework, when two countries open to trade, their average output multipliers become more similar. This suggests that trade openness could act as a force towards cross-country convergence in growth rates through changes in the production network, a result echoing previous arguments in the economic growth literature \cite{BenDavid1993}.  There is a large literature on trade, growth and convergence, and concatenating insights from that literature with our production networks environment is a potential avenue for further research.}


\jm{Finally, we can obtain a rough empirical assessment of the effect of trade on countries by zeroing out international trade entries in the matrix of intermediate payments, forcing subsequent computations to be done using only domestic input coefficients.  Doing this we observe two intuitive tendencies.  First, countries with large output multipliers, like China, Slovakia, and the Czech Republic, see their output multipliers go up further when `shutting down trade'.  This makes sense in light of the results above since we are now ejecting the input dependence of these countries on others with lower output multipliers.  Correspondingly, countries with below-average output multipliers see their output multipliers sink further in the absence of trade.  Second, the size of the effect depends on the scale of imports.  Countries with a large import share of GDP see large changes in output multiplier before and after turning off trade entries.}

\clearpage
\section{Supplementary results}
\subsection{Industry and country summary statistics}
See Tables \ref{tab_industry_table} and \ref{tab_country_empirical}.

\subsection{Persistence of output multipliers}\label{sec_persistence}
\begin{figure*}[h]
\center
{\sf\large A }\includegraphics[width=0.4\linewidth,align=t]{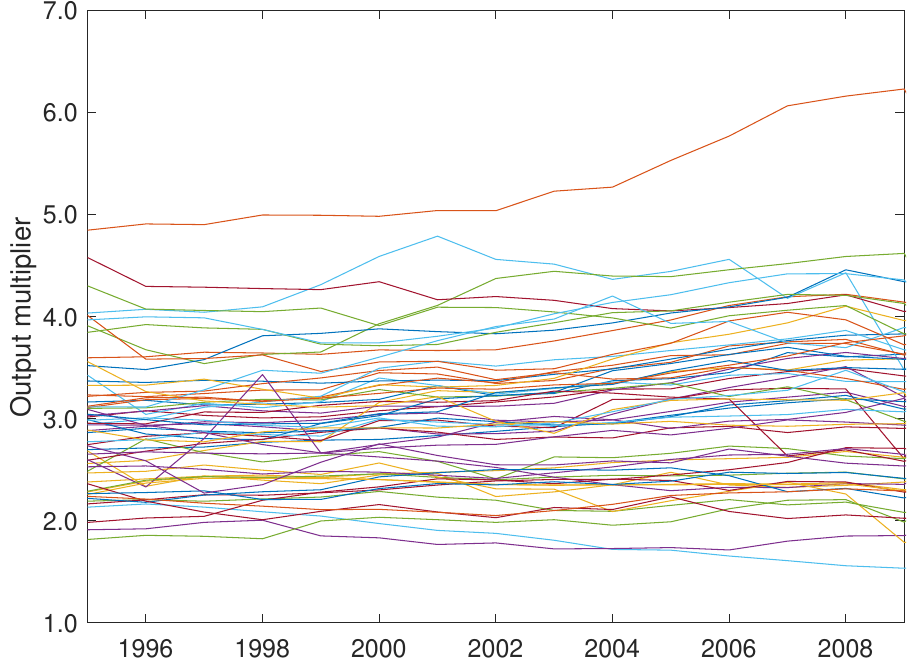}
\hfil
{\sf\large B }\includegraphics[width=0.4\linewidth,align=t]{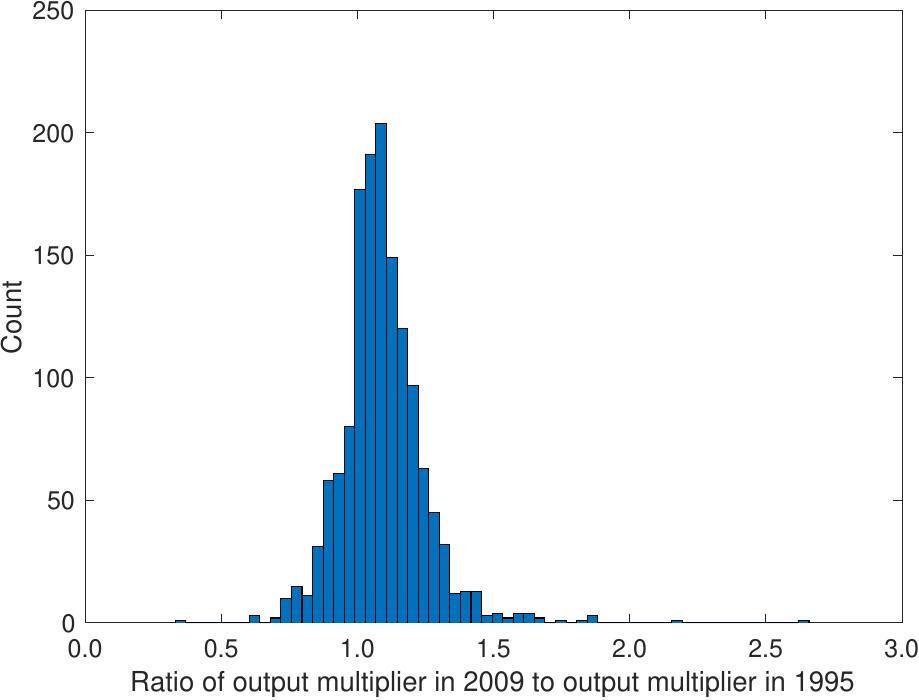}\\
\center
{\sf\large C }\includegraphics[width=0.4\linewidth,align=t]{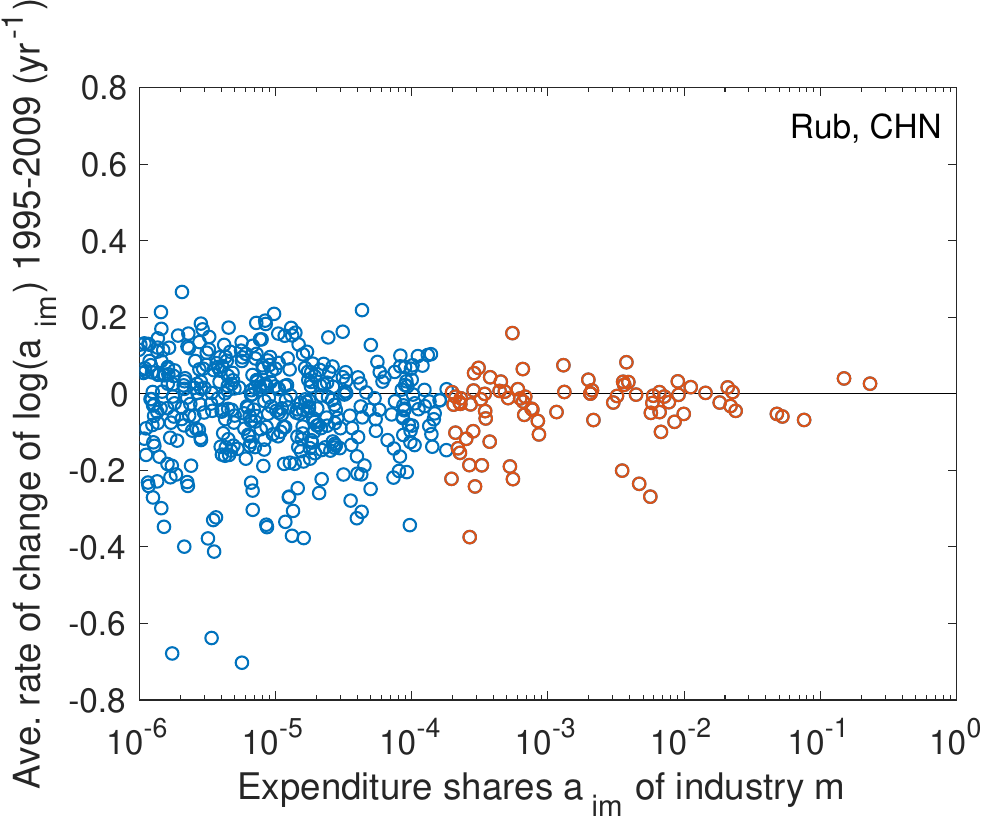}
\hfil
{\sf\large D }\includegraphics[width=0.4\linewidth,align=t]{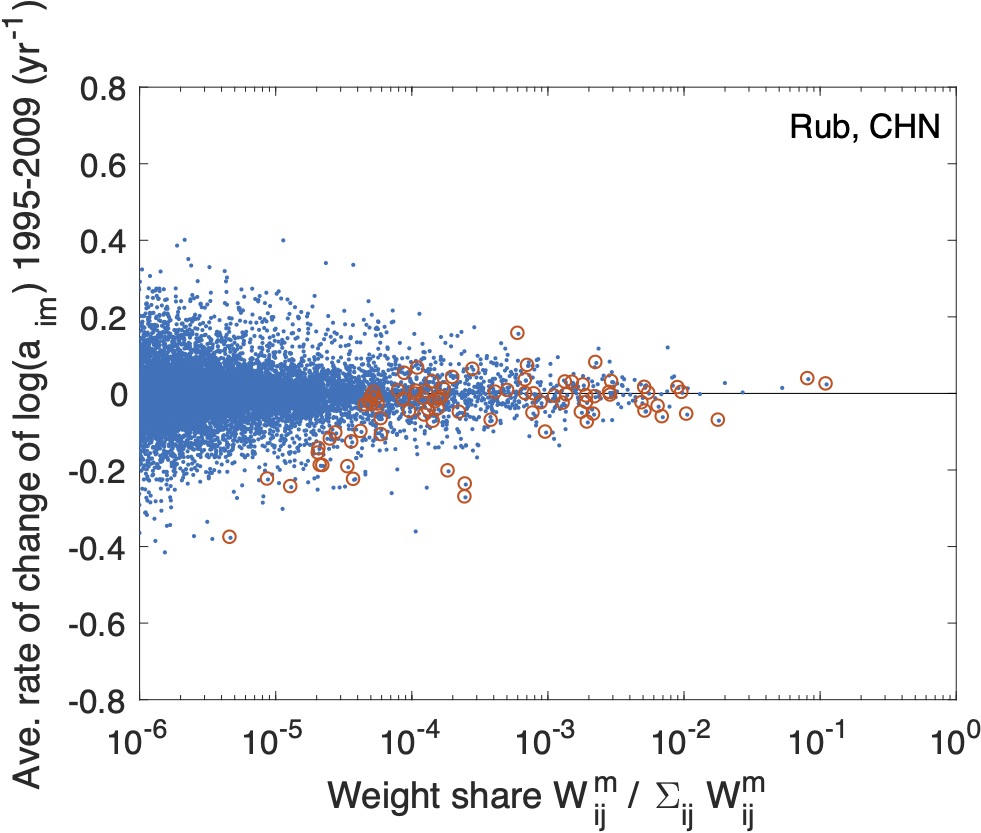}

\caption{\textbf{Persistence of industry output multipliers over time.} (A) Output multipliers over time for a random sample of 60 industries from the WIOD. (B) Histogram of the ratio of each industry's output multiplier in 2009 to its output multiplier in 1995, for all industries in the WIOD. (C) Rate of change of input coefficients versus input coefficient size for the Rubber and Plastics industry in China.  \jmTwo{The highlighted circles are the set of largest input coefficients that together represent 99\% of the industry's expenditures on intermediate goods.  Here we compute the expenditure shares $a_{im}$on the horizontal axis as the geometric mean of initial and final values, i.e. $\sqrt{a_{im}(1995) a_{im}(2009)}$.}  (D)  Rate of change of input coefficients versus influence on the input multiplier of the Rubber and Plastics industry in China.  \jmTwo{Highlighted circles show the locations of the same direct input coefficients highlighted in panel (C).}}
\label{fig_persistence}
\end{figure*}

Output multipliers change as each industry's dependence on inputs evolves with time.  Despite this, the values of output multipliers show persistence (Fig. \ref{fig_persistence}a-b).  This persistence is important because without it output multipliers could not dependably characterize an industry's or country's production chains.

One way to see this persistence is to compare the time variation in output growth, productivity improvement, and output multipliers.  For each time series $X(t)$ we compute the coefficient of variation (CV) $\sigma_X / \mu_X$, where $\sigma_X$ is the time standard deviation (volatility) and $\mu_X$ is the time average from 1995 to 2009.  Typical CVs (geometric mean across countries) are 1.4 for output growth, 2.3 for average productivity growth, and 0.041 for the average output multiplier.  By this measure average output multipliers have about $1.4/0.041 \sim 34$ times less variation over time than growth rates do, and $2.3/0.041 \approx 56$ times less variation than productivity improvement rates.  Similarly we find that industry-level output multipliers typically have about $2.47/0.057 \approx 43$ times less variation than price changes and $3.8/0.057 \approx 66$ times less variation than productivity improvement.  \jm{The mean absolute rate of change of industry level output multipliers is 0.86\% per year, while that of productivity growth rates is a much larger 18\% per year.  Output multipliers in the first and last years have a Pearson correlation 0.85, while productivity growth rates have a correlation of just -0.04.}

\jm{In terms of our price predictions, the persistence of output multipliers is also reflected in the fact that we can use values of the output multipliers in the initial year 1995 in our analyses.  For comparison, using output multipliers that are time-averaged over the full period 1995-2009 yields very similar results, with $R^2 = 0.70$ ($p \sim 10^{-7}$) and nearly the same slope (-1.5\% per year).}

A critical reason for this persistence is that the inputs with the strongest influence on an industry's output multiplier typically have much slower rates of change than other inputs.  Fig. \ref{fig_persistence}c-d shows an example using the Rubber and Plastics industry in China.  The direct inputs to this industry are captured by the 1400 elements of the column in the input matrix $A$ corresponding to this industry (Fig. \ref{fig_persistence}c).  Many of these elements change rapidly, with average rates of change in the 10's of percentage points per year.  However, these high rates of change are found mainly in matrix elements that constitute small fractions of the industry's expenses.  Among inputs whose shares are large (5\% or greater), rates of change are far smaller.  This tendency is typical across industries, and not necessarily surprising given that most industries rely on a few key inputs for production.

An output multiplier is determined not just by an industry's direct input requirements but by all indirect requirements as well.  Thus it is not just changes in one column of $A$ that matter but changes in all $1400^2$ matrix elements.  To show how changes in these elements influence a given output multiplier, note that the output multipliers are given by $\vec{\L} = H^T \vec{1} = (I - A^T)^{-1} \vec{1}$ and that taking the derivative with respect to time leads to $\dot{\vec{\L}} = (I - A^T)^{-1} \dot{A}^T (I - A^T)^{-1} \vec{1} = H^T \dot{A}^T \vec{\L}$.  In index form this is $\dot{\L}_m = \sum_{i,j} \L_i \dot{a}_{ij} H_{jm}$, or
\begin{align}
\widehat{\L}_m = \sum_{i,j} \widehat{a}_{ij} \left( \L_i a_{ij} H_{jm} \frac{1}{\L_m} \right).
\label{eq_OMchangeSummation}
\end{align}
The factor in parentheses $W_{ij}^m \equiv \L_i a_{ij} H_{jm} / \L_m$ weights changes in the $(i,j)$th element of $A$ ($\widehat{a}_{ij} = \dot{a}_{ij}/a_{ij}$) for changes to the $m$th output multiplier, indicating the matrix elements whose changes most strongly impact $\L_m$.  In Fig. \ref{fig_persistence}d we plot the rate of change of the $(i,j)$th element, $\widehat{a}_{ij}$, against its share of the total of all weights affecting industry $m$, $S_{ij}^m \equiv W_{ij}^m / \sum_{k,l} W_{kl}^m$. The results for the Rubber and Plastics industry in China are typical, and show that the most influential elements of $A$ are also the steadiest.  \jmTwo{The influence weights $W_{ij}^m$ span many orders of magnitude, and thus Eq. \eqref{eq_OMchangeSummation} indicates that most input coefficients (including those with the highest rates of change) will be heavily diminished in their contribution to $\widehat{\L}_m$.  In our data, we find that on the order of $10^2$ - $10^3$ elements of the input matrix, summed in order from highest $W_{ij}^m$ to lowest, are typically needed for a partial sum of terms in Eq. \eqref{eq_OMchangeSummation} to reach a value near its final total.  Of this very small fraction ($\sim0.005$ - $0.05\%$) of all $1400^2$ input coefficients that contribute significantly to the sum, many of the most significant ones correspond to the direct inputs to an industry (Fig. \ref{fig_persistence}c-d).}

\jmTwo{These observations point to two reasons why persistence of output multipliers can be consistent with time variation in individual input coefficients.  First, many of the largest contributions to an output multiplier's change come from an industry's direct inputs.  Most industries depend on a few key inputs for production (e.g. \cite{Carvalho2008}).  It would not be surprising for the largest inputs of many industries to remain large over long periods for technological reasons.}

\jmTwo{In addition, persistence in output multipliers can be consistent with changes in input coefficients because of a kind of averaging effect.  An output multiplier is a weighted sum of many coefficients, and changes in these coefficients are uncorrelated with output multipliers.  For intuition, first consider the recursion equation (Eq. \eqref{eq_trophic_level_ecology}) for an industry $m$'s output multiplier, $\L_m = 1 + \sum_j \L_j a_{jm}$.  Suppose that only industry $m$’s direct input coefficients can change, so that changes to $\L_m$ must come from a redistribution of input coefficients across inputs $j$.  This is a special intuitive case but, as we will see, the calculations it implies will extend to the general case.  The direct input coefficients (including the labor coefficient $\ell_m$) are constrained to sum to 1, $\sum_j a_{jm} + \ell_m = 1$, and so for $\L_m$ to change, the changes in $m$'s input coefficients must be correlated with the output multipliers of its inputs.  For $\L_m$ to rise, say, $m$'s direct input coefficients must generally rise in industries with high output multipliers and fall in industries with low output multipliers.  Without such correlations the requirement for input coefficients to sum to 1 will tend to make their contributions to change in $\L_m$ offset one another.  Theoretically, there is no obvious reason for changes in an industry's input coefficients to be correlated with the output multipliers of its inputs.  Producers do not consider output multipliers in deciding how much of each input to use, and there is no clear reason for technical change to be biased toward products with particular values of output multipliers.  Empirically the data support this intuition.  Pearson correlations between $\L_j$ and $\dot{a}_{jm}$, are smaller than 0.15 in magnitude for 97\% of industries $m$, and we find an overall correlation across industries of just 0.014.}

\jmTwo{To make this point more formally, and extend it to changes in indirect inputs to $m$, we can rewrite Eq. \eqref{eq_OMchangeSummation} in terms of covariances between output multipliers and changes in input coefficients.  First, we write Eq. \eqref{eq_OMchangeSummation} as $\dot{\L}_m = \sum_i (\sum_j \L_i \dot{a}_{ij}) H_{jm}$.  The sums in this equation run over only intermediate inputs, and to exploit the normalization constraint $\sum_j a_{jm} + \ell_m = 1$ it is useful to expand the sum over $j$ to run over the labor coefficient as well.  This can be done easily without large changes to the equation by defining $a_{0m} \equiv \ell_m$, and defining the output multiplier of households to be zero, $\L_0 \equiv 0$.  Then Eq. \eqref{eq_OMchangeSummation} can be written
\begin{align}
\dot{\L}_m 
&= \sum_i \left(\sum_{j=0}^n \L_i \dot{a}_{ij}\right) H_{jm} \nonumber\\
&= \sum_i (n+1) E\left(\L_i \dot{a}_{ij}\right) H_{jm} \nonumber\\
&= (n+1)\sum_i \left[ E(\L_i) E(\dot{a}_{ij}) + \Cov(\L_i,\dot{a}_{ij})\right] H_{jm} \nonumber\\
&= (n+1)\sum_i \Cov(\L_i,\dot{a}_{ij}) H_{jm}
\label{eq_OMchange_and_covariances}
\end{align}
where expectation values are taken across industries $j$.  To obtain the last line we use the fact that changes in input coefficients must always sum to zero: $E(\dot{a}_{ij}) = \frac{1}{n+1} \sum_{j=0}^n \dot{a}_{jm} = 0$. Eq. \eqref{eq_OMchange_and_covariances} extends the intuition of the case where only $m$'s direct input coefficients can change and shows that, intuitively, $\dot{\L}_m$ depends on $m$’s total exposure to each industry $j$ and the covariance of changes in $j$’s input coefficients with the output multipliers of its inputs.  Yet as noted already, these covariances are small.  That is, Eq. \eqref{eq_OMchange_and_covariances} connects the rate of change of $m$'s output multiplier to changes in $j$'s input structure through the elements $H_{jm}$.  We then fall back on the observation that, across industries $j$, changes in direct input coefficients have low correlations with the output multipliers of these inputs.}



\jm{From a theory perspective, a sufficient condition for stable industry output multipliers is that production functions have unit elasticity of substitution (i.e. Cobb-Douglas).  In this case the input coefficients, and any function of them, will be unchanged by changes in productivity.  If in addition the structure of final demand is fixed (e.g. we consider a closed economy with homothetic preferences, removing income effects on consumption) then the average output multiplier will also be stable.   Although output multipliers are much more stable relative to productivity improvements, empirically we do see (i) that many small input coefficients change by large amounts, and (ii) that many large input coefficients change by smaller amounts. This suggests that though persistence in output multipliers obtains in data, micro-level changes of input coefficients might not conform in obvious ways with standard Cobb-Douglas theoretical setups. For example, large changes in output multipliers are particularly likely when the expenditure share of labor relative to intermediate goods shifts significantly because the labor expenditure shares tend to be large, having an outsize influence on the output multiplier.  Such changes can occur as a result of biased technological change \cite{Hicks1932} under more general production functions, and may be an important direction for further research.}

%

\subsection{Correlations of price changes with direct and inherited components}\label{sec_decompositionOfPriceChanges}
Here we further examine the passing on of price reductions in the network and the correlations between price changes and productivity improvement noted in the main text.  Multiplying Eq. \eqref{eq_realPriceChanges_gammaVec} by $(I-A)^T$, price changes can be decomposed as
\begin{align}
r_i = -\gamma_i + \sum_j r_j a_{ji} = -\gamma_i  - \sum_{k=1}^\infty \sum_j \gamma_j \,(A^k)_{ji}.
\label{eq_own_v_inherited}
\end{align}
The price reduction of node $i$ has two components: a direct improvement due to $-\gamma_i$, and the indirect or inherited effects of improvements from the inputs that $i$ consumes.  Other nodes undergo productivity improvements, whose effects are transmitted to $i$ through input price changes.  In this sense the second term captures improvement inherited by node $i$.%
\footnote{Note that the $-\gamma_i$ term alone does not account for the entirety of the benefits of the improvement at node $i$, because $\gamma_i$ also appears in the summation in the second term.  This reflects the fact that $i$ may buy its own good as an input, as well as other goods that use $i$ for production, leading to indirect benefits to $i$ from its own improvement.  These are also inherited benefits, since they come through $i$'s inputs.  An alternate decomposition that fully isolates the effects of $\gamma_i$ from that of all other $\gamma$'s is $r_i = \gamma_i H_{ii} + \sum_{j\neq i} \gamma_j H_{ji}$.}%

Fig. 2b of the main text shows that both terms in Eq. \eqref{eq_own_v_inherited} are important, with each contributing a similar magnitude to cost change on average.  Inherited price reductions tend to contribute more, with a mean value -1.65 \% yr$^{-1}$, as compared with the direct improvements, with a mean value of -1.06 \% yr$^{-1}$.  To understand these contributions further, we study the correlations among the price changes $r_i$, the direct improvement term $-\gamma_i$, and the inherited improvement term $\Sigma_i \equiv \sum_j r_j a_{ji}$.

The correlation with $r_i$ is high for both terms, 0.92 for $\gamma_i$ and 0.71 for $\Sigma_i$.  A high correlation for both terms is expected, since these quantities are related by $r_i = -\gamma_i + \Sigma_i$.  To understand the effects of this relationship on their correlations we treat $-\gamma_i$ and $\Sigma_i$ as random variables and consider the case of three variables $X$, $Y$, and $Z$ that are related by $X + Y = Z$.  By definition, the Pearson correlation between $Z$ and the summand $X$ is
\begin{align}
\rho_{ZX} \equiv \frac{\Cov(Z,X)}{\sigma_Z \sigma_X} = \frac{E[ZX] - E[Z] E[X]}{\sigma_Z \sigma_X},
\label{eq_correlation_ZX}
\end{align}
where $\sigma_Z$ and $\sigma_X$ are the standard deviations of $Z$ and $X$.  The numerator in Eq. \eqref{eq_correlation_ZX} is
\begin{align}
\Cov(Z,X)
&= E[(X+Y)X] - E[X+Y] E[X]\nonumber\\
&= E[X^2] + E[XY] - E[X]^2 + E[Y] E[Y] \nonumber\\
&= \sigma_X^2 + \Cov(X,Y)
\end{align}
while the denominator is $\sigma_Z \sigma_X = \sigma_{X+Y} \sigma_X = \sigma_X \sqrt{\sigma_X^2 + \sigma_Y^2 + 2 \Cov(X,Y)}$, resulting in a correlation
\begin{align}
\rho_{ZX}
= \frac{\sigma_X + \Cov(X,Y)/\sigma_X}{\sqrt{\sigma_X^2 + \sigma_Y^2 + 2 \Cov(X,Y)}}.
\label{eq_rho_ZX}
\end{align}
Eq. \eqref{eq_rho_ZX} gives the correlation between the sum variable $Z$ and the summand $X$ in terms of the variances and covariances of the summands $X$ and $Y$.

We note three special cases of Eq. \eqref{eq_rho_ZX} that help separate the effects influencing the correlation $\rho_{ZX}$.  If the covariance between the summands is negligible, Eq. \eqref{eq_rho_ZX} simplifies to $\rho_{ZX} = \sigma_X /\sqrt{\sigma_X^2 + \sigma_Y^2}$.  In this case, differences in the variances of $X$ and $Y$ drive the correlation -- the variable with the larger variance attains the larger correlation with $Z$.  If the covariance is not negligible, and the variances of $X$ and $Y$ are the same, then $\rho_{ZX} = \sqrt{1 + \Cov(X,Y) /\sigma_X^2} / \sqrt{2} = \sqrt{1 + \Cov(X,Y) /\sigma_Y^2} / \sqrt{2}$.  In this case, both variables have the same correlation with $Z$, and its strength depends on the ratio of the covariance between $X$ and $Y$ with their shared variance $\sigma_X^2 = \sigma_Y^2$.  Finally, when both the covariance is negligible and the variances are the same, $\rho_{ZX} = 1 / \sqrt{2} \approx 0.707$. This last case shows that the correlation of $X$ with $Z$ will tend to be large simply because $X$ is a summand to $Z$.

We use these formulas to compute the correlations of $-\gamma_i$ and $\Sigma_i$ with $r_i$ based on the measured values of their standard deviations and covariance.  The large correlation $\rho_{r\gamma} = 0.92$ is driven by  three effects.  The largest effect is the one just noted, that $-\gamma_i$ and $\Sigma_i$ are related to $r_i$ by $r_i = -\gamma_i + \Sigma_i$.  This gives both variables a starting correlation with $r_i$ of $1/\sqrt{2} \approx 0.707$.  The next largest effect is that $-\gamma_i$ has a larger variance than $\Sigma_i$.  Intuitively, the larger variance in $-\gamma_i$ causes it to explain more of the variation in $r_i$, even though it is actually responsible for somewhat less price change on average.  Third, the covariance between $-\gamma_i$ and $\Sigma_i$ increases both variables correlation with $r_i$.  These two variables have a correlation with each other of 0.37 ($p \sim 10^{-47}$).

\subsection{Price changes and output multipliers after shuffling productivity growth rates}
\begin{figure*}[h]
\center
{\sf\large A }\includegraphics[width=0.4\linewidth,align=t]{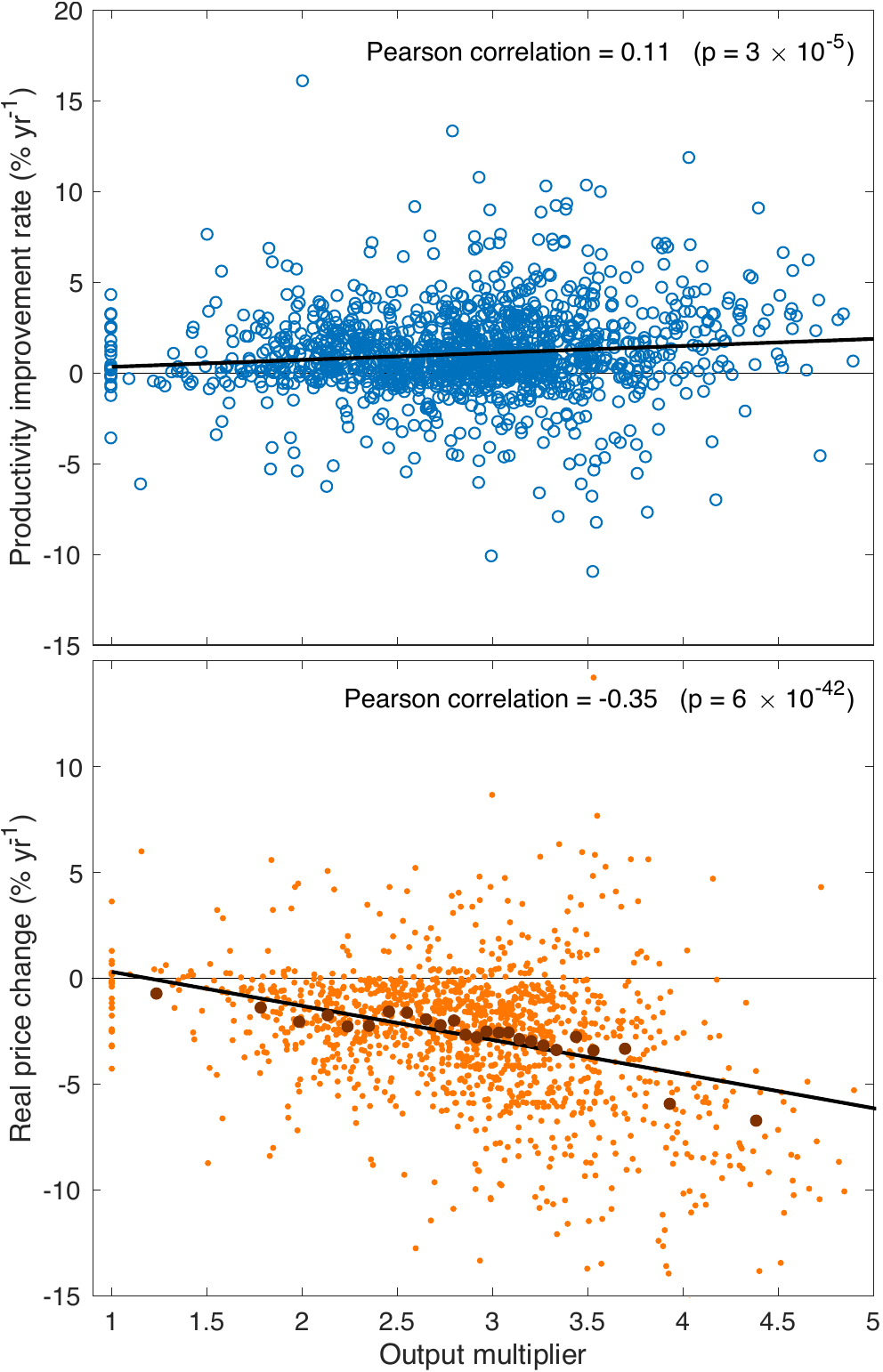}
\hfil
{\sf\large B }\includegraphics[width=0.4\linewidth,align=t]{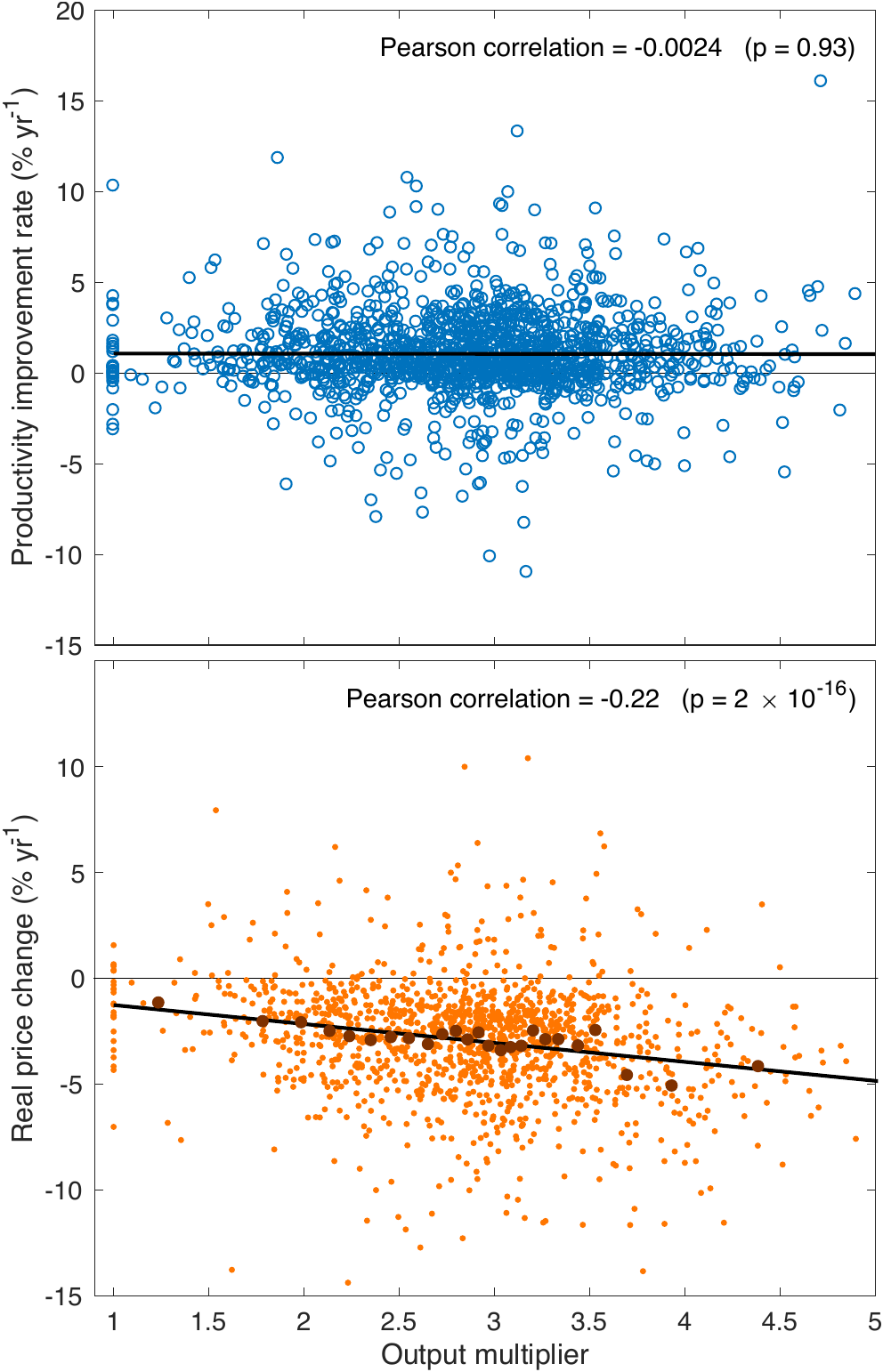}
\caption{\textbf{Correlation between price changes and output multipliers before and after shuffling improvement rates.} (A) Observed productivity improvement rates and price changes versus output multipliers.  Improvement rates have a small positive correlation with output multipliers.  (B) Productivity improvement rates after shuffling across industries to remove the correlation with the output multipliers. Resulting industry price changes were then computed with these improvement rates using the model.  Results vary from one shuffle to another, with those shown above being typical.}
\label{fig_correlation_after_shuffling}
\end{figure*}

See Fig. \ref{fig_correlation_after_shuffling}.  As noted in the main text, the difference between the observed regression slope $-1.6\%$ per year in Fig. 3a and the predicted slope $-\bar{\gamma} = -1.0\%$ per year stems from a positive correlation between output multipliers $\L_i$ and productivity improvement rates $\gamma_i$, which have a Pearson correlation 0.11 ($p \sim 10^{-5}$).  Productivity improvement rates tend to be greater for industries with higher output multipliers, increasing the magnitude of the slope in Fig. 3a.  To see whether this correlation drives the relationship between price changes and output multipliers we shuffle improvement rates across industries to remove the correlation with the output multipliers (Fig. \ref{fig_correlation_after_shuffling}), finding that the output multipliers retain a highly significant correlation with price changes even with this effect removed.

\subsection{Price changes within industry categories}
As seen in Fig. 3b-c of the main text, manufacturing industries in general tend to realize faster price reduction than services, and in the same figure it can be seen that manufacturing industries tend to have higher output multipliers.  Thus, the good correlation between faster price reduction and higher output multipliers could simply be the result of these two empirically observed effects.  However, by tying rates of price reduction to the output multiplier, the theory here makes an even more specific prediction -- even within a given industry category, variation in the output multipliers should predict variation in price changes.  We examine this by looking within each of the 35 industry categories in the WIOD data, considering observed price changes over the years 1995 - 2009 for the 40 instances of this industry category across countries.  We regress these 40 price changes against the corresponding output multipliers in the year 1995.  The results here show a dramatic agreement with the theory, with 34 of 35 industry categories having a negative slope as expected, which is statistically significant in most cases (Table \ref{tab_sameIndustryCrossCountriesVariation}).  The main text discusses the pooled version of this test, with results shown in Fig. 3d.

\subsection{Predictive ability of output multipliers and industry groupings}
We find that an industry's output multiplier is more informative of price changes than the industry's type, whether broadly- (e.g. manufacturing) or narrowly-defined (e.g. Rubber and Plastics).  We first regress real price changes against dummy indicators of whether an industry is an agriculture, manufacturing, or services industry (Table \ref{tab_regressionIndustryDummies}).  These industry labels are highly significant.  Coefficients are negative (real prices across industries are mostly decreasing) and larger in manufacturing than services, confirming our expectation that being a manufacturer is associated with relatively faster price reduction.

Nevertheless, these coefficients are not robust when output multipliers are included.  The signs of dummy variable coefficients are flipped, while the coefficient on the output multiplier is little affected compared to a regression without these indicators.  We obtain similar results when using separate indicators for each of the 35 industry categories in the WIOD.

\subsection{Cross-industry variation in price changes}\FloatBarrier
Here we examine the predictions of Eq. \eqref{eq_priceSigmaWithTime_SM} for the variation in price changes across industries with a given output multiplier.  First, note that industries with larger output multipliers empirically show greater variation in productivity growth rates.  Like the correlation of the directional changes in productivity with the output multipliers, this increase in the cross-industry variation of productivity growth is an effect outside the theory presented here, though is not inconsistent with it.  The effect contributes to the increase in $\sigma_{r_i|\L_i}$ as the output multiplier rises.  However, the slope of this relation is too small to account for the observed increase in $\sigma_{r_i|\L_i}$ with $\L_i$.  As a rough indicator, a linear regression of productivity growth rates on the output multiplier yields a slope that is only about 60\% of the value needed to explain the $\L_i$-dependence of $\sigma_{r_i|\L_i}$.

\begin{figure*}[t]
\center
\includegraphics[width=1\linewidth]{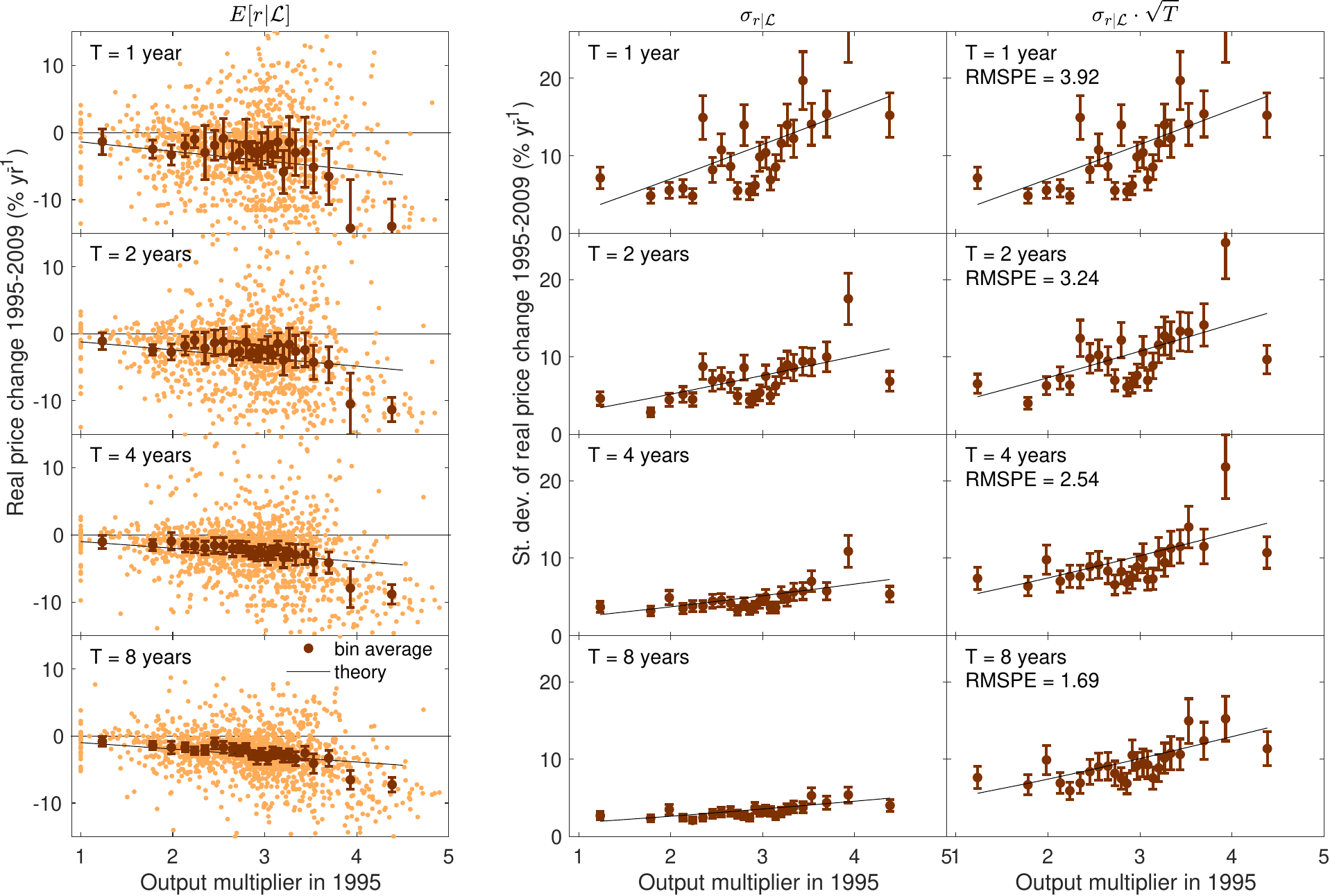}
\caption{\textbf{Cross-country mean and standard deviation of price changes over different time horizons.}   We examine the average rate of price change $r_i(t,t+\Delta t)$ for industries over time horizons of $\Delta t = $1, 2, 4, and 8 years.  Industries with similar output multipliers were grouped into bins.  The far left column shows the price changes of individual industries (light orange dots) and the average price change for all industries in a given bin (brown dots).  The black line shows the theory prediction for the bin averages.  The middle column shows the standard deviation of price changes for industries with similar output multipliers (brown dots) and the theory prediction (black line).  Eq. \eqref{eq_priceSigmaWithTime_SM} predicts that the standard deviations will shrink at a rate $\sqrt{T}$.  To see how tightly the data clusters around this prediction, in the right column we show the standard deviations and theory prediction after multiplying by the predicted shrinkage factor $\sqrt{T}$.  In these panels we also show the root mean square prediction error per bin (RMSPE), which decreases with longer time horizons.}
\label{fig_priceChanges_v_OMs_full}
\end{figure*}

In Fig. \ref{fig_priceChanges_v_OMs_full} we compare Eq. \eqref{eq_priceSigmaWithTime_SM} with observations from the WIOD over the time horizons $T =$ 1, 2, 4, and 8 years.  To take the correlation above into account, for each horizon we build a linear model of $\sigma_{\gamma,\text{direct}}$'s dependence on $\L_i$.  We then use this within Eq. \eqref{eq_priceSigmaWithTime_SM}, showing the standard deviation of price changes in two ways.  First, to better observe the shrinkage of $\sigma_{r_i(t,t+T) | \L_i}$ with time horizon $T$, we show $\sigma_{r_i(t,t+T) | \L_i}$ without adjusting for the length of the time period.  As the period $T$ becomes longer, the cross-industry variation in price changes shrinks across all output multiplier bins.  Second, to better observe the relationship between $\sigma_{r_i(t,t+T) | \L_i}$ and $\L_i$, we show $\sigma_{r_i(t,t+T) | \L_i}$ adjusted by multiplying by the predicted shrinkage factor $\sqrt{T}$.  As time passes, the time-adjusted standard deviations become more narrowly defined and are better predicted by Eq. \eqref{eq_priceSigmaWithTime_SM}.  For reference, we also show the price changes over each horizon.  The near-linear increase in $\sigma_{r_i(t,t+T) | \L_i}$ with the output multiplier accounts for the triangular appearance of the price changes in Fig. \ref{fig_priceChanges_v_OMs_full}, with price changes fanning out at larger $\L_i$.  As time passes, the triangle becomes narrower as variation in price changes around the mean $E[r_i | \L_i]$ shrinks.

\subsection{Industry price changes and consumption growth}\label{sec_consumptionGrowth_v_priceReturns}
\begin{figure*}[h]
\center
\includegraphics[width=0.5\textwidth]{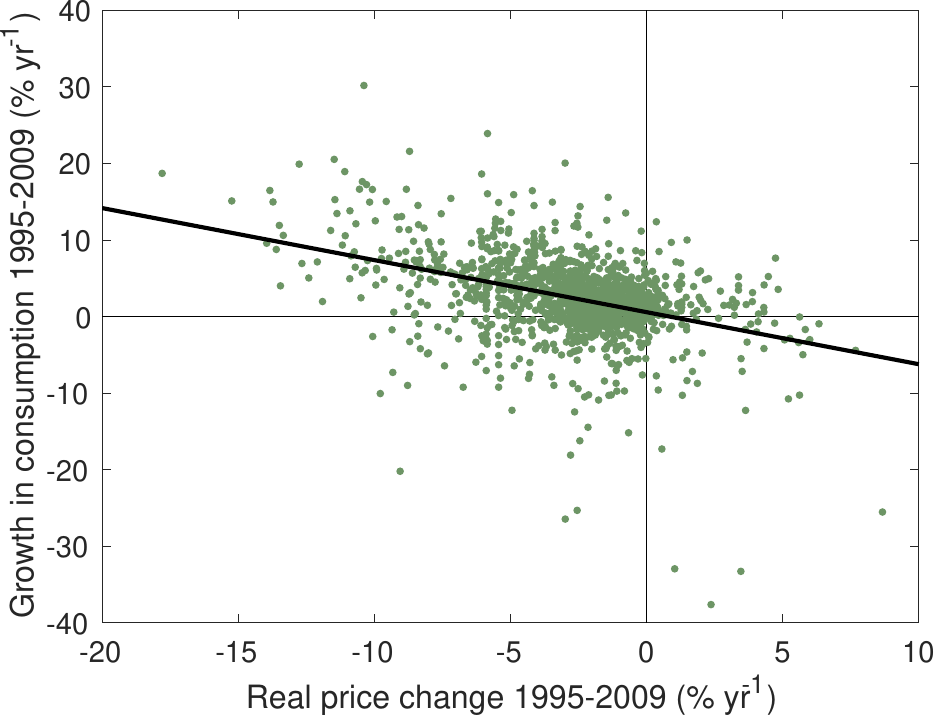}
\caption{Consumption growth and price change for each industry in the WIOD.  The black line is a regression fit with slope $\sigma = -0.68$.}
\label{fig_ConsumptionReturns_v_priceReturns}
\end{figure*}

The relationship of output multipliers with price reductions suggests a relationship with growth,  since the economy may enjoy real price decreases by consuming more.  In our data, we can see the translation of price reductions into output growth by comparing the change in consumption of each good to its change in price (Fig. \ref{fig_ConsumptionReturns_v_priceReturns}).  Although it would be surprising to see otherwise, we observe that industries with faster price reduction realize faster consumption growth.  A fit to a line yields a slope of -0.68 ($R^2 = 0.16$, $p = 2\times10^{-52}$).

Through the lens of the general equilibrium framework, this slope is consistent with previous findings about demand behavior.  Under a CES model with a final demand substitution elasticity $\sigma$, the consumption growth of a good $i$ falls with price changes as $-\sigma (r_i - \vec{\theta} \cdot \vec{r})$, and thus the slope in Fig. \ref{fig_ConsumptionReturns_v_priceReturns} corresponds to the negative of the substitution elasticity, $\sigma = 0.68$.  This places the best-fit model between completely inelastic Leontief demand ($\sigma = 0$) and unit-elasticity Cobb-Douglas demand ($\sigma = 1$), with the data closer to the latter.  While our simple regression does not address endogeneity concerns arising from the co-determination of prices and quantities, the implied demand behavior is consistent with the findings in other studies, reporting that consumers' elasticity of substitution across industries' products is near to and somewhat less than $\sigma = 1$ \cite{Ngai2007,Herrendorf2013,Oberfield2014}.  It is well appreciated that when $\sigma = 1$ the change in consumption of a good precisely offsets changes in price, leaving its share of expenditures constant.  The data here and other findings therefore suggest inertia in the consumption shares $\vec{\theta}$ as relative prices change. 

The deviation of households away from Cobb-Douglas behavior towards inelastic demand means that the expenditure shares of slower-improving goods will eventually tend to rise in the basket $\vec{\theta}$, as households only partially offset price rises by substituting.  This further suggests that, given enough time, there may be a drag on growth as the shares of faster-improving goods become smaller in the consumption basket over time, lowering the growth impact of further cost reductions in these goods.  This is the second-order effect described by Eq. \eqref{eq_outputSecondOrder} shown in Section \ref{sec_evolutionGrowthRates}.  The change also raises the question whether one should expect the average output multiplier $\bar{\L} = \vec{\theta} \cdot \vec{\L}$, computed using consumption shares in a given initial year, to shed its predictive ability over time because the consumption shares change.  However, numerical estimates argue against this, suggesting that the growth rate changes caused by changes in the consumption shares are modest.  To get a feel for the growth reduction from changes in $\vec{\theta}$, we make a rough estimate of the $\dot{\vec{\theta}}$ term in Eq. \eqref{eq_outputSecondOrder}, $-(1-\sigma) \text{Var}_{\vec{\theta}}(\vec{r})$.  We take consumers' substitution elasticity to be $\sigma = 0.68$, and compute the price change dispersion $\text{Var}_{\vec{\theta}}(\vec{r})$ using the world consumption basket $\vec{\theta}$ obtained from the vector of world final demands in the WIOD data.  The consumption-weighted variance of price changes from 1995 to 2009 is $\text{Var}_{\vec{\theta}}(\vec{r}) = 0.039 \,\%/\text{year}^2$.  (Note that since we are computing a rate of change of growth, the units in the denominator are years squared.)  By Eq. \eqref{eq_outputSecondOrder}, the growth deceleration from changes in $\vec{\theta}$ is then $-(1-\sigma) \text{Var}_{\vec{\theta}}(\vec{r}) = -0.0125\,\%/\text{year}^2$.  Under this deceleration rate, after 10 years, shifts in $\vec{\theta}$ would cause a fall in a country's growth rate on the order of $(10 \,\text{years}) \times (0.0125\,\%/\text{year}^2) \sim0.1 \%/\text{year}$, a modest decrease.

\subsection{Insensitivity of average output multiplier to coarse-graining}
As noted in the main text, data on production networks varies in level of aggregation, ranging from a few industries to hundreds of industries.  This raises the concern that the average output multiplier might have different values depending on the granularity of the underlying industry data.  However, it has been shown in Fally (2012) that the average output multiplier of a closed economy is independent of the level of aggregation, and equal to the ratio of gross output to net output $\mathcal{O}/Y$ \cite{Fally2012}.  Interestingly, an equivalent result was obtained in input-output ecology by Finn (1976), showing that the average path length of an energy input to an ecosystem equals the ratio of total energy throughput to energy input \cite{Finn1976}.  Both results arise from the aggregational properties of Markov chains and we believe a corresponding result may exist in that literature, though we currently do not have a reference.

\begin{figure*}[t!]
\center
\includegraphics[width=0.5\textwidth]{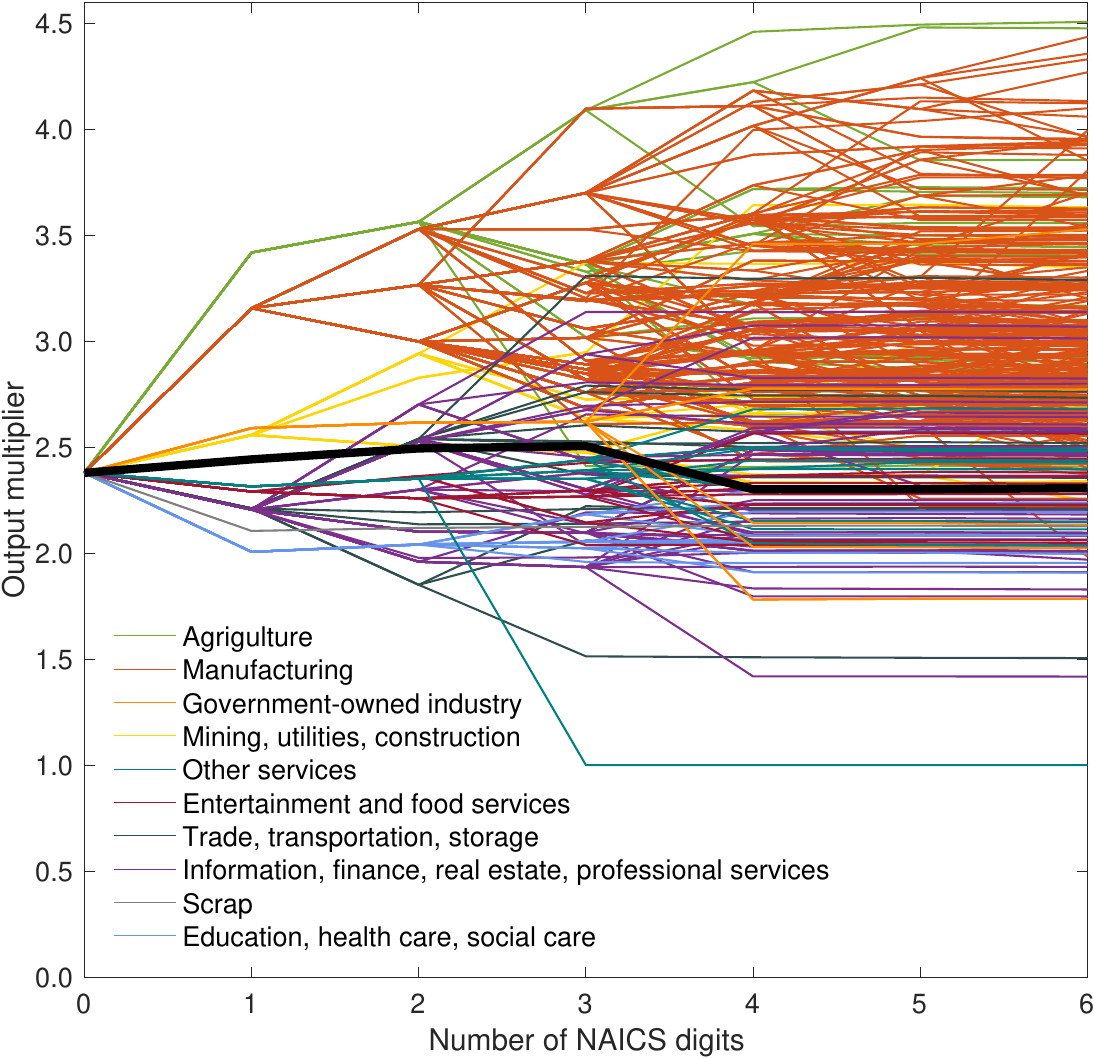}
\caption{\textbf{Industry output multipliers and country average output multiplier at varying levels of network aggregation.}  Using data from the U.S. Bureau of Economic Analysis \cite{BEA}, industries were merged to produce increasingly coarse-grained representations of the U.S production network.  At each level of aggregation, we compute the industry output multipliers (colored lines) and average output multiplier (thick black line).  Industries are merged in an order based on their 6-digit North American Industry Classification (NAICS) codes.  At the 6-digit level of aggregation (far right) all 427 industries are distinguished.  As the number of digits $n$ descends from 5 to 1, industries sharing the first $n$-digits of their NAICS codes are combined, producing a coarser production network.  In the 0-digit case industries are merged into one node.}
\label{fig_coarse_graining}
\end{figure*}

Fally (2012) also performs a test of the sensitivity of $\bar{\L}$ to aggregation in the practical context of an open economy, using data from the U.S. economy at different levels of network resolution.  Because of its relevance we repeat this test here, with the same finding that the average output multiplier is insensitive to the level of coarse-graining.  We use the 2002 benchmark input-output table from the U.S. Bureau of Economic Analysis \cite{BEA}, which distinguishes 427 industries.  An advantage of this data set is that industries are hierarchically indexed with 6-digit North American Industry Classification (NAICS) codes.  These codes let one merge industries into larger groupings, generating a series of coarser representations of the U.S production network.  At each level of aggregation, we compute the average output multiplier.  At the 6-digit level (far right in Fig. \ref{fig_coarse_graining}), all 427 industries are distinguished, and as the number of digits $n$ descends from 5 to 1, industries sharing the first $n$-digits of their NAICS codes are combined, producing networks with 308, 205, 78, 24, and 10 nodes.  In the 0-digit case all industries are merged into one node.  At each level, industry output multipliers and the GDP-weighted average output multiplier were computed on the resulting coarse-grained network.  Despite the fact that the U.S. is not a closed economy, the average output multiplier changes little over a wide range of levels of aggregation.  \jmTwo{(For further theoretical discussion of this result and the sufficiency of the average output multiplier as a macro-level statistic, see Section \ref{sec_aggregation}.)}


\subsection{Robustness of results to variations in empirical approach}
\begin{figure*}[h]
\center
{\sf\large A }
\includegraphics[height=0.17\textheight,align=t]{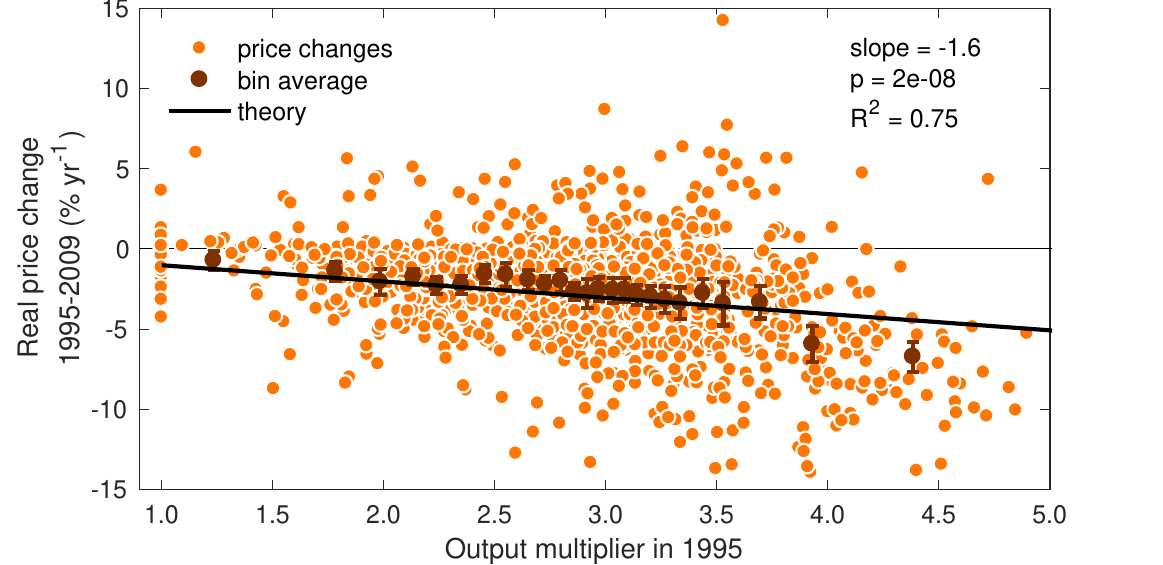}
\includegraphics[height=0.17\textheight,align=t]{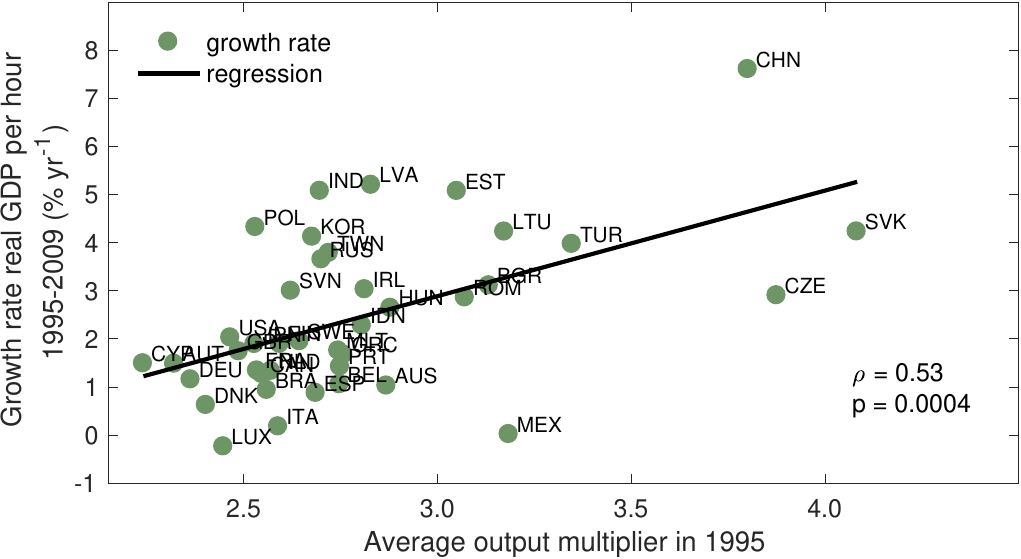}\\
{\sf\large B }
\includegraphics[height=0.17\textheight,align=t]{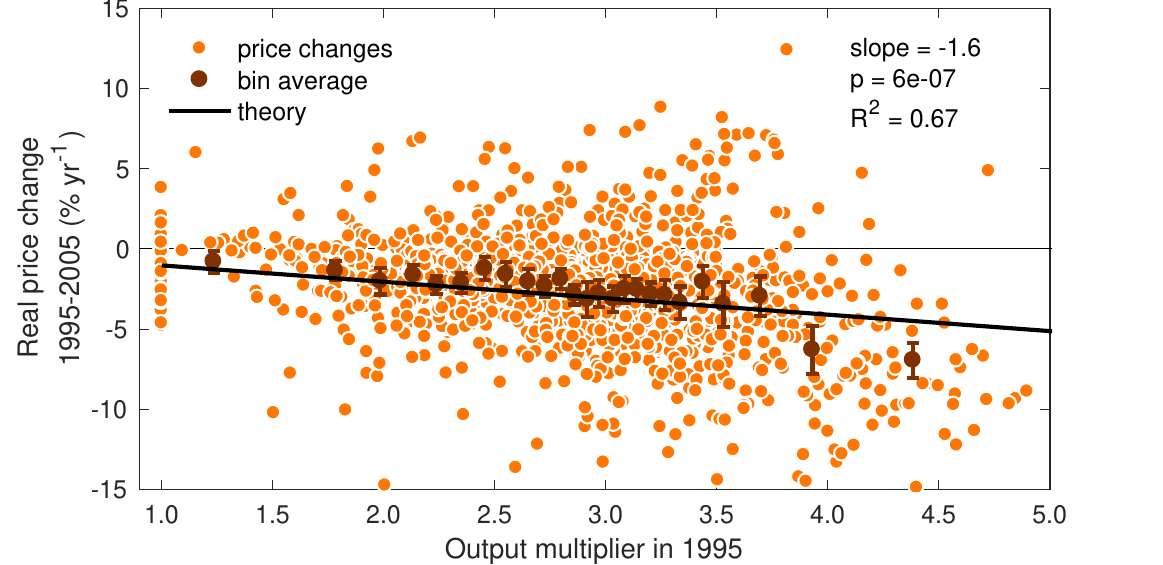}
\includegraphics[height=0.17\textheight,align=t]{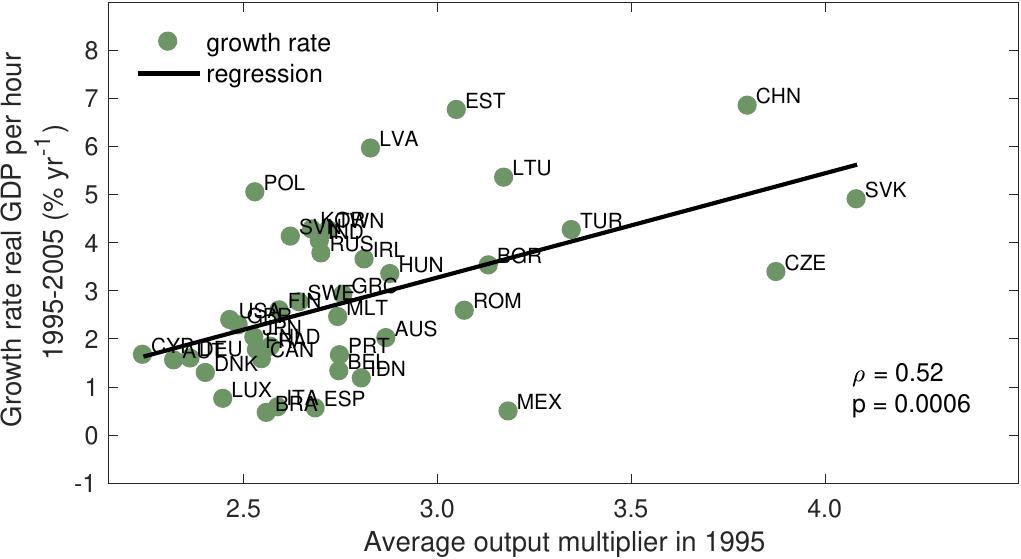}\\
{\sf\large C }
\includegraphics[height=0.17\textheight,align=t]{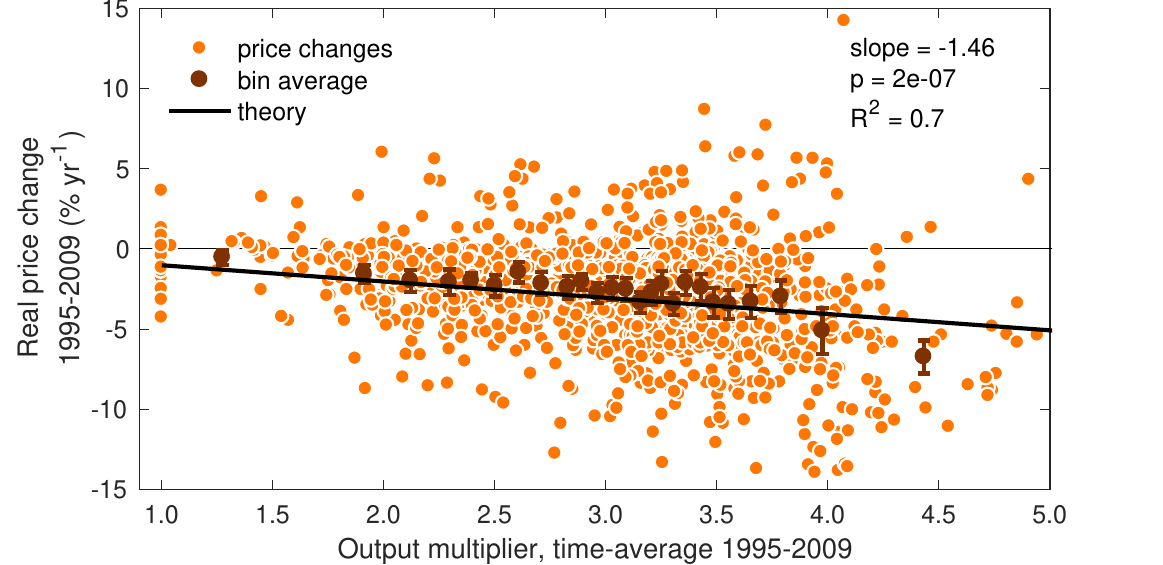}
\includegraphics[height=0.17\textheight,align=t]{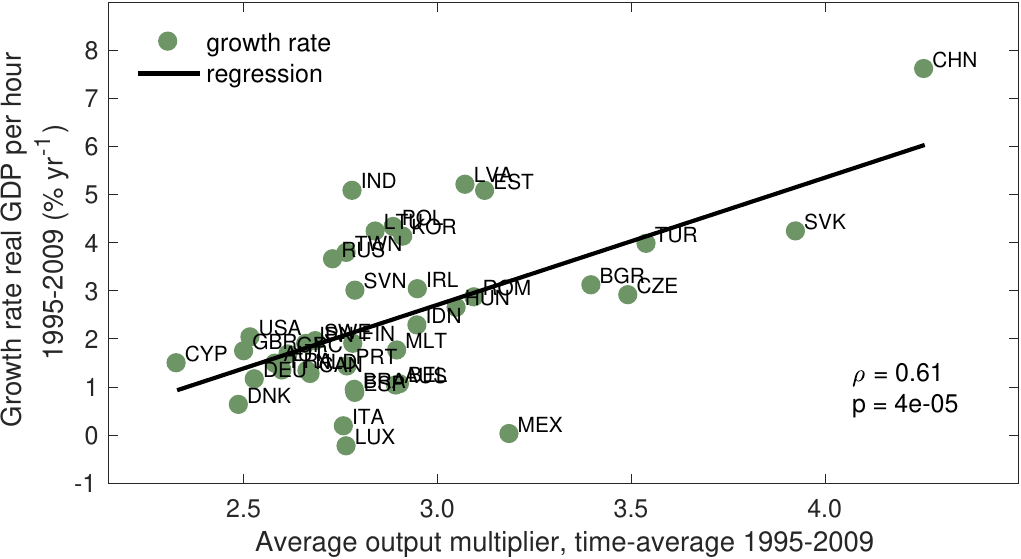}\\
{\sf\large D }
\includegraphics[height=0.17\textheight,align=t]{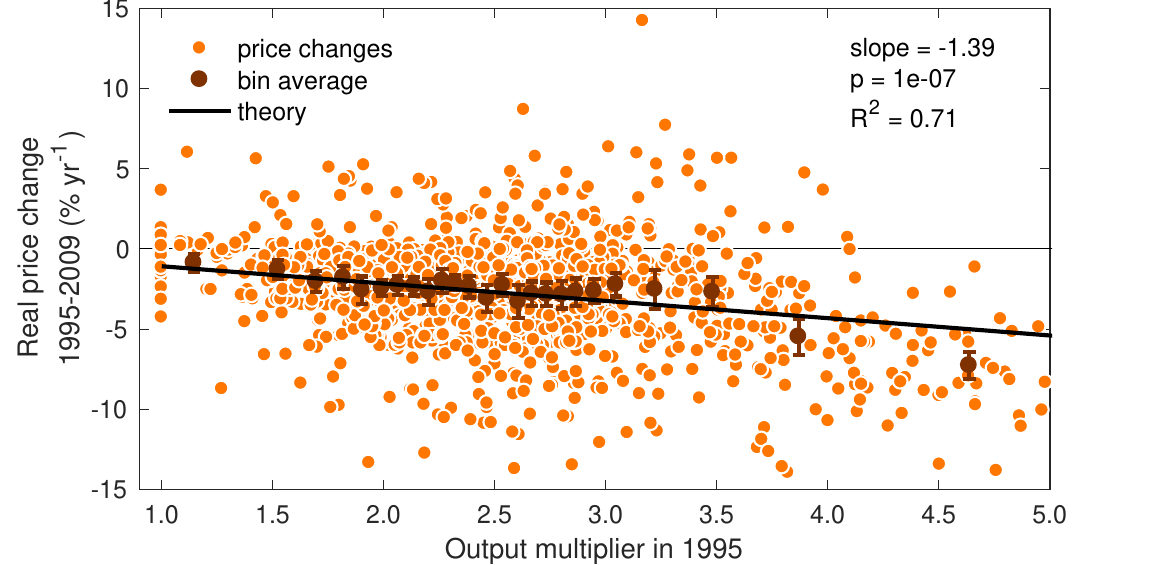}
\includegraphics[height=0.17\textheight,align=t]{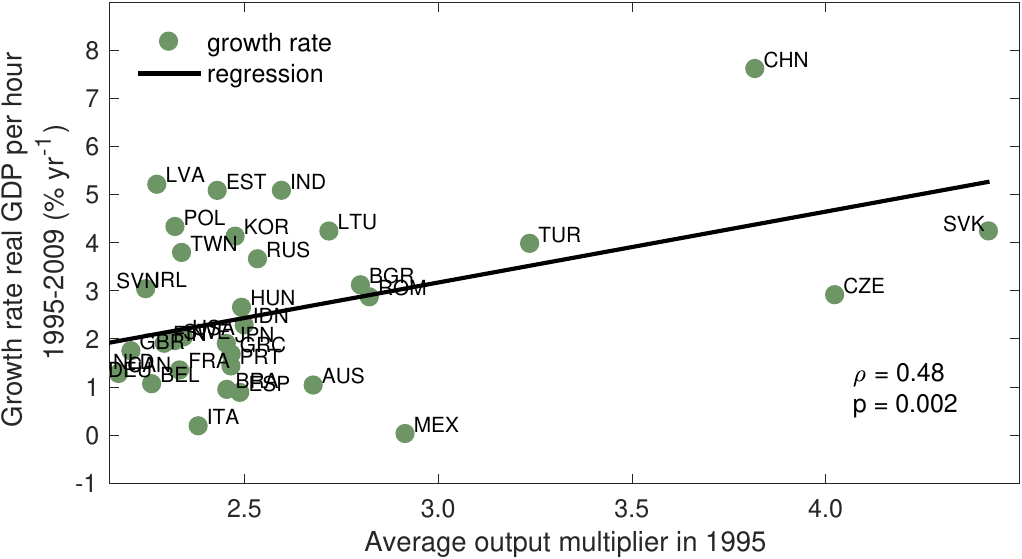}\\
\caption{\jmTwo{\textbf{Robustness to variations in prediction method.}} \jmTwo{(A) Price changes and growth rates using the assumptions described in the main text, where output multipliers in the initial year 1995 are plotted on the x-axis and either prices changes or growth rates from the period 1995-2009 are plotted on the y-axis.  These panels present the same results as Figs. 3A and 6A from the main text, and display the statistics quoted in the main text, i.e. for prices, the regression slope, $p$-value, and $R^2$ from regressing bin average price changes against output multipliers, and for growth rates the Pearson correlation $\rho$ and $p$-value.  Panels (B) - (D) show the results when (B) examining the alternate time period 1995-2005 that excludes the years of the Great Recession (2007-2009); (C) using time-averages of output multipliers instead of initial year values; and (D) computing output multipliers after artificially shutting down trade entries in the input-output matrix.}}
\label{fig_robustness}
\end{figure*}

\FloatBarrier
\jmTwo{We explore several variations in our empirical approach in order to see how they impact our results (Fig. \ref{fig_robustness}).  First, a number of results in the main text examine price changes or growth rates over an observation period that includes the years 2007-2009, a period that is exceptional because of the tumultuous productivity change that occurred due to the Great Recession.  We therefore explored the effect of changing our observation period to exclude these years.  We find that excluding these years has little effect.  For concreteness, we focus on the results of Figs. 3A and 6A from the main text, which plot industry price changes and country growth rates against the output multipliers.  (Note that some other figures, such as Fig. 4 of the main text, already exclude the period of the Great Recession, and show results in agreement with our theoretical predictions.)  The period of the Great Recession includes years of slower productivity growth overall, and we see in our data that productivity growth rates fell in the years 2007 and 2008.  (We use forward differences to define variables that capture changes between years.  Thus, the productivity growth rate we assign to e.g. 2007 refers to the change in productivity in our data between 2007 and 2008.)  Nevertheless, we find that including or excluding these years has a relatively small impact on the average rate of productivity growth, price change, or output growth over our observation period.  More importantly, although average rates of change will differ for any given observation period, our theoretical results indicate that we should continue to see a relationship between output multipliers on the one hand and price changes or growth rates on the other (potentially with different slopes), and we continue to see such a relationship when examining different observation windows that exclude the Great Recession.}

\jmTwo{In addition, as already noted in Section \ref{sec_persistence}, the relative persistence of output multipliers suggests that we should obtain similar results whether we compare price changes and growth rates against the time averages of output multipliers instead of the initial year values.  Making this change, we also find very similar results for both price changes and growth rates.}

\jmTwo{Finally,} the model here is developed assuming a closed economy, while the empirical analyses were carried out on economies that are open to trade.  As a check on the robustness of our results, we assess the effect of closing off countries to trade by zeroing out international trade entries in the matrix of intermediate payments.  This forces all subsequent computations to be done using only the input coefficients $a_{ij}$ derived from countries' domestic purchases.  Zeroing out trade payments breaks the balance of payments through nodes, but input coefficients can be computed in the usual way by dividing the resulting domestic input payments by industries' new total expenditures after the change.  Across the board, the results for tests presented in the main text and Supplementary Materials are similar.  The computed industry output multipliers and country average output multipliers are also similar.  This suggests that the results for a given country are driven primarily by the structure of production within the country, rather than by aspects of the global structure of trade.

\clearpage
\begin{table}
\center
\resizebox{0.95\textwidth}{!}{	
\begin{threeparttable}
\caption{\textbf{Cross-country average properties of industries from the WIOD dataset.}}
\label{tab_industry_table}
\begin{tabular}{lp{4in}>{$}r<{$}>{$}r<{$}>{$}r<{$}}											
\hline
\textbf{Code}	&	\textbf{Industry}	&	\textbf{Average $\boldsymbol{\mathcal{L}_i}^\dag$}				&	\textbf{Average $  \boldsymbol{{\gamma}_i}^\dag$} 				&	\textbf{Average  $\boldsymbol{r_i}^\dag$}				\\
	&		&					&	\text{(\% yr$^{-1}$)}				&	\text{(\% yr$^{-1}$)}				\\
\hline
Cok	&Coke, Refined Petroleum and Nuclear Fuel	&3.66 (\pm0.30)	&-1.05 (\pm4.79)	&0.32 (\pm4.95)\\
Tpt	&Transport Equipment	&3.63 (\pm0.53)	&1.93 (\pm2.31)	&-4.12 (\pm3.14)\\
Chm	&Chemicals and Chemical Products	&3.60 (\pm0.39)	&1.86 (\pm2.21)	&-3.77 (\pm2.95)\\
Elc	&Electrical and Optical Equipment	&3.55 (\pm0.49)	&2.66 (\pm2.30)	&-5.66 (\pm3.39)\\
Met	&Basic Metals and Fabricated Metal	&3.54 (\pm0.44)	&0.78 (\pm3.01)	&-2.35 (\pm4.04)\\
Rub	&Rubber and Plastics	&3.48 (\pm0.44)	&2.02 (\pm2.18)	&-4.19 (\pm3.06)\\
Ele	&Electricity, Gas and Water Supply	&3.47 (\pm0.48)	&0.22 (\pm2.28)	&-1.51 (\pm2.87)\\
Fod	&Food, Beverages and Tobacco	&3.44 (\pm0.29)	&0.27 (\pm1.85)	&-3.05 (\pm2.90)\\
Mch	&Machinery, Nec	&3.37 (\pm0.41)	&1.87 (\pm2.84)	&-3.84 (\pm3.41)\\
Ait	&Air Transport	&3.36 (\pm0.43)	&1.69 (\pm3.32)	&-3.59 (\pm4.02)\\
Wtt	&Water Transport	&3.33 (\pm0.33)	&0.81 (\pm2.32)	&-2.56 (\pm3.11)\\
Lth	&Leather, Leather and Footwear	&3.31 (\pm0.41)	&1.77 (\pm1.89)	&-3.94 (\pm2.81)\\
Tex	&Textiles and Textile Products	&3.30 (\pm0.40)	&2.07 (\pm1.73)	&-4.43 (\pm2.62)\\
Mnf	&Manufacturing, Nec; Recycling	&3.30 (\pm0.44)	&1.76 (\pm2.19)	&-3.77 (\pm3.24)\\
Pup	&Pulp, Paper, Paper , Printing and Publishing	&3.30 (\pm0.40)	&1.90 (\pm1.49)	&-4.23 (\pm2.71)\\
Est	&Real Estate Activities	&3.30 (\pm0.60)	&0.04 (\pm1.63)	&-1.59 (\pm1.92)\\
Wod	&Wood and Products of Wood and Cork	&3.29 (\pm0.38)	&1.05 (\pm2.31)	&-3.38 (\pm3.59)\\
Omn	&Other Non-Metallic Mineral	&3.29 (\pm0.37)	&1.77 (\pm1.51)	&-3.48 (\pm2.42)\\
Cst	&Construction	&3.19 (\pm0.46)	&-0.48 (\pm1.42)	&-1.23 (\pm2.25)\\
Otr	&Other Supporting and Auxiliary Transport Activities; Activities of Travel Agencies	&3.02 (\pm0.58)	&0.80 (\pm1.96)	&-2.36 (\pm2.35)\\
Min	&Mining and Quarrying	&3.01 (\pm0.36)	&-0.93 (\pm2.94)	&-0.38 (\pm3.24)\\
Pst	&Post and Telecommunications	&2.85 (\pm0.44)	&2.69 (\pm2.96)	&-4.77 (\pm3.17)\\
Ldt	&Inland Transport	&2.83 (\pm0.45)	&0.91 (\pm1.12)	&-2.33 (\pm1.85)\\
Htl	&Hotels and Restaurants	&2.83 (\pm0.39)	&0.27 (\pm1.48)	&-1.94 (\pm2.10)\\
Sal	&Sale, Maintenance and Repair of Motor Vehicles and Motorcycles; Retail Sale of Fuel	&2.68 (\pm0.47)	&1.23 (\pm2.08)	&-2.62 (\pm2.58)\\
Agr	&Agriculture, Hunting, Forestry and Fishing	&2.66 (\pm0.53)	&2.72 (\pm1.77)	&-4.49 (\pm2.73)\\
Whl	&Wholesale Trade and Commission Trade, Except of Motor Vehicles and Motorcycles	&2.66 (\pm0.44)	&1.38 (\pm1.32)	&-2.91 (\pm2.01)\\
Ocm	&Other Community, Social and Personal Services	&2.63 (\pm0.40)	&-0.11 (\pm1.41)	&-1.15 (\pm1.75)\\
Obs	&Renting of M\&Eq and Other Business Activities	&2.61 (\pm0.45)	&-0.03 (\pm1.23)	&-1.40 (\pm1.61)\\
Fin	&Financial Intermediation	&2.57 (\pm0.39)	&1.87 (\pm2.43)	&-3.38 (\pm3.23)\\
Rtl	&Retail Trade, Except of Motor Vehicles and Motorcycles; Repair of Household Goods	&2.43 (\pm0.43)	&1.54 (\pm1.30)	&-2.78 (\pm1.95)\\
Hth	&Health and Social Work	&2.29 (\pm0.46)	&-0.59 (\pm1.71)	&-0.51 (\pm1.83)\\
Pub	&Public Admin and Defence; Compulsory Social Security	&2.17 (\pm0.38)	&0.17 (\pm1.63)	&-1.18 (\pm1.90)\\
Edu	&Education	&1.75 (\pm0.38)	&-0.25 (\pm2.09)	&-0.45 (\pm2.13)\\
Pvt	&Private Households with Employed Persons	&1.07 (\pm0.27)	&0.57 (\pm1.57)	&-0.65 (\pm1.66)\\
\hline
\end{tabular}

\begin{tablenotes}
\item \dag~ Averages are over countries and the period 1995 - 2009.  Numbers in parentheses give standard deviations across countries.
\end{tablenotes}
\end{threeparttable}}
\end{table}

\begin{table}
\center
\resizebox{0.95\textwidth}{!}{	
\begin{threeparttable}
\caption{\textbf{Summary statistics for countries in the WIOD dataset for 1995 - 2009.}}
\label{tab_country_empirical}										
\begin{tabular}{llrrrr}															
\hline
\textbf{Code}	&	\textbf{Country}	&	\textbf{GDP per}	&	\textbf{Ave. growth per}	&	\textbf{Ave. improvement}	&	\textbf{Ave. output multiplier}\\
	&		&	\textbf{cap. in 1995}	&	\textbf{cap. (1995 - 2009)}	&	\textbf{rate $\boldsymbol{\tilde{\gamma}_c}$ (1995 - 2009)}	&	\textbf{$\boldsymbol{\bar{\L}_c}$ (1995 - 2009)}\\
	&	&(2011 PPP\$)	& (\% yr$^{-1}$)	&(\% yr$^{-1}$)	&	\\
\hline
AUS	&Australia	&30,347	&2.18	&0.12	&2.89\\
AUT	&Austria	&33,544	&1.65	&0.25	&2.58\\
BEL	&Belgium	&32,361	&1.51	&0.16	&2.90\\
BGR	&Bulgaria	&8,434	&4.19	&0.30	&3.40\\
BRA	&Brazil	&11,012	&1.47	&0.55	&2.79\\
CAN	&Canada	&32,100	&1.55	&0.71	&2.67\\
CHN	&China	&2,550	&8.65	&1.74	&4.26\\
CYP	&Cyprus	&26,444	&1.64	&0.74	&2.33\\
CZE	&Czech Republic	&19,093	&2.62	&2.30	&3.49\\
DEU	&Germany	&33,849	&1.01	&0.31	&2.53\\
DNK	&Denmark	&36,670	&1.05	&0.48	&2.49\\
ESP	&Spain	&25,630	&1.83	&-0.10	&2.79\\
EST	&Estonia	&11,068	&4.78	&2.81	&3.12\\
FIN	&Finland	&27,303	&2.45	&0.65	&2.78\\
FRA	&France	&30,822	&1.15	&0.62	&2.60\\
GBR	&United Kingdom	&28,513	&1.64	&0.92	&2.50\\
GRC	&Greece	&21,641	&2.57	&1.48	&2.62\\
HUN	&Hungary	&15,136	&2.65	&0.92	&3.05\\
IDN	&Indonesia	&6,022	&2.09	&0.40	&2.95\\
IND	&India	&2,058	&4.91	&1.83	&2.78\\
IRL	&Ireland	&26,002	&3.88	&1.04	&2.95\\
ITA	&Italy	&32,730	&0.53	&-0.06	&2.76\\
JPN	&Japan	&31,224	&0.37	&0.35	&2.66\\
KOR	&Korea, Republic of	&16,798	&3.83	&1.19	&2.91\\
LTU	&Lithuania	&9,229	&5.53	&2.62	&2.84\\
LUX	&Luxembourg	&64,018	&2.23	&-0.32	&2.77\\
LVA	&Latvia	&8,145	&5.78	&1.90	&3.07\\
MEX	&Mexico	&12,609	&1.16	&-0.41	&3.19\\
MLT	&Malta	&20,720	&1.86	&0.78	&2.90\\
NLD	&Netherlands	&35,005	&1.86	&0.42	&2.67\\
POL	&Poland	&11,149	&4.40	&0.17	&2.89\\
PRT	&Portugal	&21,974	&1.44	&0.30	&2.77\\
ROM	&Romania	&10,271	&3.76	&1.85	&3.10\\
RUS	&Russia	&12,012	&3.90	&1.74	&2.73\\
SVK	&Slovak Republic	&12,876	&4.25	&2.46	&3.93\\
SVN	&Slovenia	&18,244	&3.10	&1.04	&2.79\\
SWE	&Sweden	&31,044	&1.96	&0.68	&2.69\\
TUR	&Turkey	&11,530	&2.10	&2.24	&3.54\\
TWN	&Taiwan	&no data	&3.22	&1.32	&2.77\\
USA	&United States	&39,476	&1.48	&0.73	&2.52\\
RoW	&Rest of World	&9,139	&N.A.	&N.A.	&2.87\\
\hline
\end{tabular}

\begin{tablenotes}
\item The first column gives the ISO code of each country.  GDP per capita data is from the World Bank. \cite{WorldBank}
\end{tablenotes}
\end{threeparttable}}
\end{table}

\FloatBarrier

\FloatBarrier

\begin{table}\centering
\resizebox{\textwidth}{!}{
\begin{threeparttable}
\caption{Regressions of price changes (1995-2009) against output multipliers (1995) within industry categories.}
\label{tab_sameIndustryCrossCountriesVariation}

\begin{tabular}{p{0.7in}lr>{$}l<{$}l}
\textbf{Industry code}	&\textbf{Industry name}	&\textbf{Slope}	&\textbf{$p$-value}	&\textbf{}\\
\hline
Agr	&Agriculture, Hunting, Forestry and Fishing	&-2.21	&1.08 \times 10^{-3}	&**\\
Min	&Mining and Quarrying	&-0.28	&0.83	&\\
Fod	&Food, Beverages and Tobacco	&-3.66	&5.51 \times 10^{-4}	&***\\
Tex	&Textiles and Textile Products	&-2.94	&1.82 \times 10^{-3}	&**\\
Lth	&Leather, Leather and Footwear	&-2.85	&2.77 \times 10^{-3}	&**\\
Wod	&Wood and Products of Wood and Cork	&-4.61	&1.44 \times 10^{-4}	&***\\
Pup	&Pulp, Paper, Paper, Printing and Publishing	&-3.77	&1.21 \times 10^{-5}	&***\\
Cok	&Coke, Refined Petroleum and Nuclear Fuel	&-0.08	&0.98	&\\
Chm	&Chemicals and Chemical Products	&-3.36	&1.70 \times 10^{-3}	&**\\
Rub	&Rubber and Plastics	&-3.83	&1.84 \times 10^{-4}	&***\\
Omn	&Other Non-Metallic Mineral	&-4.41	&2.61 \times 10^{-8}	&***\\
Met	&Basic Metals and Fabricated Metal	&-2.29	&0.11	&\\
Mch	&Machinery, Nec	&-4.36	&3.08 \times 10^{-4}	&***\\
Elc	&Electrical and Optical Equipment	&-3.88	&6.68 \times 10^{-4}	&***\\
Tpt	&Transport Equipment	&-2.25	&9.43 \times 10^{-3}	&**\\
Mnf	&Manufacturing, Nec; Recycling	&-4.53	&4.78 \times 10^{-7}	&***\\
Ele	&Electricity, Gas and Water Supply	&-0.69	&0.41	&\\
Cst	&Construction	&-2.00	&8.19 \times 10^{-3}	&**\\
Sal	&Sale, Maintenance and Repair of Motor Vehicles and Motorcycles; Retail Sale of Fuel	&-1.78	&0.02	&*\\
Whl	&Wholesale Trade and Commission Trade, Except of Motor Vehicles and Motorcycles	&-1.78	&4.02 \times 10^{-3}	&**\\
Rtl	&Retail Trade, Except of Motor Vehicles and Motorcycles; Repair of Household Goods	&-2.08	&1.26 \times 10^{-4}	&***\\
Htl	&Hotels and Restaurants	&-2.27	&3.39 \times 10^{-3}	&**\\
Ldt	&Inland Transport	&-1.87	&1.02 \times 10^{-3}	&**\\
Wtt	&Water Transport	&-0.68	&0.57	&\\
Ait	&Air Transport	&-4.32	&6.18 \times 10^{-4}	&***\\
Otr	&Other Supporting and Auxiliary Transport Activities; Activities of Travel Agencies	&-0.59	&0.38	&\\
Pst	&Post and Telecommunications	&0.74	&0.42	&\\
Fin	&Financial Intermediation	&-0.65	&0.55	&\\
Est	&Real Estate Activities	&-1.04	&0.03	&*\\
Obs	&Renting of M\&Eq and Other Business Activities	&-1.02	&0.09	&\\
Pub	&Public Admin and Defence; Compulsory Social Security	&-1.54	&0.01	&*\\
Edu	&Education	&-2.30	&6.30 \times 10^{-3}	&**\\
Hth	&Health and Social Work	&-1.02	&0.08	&\\
Ocm	&Other Community, Social and Personal Services	&-1.34	&0.05	&\\
Pvt	&Private Households with Employed Persons	&-1.69	&0.07	&\\
\hline
\end{tabular}
\begin{tablenotes}
\item Significance levels: * = 0.05, ** = 0.01, *** = 0.001.
\end{tablenotes}
\end{threeparttable}}
\end{table}

\begin{table}\centering
\caption{Regression of price changes against industry labels and output multipliers. ($p$-values shown below each variable.)}
\label{tab_regressionIndustryDummies}

\begin{tabular}{l>{$}c<{$}>{$}c<{$}>{$}c<{$}}
	&\textbf{(1)}	&\textbf{(2)}	&\textbf{(3)}\\
\hline
Output multipliers	&	&-0.016	&-0.017\\
	&	&5.75 \times 10^{-42}	&1.20 \times 10^{-34}\\
Agriculture	&-0.038	&	&0.012\\
	&1.89 \times 10^{-26}	&	&1.90 \times 10^{-2}\\
Manufacturing	&-0.031	&	&0.022\\
	&9.91 \times 10^{-118}	&	&4.78 \times 10^{-7}\\
Services	&-0.022	&	&0.020\\
	&1.06 \times 10^{-64}	&	&8.54 \times 10^{-9}\\
Constant	&	&0.019	&\\
	&	&1.84 \times 10^{-8}	&\\
$R^2$	&0.030	&0.125	&0.131\\
$n$	&1435	&1435	&1435\\
\hline
\end{tabular}
\end{table}

\FloatBarrier

\bibliography{MasterBibliography.bib}
\bibliographystyle{naturemag}